\documentclass[12pt]{article}
\usepackage{rotating}%
\usepackage{amsmath} %
\usepackage{amsfonts} %
\usepackage{mathrsfs} %
\usepackage{fancyhdr} %
\usepackage{epsfig,morefloats}%
\usepackage{amssymb}%
\usepackage{color}%
\usepackage{setspace}
\usepackage{caption}
\usepackage{subfigure}
\usepackage{graphicx}
\usepackage[compact]{titlesec}
\usepackage[pagewise]{lineno}
\usepackage{cancel}
\usepackage{multirow}

\usepackage{setspace}

\usepackage[top=0.9in,bottom=0.9in,left=0.8in,right=0.8in]{geometry}
\usepackage{lscape}%
\usepackage{amsthm}%
\usepackage{enumerate}
\usepackage{threeparttable}
\usepackage{diagbox}
\usepackage{makecell}
\usepackage{authblk}
\usepackage{enumerate}
\usepackage{subfig}

\allowdisplaybreaks[4]
\usepackage[square,sort,comma,numbers,sectionbib,round]{natbib}
\def\be{\begin{equation}}
\def\ee{\end{equation}}
\def\ben{\begin{equation*}}
\def\een{\end{equation*}}
\def\bea{\begin{eqnarray}}
\def\eea{\end{eqnarray}}
\def\bd{\begin{displaymath}}
\def\ed{\end{displaymath}}
\def\bda{\begin{eqnarray*}}
\def\eda{\end{eqnarray*}}

\def\ha1{\hat \beta_1}

\def\bsc{\begin{scriptsize}}
\def\esc{\end{scriptsize}}

\def\bnt{\begin{enumerate}}
\def\ent{\end{enumerate}}
\def\T{{ \mathrm{\scriptscriptstyle T} }}
\def\p{{ \mathrm{p} }}
\def\c{{\mathrm{c}}}
\def\EL{{ \mathrm{\scriptscriptstyle EL} }}
\def\PEL{{ \mathrm{\scriptscriptstyle PEL} }}
\def\calT{{ \mathcal{\scriptscriptstyle (T)} }}
\def\calI{{ \mathcal{\scriptscriptstyle (I)} }}
\def\calD{{ \mathcal{\scriptscriptstyle (D)} }}
\def\calH{{ \mathcal{\scriptscriptstyle (H)} }}
\def\calA{{ \mathcal{\scriptscriptstyle (A)} }}

\def\calL{{ \mathcal{\scriptscriptstyle L} }}
\def\calF{{ \mathcal{\scriptscriptstyle F} }}
\def\cali{{ \mathcal{\scriptscriptstyle I} }}
\def\calR{{ \mathcal{\scriptscriptstyle R} }}
\def\cala{{ \mathcal{\scriptscriptstyle A} }}
\def\calM{{ \mathcal{\scriptscriptstyle M} }}
\def\cald{{ \mathcal{\scriptscriptstyle D} }}
\def\calp{{ \mathcal{\scriptscriptstyle P}}}
\def\calS{{ \mathcal{\scriptscriptstyle S} }}
\def\calG{{ \mathcal{\scriptscriptstyle G} }}
\def\calF{{ \mathcal{\scriptscriptstyle F} }}
\def\calt{{ \mathcal{\scriptscriptstyle T} }}
\def\calB{{ \mathcal{\scriptscriptstyle B} }}

\newtheorem{theorem}{Theorem}[section]

\newtheorem{lemma}{Lemma}[section]

\newtheorem{proposition}{Proposition}[section]

\theoremstyle{definition}
\newtheorem{as}{Condition}

\newtheorem{example}{Example}
\newtheorem{remark}{Remark}

\makeatletter
\newcommand{\figcaption}{\def\@captype{figure}\caption}
\newcommand{\tabcaption}{\def\@captype{table}\caption}
\makeatother

\newcommand{\sgn}{\mbox{\rm sgn}}

\newcommand{\supp}{\mathrm{supp}}

\newcommand{\bA}{{\mathbf A}}
\newcommand{\bB}{{\mathbf B}}

\newcommand{\bJ}{{\mathbf J}}
\newcommand{\bG}{{\mathbf G}}

\newcommand{\bI}{{\mathbf I}}

\newcommand{\bM}{{\mathbf M}}
\newcommand{\bQ}{{\mathbf Q}}

\newcommand{\bR}{{\mathbf R}}
\newcommand{\bS}{{\mathbf S}}
\newcommand{\bU}{{\mathbf U}}
\newcommand{\bV}{{\mathbf V}}
\newcommand{\bW}{{\mathbf W}}
\newcommand{\bX}{{\mathbf X}}

\newcommand{\ba}{{\mathbf a}}
\newcommand{\bb}{{\mathbf b}}

\newcommand{\bfe}{{\mathbf e}}
\newcommand{\bff}{{\mathbf f}}
\newcommand{\bh}{{\mathbf h}}
\newcommand{\bg}{{\mathbf g}}

\newcommand{\bm}{{\mathbf m}}

\newcommand{\bu}{{\mathbf u}}

\newcommand{\bw}{{\mathbf w}}

\newcommand{\bz}{{\mathbf z}}
\newcommand{\balpha} {\boldsymbol{\alpha}}
\newcommand{\bbeta}  {\boldsymbol{\beta}}
\newcommand{\bfeta}  {\boldsymbol{\eta}}

\newcommand{\bdelta} {\boldsymbol{\delta}}
\newcommand{\blambda}{\boldsymbol{\lambda}}
\newcommand{\bepsilonb}{\boldsymbol{\varepsilon}}

\newcommand{\bfvarsigma}{{\boldsymbol{\varsigma}}}

\newcommand{\bgamma}{\boldsymbol{\gamma}}

\newcommand{\bpi}{\boldsymbol{\pi}}

\newcommand{\bTheta} {\boldsymbol{\Theta}}

\newcommand{\bPsi} {\boldsymbol{\Psi}}
\newcommand{\bpsi}{\boldsymbol{\psi}}

\newcommand{\btheta} {\boldsymbol{\theta}}
\newcommand{\bxi} {\boldsymbol{\xi}}
\newcommand{\bXi} {\boldsymbol{\Xi}}

\newcommand{\bzeta} {\boldsymbol{\zeta}}
\newcommand{\bGamma} {\boldsymbol{\Gamma}}

\newcommand{\bC}{{\mathbf C}}
\newcommand{\bD}{{\mathbf D}}
\newcommand{\bzero}{{\mathbf 0}}

\newcommand{\bchi}{\boldsymbol{\chi}}

\allowdisplaybreaks[4]

\numberwithin{equation}{section}
\numberwithin{remark}{section}

\renewcommand{\thetable}{\arabic{table}}

\numberwithin{equation}{section}

\title{\bf Culling the Herd of Moments with Penalized Empirical Likelihood
} 
\author[a]{Jinyuan Chang}
\author[b,c]{Zhentao Shi}
\author[a]{Jia Zhang}
\affil[a]{\it \small  Joint Laboratory of Data Science and Business
Intelligence, Southwestern University of Finance and Economics, Chengdu, China
}
\affil[b]{\it \small Department of Economics, the Chinese University of Hong Kong, Sha Tin, New Territories, Hong Kong Special Administrative Region of China}
\affil[c]{\it \small
School of Economics, Georgia Institute of Technology, Atlanta, GA, U.S.A.
}

\date{}

\begin{document}

\bibliographystyle{agsm}
\bibpunct{(}{)}{,}{a}{}{;}

\maketitle
\begin{abstract}
Models defined by moment conditions are at the center of structural econometric estimation, but economic theory is mostly agnostic about moment selection. While a large pool of valid moments can potentially improve estimation efficiency, in the meantime a few invalid ones may undermine consistency. This paper investigates the empirical likelihood estimation of these moment-defined models in high-dimensional settings. We propose a penalized empirical likelihood (PEL) estimation and establish its oracle property with consistent detection of invalid moments. The PEL estimator is asymptotically normally distributed, and a projected PEL procedure further eliminates its asymptotic bias and provides more accurate normal approximation to the finite sample behavior. Simulation exercises demonstrate excellent numerical performance of these methods in estimation and inference.

\end{abstract}

\noindent%

\vfill

\thispagestyle{empty}

\newpage

\setcounter{page}{1}

\begin{spacing}{1.5}

\section{Introduction} \label{s1}

Economists' perennial pursuit of structural mechanisms leads to
models defined by moments.
These models can be written in a semiparametric form
$\mathbb{E}\{\bg (\bX_i;\btheta_0)\}=\bzero,$ where
$\bg = (g_j)_{j \in \{1,\ldots, r\} }$ is a vector of $r$ estimating functions,
$\btheta_0$ is a vector of unknown parameters,  and $\bX_i$ is observed data.
To estimate these models, the most popular method is
\emph{generalized method of moments}
(GMM) \citep{hansen1982large}.
\emph{Empirical likelihood} (EL) \citep{QinLawless1994} is a competitive alternative to GMM, thanks to its nice statistical properties.
Both GMM and EL are essential building blocks of modern econometrics
\citep{anatolyev2011methods}.

Ideally, economists count on economic theory to guide the choice of variables and moments.
However, the truth is that most economic theories are parsimonious abstractions and rarely pinpoint these choices in data-rich environments.
The indeterminacy of moment selection brings about three related issues.
The first is \emph{weak moments} \citep{stock2012survey}, which
threatens identification of the true value $\btheta_0$
when multiple $\btheta$'s satisfy $\mathbb{E}\{\bg (\bX_i;\btheta)\} \approx \bzero$.
Practitioners respond to the concerns of weak moments by adding more moment conditions in the hope
to strengthen identification, causing the second issue of
\emph{many moments} \citep{roodman2009note}.
The hazard of many moments is the possible inclusion of \emph{invalid moments}, meaning
 $\mathbb{E}\{g_j (\bX_i;\btheta_0)\}\neq 0$ for some $j\in \{1,\ldots,r\}$ \citep{murray2006avoiding},
 which is the third issue.
These cited survey papers highlight the unease incurred by the three challenges
and econometricians' efforts in coping with them.

When an underlying economic theory is ambivalent, empirical results based on it can be controversial
and susceptible to cherry-picking.
In such a circumstance,
of great importance are data-driven methods to guide and discipline moment selection.
In the low-dimensional settings,
\citet{andrews2001consistent} propose the GMM information criteria, and
\citet{hong2003generalized} follow with the counterpart for EL.
Information criteria are evaluated exhaustively at all
combinations of moments, and the computation becomes infeasible when
there are many potential moments.
To overcome this challenge,
\citet{liao2013adaptive} ushers
the adaptive Lasso shrinkage into GMM to select among
a finite number of moments, and \citet{ChengLiao2015}
further extend it to deal with a diverging number of moments and accommodate invalid ones.

Unprecedented progress in computation and information technology
fuel an arms race between the sheer size of the data and the scale
of empirical models.
In the era of big data,
on the one hand the cost of data collection and processing is tremendously lowered and
rich datasets open new perspectives to inspect a myriad of problems; on the other hand
economists attempt to build general models to capture various sources of heterogeneity
in observational data.
Empirical applications abound with models of many potential moments.
For instance, \citet{eaton2011} create 1360 moments to estimate a structural trade model,
\citet{altonji2013modeling} match 2429 moments implied by a model of earning dynamics,
and early works of linear instrumental variables (IV) models produce thousands of instruments by interacting variables  \citep{angrist1992effect}.
Although the sample sizes in these examples are non-trivial,
the proliferation of moments calls for a moment selection procedure capable
of handling high-dimensional moments
at a magnitude unrestricted by the sample size.
In particular, invalid ones that jeopardize consistency must be identified
and ``culled'' from the herd of moments.

The quadratic form of the usual GMM criterion function is incompatible with
high-dimensional moments \citep{shi2016econometric, shi2016estimation},
and thus \citet{belloni2018high}  regularize GMM with the sup-norm.
In this paper, we contribute
the most general and versatile procedure for high-dimensional nonlinear settings,
to the best of our knowledge.
We first develop a \emph{penalized empirical likelihood} (PEL) solution \citep{ChangTangWu2018} to deal with valid and invalid moments simultaneously.
We neutralize the invalid moments
by an auxiliary parameter, following \citet{liao2013adaptive} and \citet{ChengLiao2015}.
We establish the rate of convergence and asymptotic normality of the PEL estimator, and show
the efficiency gain from incorporating extra valid moments.
Under suitable conditions, it transpires that PEL enjoys the oracle property
of consistent moment selection and parameter selection.

The asymptotic normal distribution of the PEL estimator involves a bias term caused by the high-dimensional moments.
To spare the estimation of the bias term,
we can take further actions to project out the influence of the high-dimensional nuisance parameter in the PEL estimator,
which is called \emph{projected PEL} (PPEL) \citep{Chang2020}.
The asymptotic normality of the PPEL estimator is free of bias,
which facilitates statistical inference
of the structural parameter as well as the validity of moments.
Invoking statistical learning to assist our decisions, our method fits well in the recent trend of machine learning for the automatic selection of moments and variables.

Although this paper follows \citet{ChangTangWu2018} for estimation and \citet{Chang2020} for inference procedures,
the key insight lies in the observations that the high-dimensional auxiliary parameter, which signifies
the magnitude of misspecification, can be incorporated in the
EL method as an additional high-dimensional parameter.
In this paper, ``high-dimensional'' means that the numbers of parameters and/or moments are larger than the sample size, which goes beyond the scope of \citet{liao2013adaptive} and \citet{ChengLiao2015}.
On the other hand, in order to adapt \citet{ChangTangWu2018} and \citet{Chang2020} to accommodate
misspecified moments, we must deal with the distinctive roles of the main parameter of interest and
the auxiliary parameter in identification and the technical challenges induced by them. 
Differences from \citet{ChangTangWu2018} are highlighted in Section \ref{se:pel} 
about the rates of converges of the two components of the parameters, and those from \citet{Chang2020}
are elaborated in Section \ref{sec:infer} about the ways of confidence region construction.

\textbf{Literature review}.
Our paper stands on strands of literature,
which are too vast to survey exhaustively.
The accumulation of moments started from the linear IV model
\citep{angrist1990lifetime, angrist1991does}. The linear IV
model motivates theoretical research on issues of many IV
\citep{bekker1994alternative}, weak IV
(\citealp{stock2005testing}, \citealp{andrews2012estimation}),
many weak IV \citep{chao2011asymptotic, hansen2014instrumental},
invalid moments and many invalid IV
\citep{kolesar2015identification,windmeijer2019use}, to name a few.
In high-dimensional contexts, \citet{belloni2012sparse} use Lasso method for IV selection
in the first stage, and \citet{belloni2014inference} deal with post-selection inference.
Utilizing the linear structure, \citet{gold2020inference} and \citet{caner2018high}
provide inferential procedures for low-dimensional parameters
in models with high-dimensional endogenous variables and high-dimensional IVs.
Our method includes the linear IV model as a special case. In particular, the case of high-dimensional
structural parameters is elaborated in Section \ref{se:hpel}.

The proliferation of moments spreads from linear IV models to nonlinear
models. For example, in empirical industrial organization
researchers bring in moments from various resources,
some of which are guided by economic theory, to
mitigate the concerns of weak identification and improve estimation efficiency
\citep{ackerberg2007econometric}.
In empirical macroeconomics, identification failure and moment misspecification are
common issues \citep{mavroeidis2005identification}.
Under the GMM framework,
inference under weak moments (\citealp{stock2000gmm}, \citealp{kleibergen2005testing}, \citealp{andrews2020optimal}),
estimation under many weak moments \citep{han2006gmm}, and robust procedures for
invalid moments (\citealp{ditraglia2016using}, \citealp{caner2018adaptive}) have been developed.

EL's attractive theoretical properties are studied extensively
(\citealp{kitamura2001asymptotic}, \citealp{otsu2010bahadur}, \citealp{matsushita2013second}, \citealp{ChangChenChen2015}).
\citet{otsu2006generalized} and \citet{newey2009generalized} deal with its inference under weak IV,
and \citet{caner2015hybrid} select instruments in linear IV models.
Penalization schemes on EL have been introduced by \citet{otsu2007penalized}, \citet{tang2010penalized} and \citet{ChangTangWu2018}.

\textbf{Organization.} The rest of the paper is organized as follows. Section \ref{s2} introduces
the model and the analytic framework.
We first derive the asymptotic properties of our estimation in the low-dimensional case in Section 3,
extend them to the high-dimensional structural parameter in Section 4,
and we further refine PEL with projection to eliminate its bias in Section 5.
The theoretical results are supported in Section 6 by Monte Carlo simulations.
The influential study of the determinants of economic outcomes after colonialism is revisited
in Section 7. Section 8 concludes the paper. 
Due to the limitations of space, the proofs and the technical details of secondary importance are relegated into the supplementary materials.

\textbf{Notations.} We conclude Introduction with notations used throughout the paper.
``Low-dimensional'' is referred to the cases that the number of parameters or moments is much smaller than the sample size $n$, whereas ``high-dimensional'' goes the opposite.
For two sequences of positive numbers $\{a_n\}$ and $\{b_n\}$, we write $a_n\lesssim b_n$ or $b_n\gtrsim a_n$
if there exists a positive constant $c$ such that $\limsup_{n\rightarrow\infty}a_n/b_n\leq c$,
and write $a_n\ll b_n$ or $b_n\gg a_n$ if $\limsup_{n\rightarrow\infty}a_n/b_n=0$.

Denote by $1(\cdot)$ the indicator function. For a positive integer $q$, we write $[q]=\{1,\ldots,q\}$. For a $q\times q$ symmetric matrix $\bM$, denote by $\lambda_{\min}(\bM)$ and $\lambda_{\max}(\bM)$ the smallest and largest eigenvalues of $\bM$, respectively. 
For a $q_1\times q_2$ matrix $\bB=(b_{i,j})_{q_1\times q_2}$, let $\bB^\T$ be its transpose, $\bB^{\otimes2}=\bB\bB^\T$, $\bB^{\circ\kappa}=(|b_{i,j}|^\kappa)_{q_1\times q_2}$ for any $\kappa>0$, $|\bB|_\infty=\max_{i\in[q_1],j\in[q_2]}|b_{i,j}|$ be the sup-norm,
and $\|\bB\|_2=\lambda_{\max}^{1/2}(\bB^{\otimes2})$ be the spectral norm.
Specifically, if $q_2=1$, we use $|\bB|_\infty = \max_{i\in[q_1]}|b_{i,1}|$, $|\bB|_1=\sum_{i=1}^{q_1}|b_{i,1}|$ and $|\bB|_2=(\sum_{i=1}^{q_1}b_{i,1}^2)^{1/2}$ to denote the $L_{\infty}$-norm, $L_1$-norm and $L_2$-norm of the $q_1$-dimensional vector $\bB$, respectively.
For two square matrices $\bM_1$ and $\bM_2$, we say $\bM_1 \le \bM_2$ if $(\bM_2 - \bM_1)$ is a positive semi-definite matrix.

The population mean is denoted by $\mathbb{E}(\cdot)$, and the sample mean by $\mathbb{E}_n(\cdot)= n^{-1}\sum_{i=1}^n \{\cdot\}$.
For a given index set $\mathcal{L}$, let $| \mathcal{L} | $ be its cardinality.
For a generic multivariate function $\bh(\cdot;\cdot)$, we denote by $\bh_{\calL}(\cdot;\cdot)$ the subvector of $\bh(\cdot;\cdot)$ collecting the components indexed by $\mathcal{L}$.
Analogously, we write $\ba_{\calL}$ as the corresponding subvector of $\ba$.
For simplicity and when no confusion arises, we use the generic notation $\bh_i(\btheta)$ as the equivalence to $\bh(\bX_i;\btheta)$,
and $\nabla_{\btheta} \bh_i(\btheta) $ for the first-order partial derivative of $\bh_i(\btheta)$ with respect to $\btheta$.
Denote by $h_{i,k}(\btheta)$ the $k$-th component of $\bh_i(\btheta)$,
and by $\nabla^2_{\btheta}h_{i,k}(\btheta) $ the second derivative of $h_{i,k}(\btheta)$ with respect to $\btheta$.
Let $\bar \bh(\btheta)=\mathbb{E}_n\{\bh_i (\btheta)\}$,
and write its $k$-th component as $\bar h_k(\btheta)= \mathbb{E}_n\{h_{i,k}(\btheta)\}$.
Analogously, let $\bh_{i,\calL}(\btheta)=\bh_{\calL}(\bX_i;\btheta)$
and $\bar{\bh}_{\calL}(\btheta)= \mathbb{E}_n \{\bh_{i,\calL}(\btheta)\}$.

\section{Empirical likelihood with a herd of moments}
\label{s2}

In this section we introduce the model and the EL estimation. Let $\bX_1,\ldots,\bX_n$ be $d$-dimensional independent and identically distributed generic observations, and $\btheta=(\theta_1,\ldots,\theta_p)^\T$ be a $p$-dimensional parameter taking values in $\bTheta \subset \mathbb{R}^p$. For a set of $r_1$ estimating functions
    $
    \bg^{\calI}(\cdot;\cdot)=\{ g_{j}^\calI (\cdot; \cdot) \}_{j \in \mathcal{I}}
    $,
the information of the model parameter $\btheta$ is collected by the unbiased moment condition
    \begin{equation}\label{eq:esteq}
    \bzero = \mathbb{E}\{\bg^{\calI}(\bX_i;\btheta_0)\} = \mathbb{E}\{\bg_i^{\calI}(\btheta_0)\} 
    \end{equation}
at the unknown true parameter $\btheta_0\in\bTheta$, where $r_1\geq p$ is necessary for identifying $\btheta_0$.
The superscript $(\mathcal{I})$ labels this \emph{initial} set, to be distinguished from the other
set $(\mathcal{D})$ in Section \ref{sec:add}.

\subsection{EL estimation with valid moments}

Motivated from empirical applications in asset pricing,
the two-step GMM \citep{hansen1982large} was the default estimating method for the moment-defined model (\ref{eq:esteq}).
Intensive theoretical studies and numerical evidence in 1980's and 90's revealed some undesirable finite-sample properties of
the two-step GMM \citep{altonji1996small}.
EL and the continuously updating GMM (CUE) \citep{hansen1996finite} emerged as competitive solutions,
and they were later unified as members of \emph{generalized empirical likelihood} \citep{newey2003higher}.

This paper focuses on EL with estimating equations:
    \begin{equation*}\label{eq:el}
    L(\btheta)=\max\bigg\{\prod_{i=1}^n\pi_i:\pi_i>0\,,~\sum_{i=1}^n\pi_i=1\,,~\sum_{i=1}^n\pi_i\bg^{\calI}_i(\btheta)=\bzero\bigg\}\,,
    \end{equation*}
proposed by \cite{QinLawless1994}  based on the seminal idea of EL \citep{Owen1988,Owen1990}.
Maximizing $L(\btheta)$ with respect to $\btheta$ delivers the EL estimator $ \hat{\btheta}_{\EL}^{\calI}=\arg\max_{\btheta\in\bTheta}L(\btheta)$, which can be carried out equivalently by solving the corresponding dual problem
    \begin{equation}\label{eq:elestk}
    \hat{\btheta}_{\EL}^{\calI}=\arg\min_{\btheta\in\bTheta}\max_{\blambda\in\hat{\Lambda}_{n}^{\calI}(\btheta)}\sum_{i=1}^n\log\{1+\blambda^\T\bg^{\calI}_i(\btheta)\}\,,
    \end{equation}
where $\hat{\Lambda}_{n}^{\calI}(\btheta)=\{\blambda\in\mathbb{R}^{r_1}:\blambda^\T\bg^{\calI}_i(\btheta)\in\mathcal {V}~\textrm{for any}~i\in[n]\}$ and  $\mathcal{V}$ is an open interval containing zero.

To fix ideas, we first study the asymptotic property of $\hat{\btheta}_{\EL}^{\calI}$ under
regularity conditions.
When the sample size $n$ grows,  we adopt the asymptotic framework of  \cite{Hjort2009} and \cite{ChangChenChen2015} to take
the observations $\{\bg_i^{\calI}(\btheta)\}_{i=1}^n$ as a multi-index array, where $r_1$, $p$ and $d$ may depend on $n$.
Proposition A.1 in the supplementary materials shows that the standard asymptotic normality for
$\hat{\btheta}_{\EL}^{\calI}$ holds under
mild regularity conditions.

\subsection{Oracle EL estimation} \label{sec:add}

In applied econometrics, there has been a tendency of assembling many IVs or creating
many moments to identify the parameter of interest.
In these applications, researchers often have some
ideas about the relative importance of moments;
in the meantime, when researchers work with more and more moments,
some invalid ones may creep in.

\begin{example}
\citet{eaton2011} deem as the key moments 128($=2^7$) combinations of the largest 7 trade partners of France,
which are more important than the other 1232 moments.
\citet{angrist1990lifetime} treats the date of birth (DOB) as the key IV and it is complemented by the IVs generated by interactions;  recently \citet{kolesar2015identification} raise the potential invalidity
among these interaction terms.\footnote{Identification of the simple linear IV model requires the IVs satisfying the orthogonality condition and the relevance condition.
However, the more relevant an IV to the endogenous variables,
the more likely it is that the so-called ``IV'' is correlated with the structural
error, thereby violating orthogonality. There is a thin line between a valid IV and an invalid one.
}
\end{example}

To put into an analytic framework a herd of extra moments with unknown validity \emph{ex ante},
suppose that $r_2$ estimating functions
$\bg^{\calD}=\{g_{j}^\calD\}_{j\in \mathcal{D}}$ are partitioned into two groups:
\[
\mathcal{A}=\{j\in \mathcal{D}:\mathbb{E}\{g^\calD_{i,j}(\btheta_0)\}=0\}~~~\textrm{and}~~~
\mathcal{A}^{\c}=\{j\in \mathcal{D}:\mathbb{E}\{g^\calD_{i,j}(\btheta_0)\}\neq 0\}\,.
\]
Here the estimating functions in the set $\mathcal{A}$ are correctly specified, which can help improve the efficiency in estimating $\btheta_0$.
On the contrary, those in $\mathcal{A}^{\c}$ are misspecified, and they can only undermine the identification of $\btheta_0$.

If there is an ``oracle'' that reveals which components of $\bg^{\calD}$ belong to $\mathcal{A}$, i.e., the valid estimating functions, we can collect all valid estimating functions indexed by $\mathcal{H}:=\mathcal{I}\cup\mathcal{A}$ to estimate $\btheta_0$.
Denote $\bg^{\calH} =\{ \bg^{\calI,\T},\bg^{\calA,\T}\}^\T$
 and $h=|\mathcal{H}|$ as the total number of valid moments. The associated EL estimation for $\btheta_0$ based on the estimating functions $\bg^{\calH}$ is given by
\begin{equation*}\label{eq:elest11}
\hat{\btheta}_{\EL}^{\calH}=\arg\min_{\btheta\in\bTheta}\max_{\blambda\in\hat{\Lambda}_{n}^{\calH}(\btheta)}\sum_{i=1}^n\log\{1+\blambda^\T\bg^{\calH}_i(\btheta)\}\,,
\end{equation*}
where $\hat{\Lambda}_{n}^{\calH}(\btheta)=\{\blambda\in\mathbb{R}^h:\blambda^\T\bg^{\calH}_i(\btheta)\in\mathcal {V}~\textrm{for any}~i\in[n]\}$. 
By the same arguments as those in Proposition A.1 in the supplementary materials,
the following Proposition \ref{cy1} verifies the asymptotic normality for $\hat{\btheta}_{\EL}^{\calH}$.

\begin{proposition}[Oracle property]\label{cy1}
Assume that $\bg^{\calH} $ satisfies the conditions {\rm(A.1)--(A.5)} in the supplementary materials associated with $\bg^{\calH}$. If $h^{3}n^{-1+2/\gamma}=o(1)$ and $h^{3}p^2n^{-1}=o(1)$, then 
$
\sqrt{n}\balpha^\T\{\bJ^{\calH}\}^{1/2}\{\hat{\btheta}_{{\EL}}^{\calH}-\btheta_0\}\xrightarrow{d}\mathcal{N}(0,1)$  
as $n\rightarrow\infty$ for any $\balpha\in\mathbb{R}^p$ with $|\balpha|_2=1$,  where
$
\bJ^{\calH}=([\mathbb{E}\{\nabla_{\btheta}\bg^{\calH}_i(\btheta_0)\}]^\T\{\bV^{\calH}(\btheta_0)\}^{-1/2})^{\otimes2}
$
with $\bV^{\calH}(\btheta_0)=\mathbb{E}\{\bg^{\calH}_i(\btheta_0)^{\otimes2}\}$.
\end{proposition}

Obviously $\hat{\btheta}_{\EL}^{\calH}$ is more efficient than $\hat{\btheta}_{\EL}^{\calI}$
thanks for the additional moment restrictions in $\mathcal{A}$, which
disclose more information about the parameter and thus tie down the estimation variability.
This is parallel to \citet{hall2007information} in GMM for finite numbers of parameters and moments.
In reality, however, we are oblivious to $\mathcal{A}$, and therefore
we cannot blindly summon all $r_2$ estimating functions in $\bg^{\calD}$
to estimate $\btheta_0$.
Following \citet{liao2013adaptive}, we introduce an auxiliary parameter
$\bxi=\mathbb{E}\{\bg^{\calD}_i(\btheta)\}$ to neutralize the misspecified moments.
Denote the augmented parameter $\bpsi=(\btheta^\T,\bxi^\T)^\T$,
and the augmented parameter space $\bPsi = \bTheta\times\mathbf{\Upsilon}$.
Let $r=r_1+r_2$ and stack these $r$ estimating functions as
$\bg^{\calT}(\bX;\bpsi)=\{\bg^{\calI}(\bX;\btheta)^\T,\bg^{\calD}(\bX;\btheta)^\T-\bxi^\T\}^\T$.
Then $\bpsi_0=(\btheta_0^\T,\bxi_0^\T)^\T$ with $\bxi_0=\mathbb{E}\{\bg^{\calD}_i(\btheta_0)\}$ can be identified by
\begin{equation}\label{eq:esteq2}
\mathbb{E}\{\bg^{\calT}_i(\bpsi_0)\}
=
\mathbb{E}
\begin{Bmatrix}
\bg^{\calI}_i(\btheta_0) \\
\bg^{\calD}_i(\btheta_0)-\bxi_0
\end{Bmatrix}
=\bzero\, .
\end{equation}

Can we directly include the auxiliary parameter $\bxi$ into the EL estimation? The answer is negative. If the auxiliary parameter is not regularized,
the EL estimators with and without $\bxi$ are the same, up to numerical errors.
\begin{proposition}\label{prop11}
If $\hat{\btheta}_{\EL}^{\calI} $ is the unique solution of {\rm\eqref{eq:elestk}}, then $\hat{\btheta}_{\EL}^{\calT} = \hat{\btheta}_{\EL}^{\calI} $, where $\hat{\btheta}_{\EL}^{\calT}$ is the associated subvector of
$\hat{\bpsi}_{\EL}^{\calT} =\arg\min_{\bpsi\in \bPsi
    }\max_{\blambda\in\hat{\Lambda}_{n}^{\calT}(\bpsi)
    }\sum_{i=1}^n\log\{1+\blambda^\T\bg^{\calT}_i(\bpsi)\}$
for the estimation of $\btheta_0$ with $\hat{\Lambda}_{n}^{\calT}(\bpsi)=\{\blambda\in\mathbb{R}^r:\blambda^\T\bg^{\calT}_i(\bpsi)\in\mathcal {V}~\textrm{for any}~i\in[n]\}$.
\end{proposition}

Proposition \ref{prop11} states the equivalence between $\hat\btheta^{\calT}_{\EL}$ and $\hat\btheta^{\calI}_{\EL}$. In the next section, we will show that desirable efficiency as in 
the oracle estimator $\hat{\btheta}_{\EL}^{\calH}$
can be achieved if we extend \citet{liao2013adaptive}'s idea of shrinking the auxiliary parameter\footnote{
In GMM estimation under a fixed $r$,  \citet{liao2013adaptive} shrinks $\bxi$ toward zero using the adaptive Lasso \citep{zou2006adaptive}.
}
by further penalizing the Lagrange multipliers associated with the high-dimensional moments.

\section{High-dimensional moments with low-dimensional $\btheta$}\label{se:pel}

We consider in this section the model \eqref{eq:esteq2} with a low-dimensional parameter $\btheta$,
low-dimensional estimating functions $\bg^{\calI}$ and
high-dimensional estimating functions $\bg^{\calD}$. In this setting, $p$ and $r_1$ are either fixed or diverge at some slow polynomial rates of $n$,  whereas $r_2$ can grow much larger than $n$.
The components in the parameter $\btheta$ are indexed by $\mathcal{P}$.
All moments together are indexed by $\mathcal{T} = \mathcal{I} \cup \mathcal{D}$, in which
the researcher knows that those in $\mathcal{I}$ are correctly specified
in the sense that $\mathbb{E}\{ \bg_i^{\calI}(\btheta_0)\} = \bzero$,
whereas she is uncertain whether those in $\mathcal{D} = \mathcal{A}\cup \mathcal{A}^{\c} $
satisfy $\mathbb{E}\{ \bg_i^{\calD}(\btheta_0)\} = \bzero $ or not.
As there is a one-to-one relationship between $\bxi$ and the uncertain moments, we use $\mathcal{D}$ to index the components in $\bxi$ as well.

We propose the following PEL
to simultaneously estimate the unknown parameter $\btheta_0$ and determine the validity of the estimating functions in
$\bg^{\calD}$:
\begin{equation}\label{eq:est11}
  (\hat{\btheta}_{\PEL}^\T,\hat{\bxi}_{\PEL}^\T)^\T
 =
 \arg\min_{\bpsi\in \bPsi}\max_{\blambda\in\hat{\Lambda}_{n}^{\calT}(\bpsi)}\bigg[\frac{1}{n}\sum_{i=1}^n\log\{1+\blambda^\T\bg^{\calT}_i(\bpsi)\} -\sum_{j\in \mathcal{D} } P_{2,\nu}(|\lambda_j|)
 +\sum_{k\in \mathcal{D} }P_{1,\pi}(|\xi_k|)\bigg]\,,
\end{equation}
where $\hat{\Lambda}_{n}^{\calT}(\bpsi)=\{\blambda\in\mathbb{R}^{r}:\blambda^\T\bg^{\calT}_i(\bpsi)\in\mathcal {V}~\textrm{for any}~i\in[n]\}$, and 
$P_{1,\pi}(\cdot)$ and $P_{2,\nu}(\cdot)$ are two penalty functions with tuning parameters $\pi$ and $\nu$, respectively.
With the penalty function $P_{2,\nu}(\cdot)$ and appropriately selected tuning parameter $\nu$, the estimator $(\hat{\btheta}_{\PEL}^\T,\hat{\bxi}_{\PEL}^\T)^\T$ is associated with a sparse Lagrange multiplier $\blambda$. 
Since the sparse $\blambda$ invokes a subset of the estimating functions $\bg^{\calT}(\cdot;\cdot)$, 
it digests the high-dimensional moments as long as the number of nonzero components in $\blambda$ is small. 
On the other hand, the penalty $P_{1,\pi}(\cdot)$ is applied to identify which components in $\bg^{\calD}(\cdot;\cdot)$ are correctly specified and can estimate $(\xi_{0,k})_{k\in\mathcal{A}}$ exactly as $0$ with high probability. 

These penalties originally appeared in \citet{ChangTangWu2018} though,
there are several important differences. 
First, given the low-dimensional $\btheta$, it is unnecessary to assume sparsity on $\btheta_0$ for identification. 
As a result, our procedure here in \eqref{eq:est11} only penalizes $\bxi$ and $\blambda_{\mathcal{D}}$, not the entire parameter $\bpsi=(\btheta^\T,\bxi^\T)^\T$ and the associated Lagrange multiplier $\blambda$.  
Were $\blambda_{\mathcal{I}}$ penalized, we would rule out some components of $\bg^{\calI}(\cdot;\cdot)$
and thus may suffer loss of estimation efficiency for $\btheta_0$. 
Second, the identification condition invoked in this paper 
deviates from that in \citet{ChangTangWu2018}. Our Condition \ref{A.1} for identification is imposed on $\btheta_0$, 
rather than the whole parameter $\bpsi_0$. 
In contrast, \citet{ChangTangWu2018} specify identification condition for the parameter entity. 
While estimators in \citet{ChangTangWu2018} share the same rate of convergence, 
we must deal with the disparate rates of convergence of the main component $\hat{\btheta}_{\PEL}$ and the auxiliary $\hat{\bxi}_{\PEL}$, 
which significantly complicates the theoretical analysis of \eqref{eq:est11},
as detailed in Proposition \ref{thm1} below.

For any penalty function $P_\tau(\cdot)$ with a tuning parameter $\tau$, let $\rho(t;\tau)=\tau^{-1}P_\tau(t)$ for any $t\in[0,\infty)$ and $\tau\in(0,\infty)$. We assume that the two penalty functions $P_{1,\pi}(\cdot)$ and $P_{2,\nu}(\cdot)$ involved in \eqref{eq:est11} belong to the following class:
\begin{equation}\label{eq:classp}
\begin{split}
\mathscr{P}=\{P_\tau(\cdot):&~\rho(t;\tau)~\mbox{is increasing in}~t\in[0,\infty)~\mbox{and has continuous derivative}\\
&~\rho'(t;\tau)~\mbox{for any}~t\in(0,\infty)~\mbox{with} ~\rho'(0^+;\tau)\in(0,\infty),~\mbox{where}\\
&~\rho'(0^+;\tau)~\mbox{is independent of}~\tau\}\,.
\end{split}
\end{equation}
This class $\mathscr{P}$, considered in \cite{LvFan2009}, is broad and general. The commonly used $L_1$ penalty, SCAD penalty \citep{FanLi2001} and MCP penalty \citep{Zhang2010} are all included in $\mathscr{P}$. When $P_\tau(\cdot)\in\mathscr{P}$, we write the associated $\rho'(0^+;\tau)$ as $\rho'(0^+)$ for simplification.

To establish the limiting distribution of $\hat{\btheta}_{\PEL}$ in \eqref{eq:est11}, we assume the following regularity conditions.

\begin{as}\label{A.1}
There exists a universal constant $K_1>0$ such that
$$\inf_{\btheta\in\bTheta:\,|\btheta-\btheta_{0}|_\infty>\varepsilon}|\mathbb{E}\{\bg^{\calI}_i(\btheta)\}|_\infty\ge K_1 \varepsilon$$ for any $\varepsilon>0$. 
\end{as}

\begin{as}\label{A.2}
There exist universal constants $K_2>0$ and $\gamma>4$ such that
$$
\max_{j\in\mathcal{I}}\mathbb{E}\bigg\{\sup_{\btheta\in\bTheta}|g^{\calI}_{i,j}(\btheta)|^\gamma\bigg\}+\max_{j\in\mathcal{D}}\mathbb{E}\bigg\{\sup_{\btheta\in\bTheta}|g^{\calD}_{i,j}(\btheta)|^\gamma\bigg\}\le K_2\,.$$
\end{as}

\begin{as}\label{A.4}
For each $j\in \mathcal{T}$, the function $g^{\calT}_{j}(\bX;\bpsi)$ is twice continuously differentiable with respect to $\bpsi \in\bPsi$ for any $\bX$. For $\gamma$ specified in Condition \ref{A.2}, $$
\sup_{\bpsi\in\bPsi}
\bigg(
    |   \mathbb{E}_n[\{\nabla_{\bpsi}  \bg^{\calT}_{i}(\bpsi)\}^{\circ2}]
    |_{\infty} + \max_{j\in \mathcal{T}}
 | \mathbb{E}_n[\{\nabla^2_{\bpsi}g^{\calT}_{i,j}(\bpsi) \}^{\circ2}]
 |_{\infty} +
| \mathbb{E}_n[\{\bg^{\calT}_{i}(\bpsi)\}^{\circ\gamma}]|_{\infty}\bigg)
 =O_\p(1)\,.$$
There is some universal constant $K_3>0$ such that
$\sup_{\btheta\in\bTheta}
|\mathbb{E} \{ \nabla_{\btheta}   \bg_{i,\cala^{\c}}^\calD(\btheta) \} |_\infty \le K_3$.
\end{as}

For any index set $\mathcal{F}\subset\mathcal{T}$ and $\bpsi\in\bPsi$, define $
\bV_{\calF}^{\calT}(\bpsi)=\mathbb{E}\{\bg_{i,\calF}^{\calT}(\bpsi)^{\otimes2}\}$. When $\mathcal{F}=\mathcal{T}$, we write $\bV^{\calT}(\bpsi)=\bV_{\calt}^{\calT}(\bpsi)$ for conciseness.

\begin{as}\label{A.3}
There exists a universal constant $K_4>1$ such that $K_4^{-1}<\lambda_{\rm min}\{\bV^{\calT}(\bpsi_0)\}\le\lambda_{\rm max}\{\bV^{\calT}(\bpsi_0)\}<K_4$.
\end{as}

Conditions \ref{A.1}--\ref{A.3} are standard regularity assumptions in the literature.
Condition \ref{A.1} is an identification assumption of the estimating equations in the known set $\mathcal{I}$.
Write $\bxi_0= (\xi_{0,k})_{k\in\mathcal{D}}$, and we continue by defining
\begin{align}\label{eq:an}
a_n=\sum_{k\in\mathcal{D}}P_{1,\pi}(|\xi_{0,k}|)~~\textrm{and}~~
\phi_n=\max\{pa_n^{1/2},pr_1^{1/2}\aleph_n,\nu\}\,
\end{align}
in this section, where $\aleph_n=(n^{-1}\log r)^{1/2}$.
Suppose:
\begin{equation}\label{eq:xi1}
\mbox{\begin{minipage}[c]{.55\linewidth}
There exist $\chi_n\to0$ and $c_n\to0$ with $\phi_nc_n^{-1}\to0$ such that $
\max_{k\in\mathcal{A}^{\c}}\sup_{0<t<|\xi_{0,k}|+c_n}P'_{1,\pi}(t)=O(\chi_n)$
\end{minipage} }
\end{equation}
to control the bias induced by $P_{1,\pi}(\cdot)$ on $\hat{\bxi}_{\PEL}$.
With the assumption $\phi_n=o(\min_{k\in\mathcal{A}^{\c}}|\xi_{0,k}|)$ that
the nonzero components of $\bxi_0$ do not diminish to zero too fast, \eqref{eq:xi1} can be replaced by
\begin{align}\label{eq:xi}
\max_{k\in\mathcal{A}^{\c}}\sup_{c|\xi_{0,k}|<t<c^{-1}|\xi_{0,k}|}P'_{1,\pi}(t)=O(\chi_n)
\end{align}
for some constant $c\in(0,1)$. If we select $P_{1,\pi}(\cdot)$ as an asymptotically unbiased penalty such as SCAD or MCP, we have $\chi_n=0$ in \eqref{eq:xi} when
\begin{align}\label{eq:signal}
\min_{k\in\mathcal{A}^{\c}}|\xi_{0,k}|\gg\max\{\phi_n,\pi\}\,.
\end{align}
To simplify the presentation,  in this section we assume that \eqref{eq:signal} holds and
 $\chi_n=0$ in \eqref{eq:xi}.\footnote{
If \eqref{eq:signal} is violated,  the asymptotic normality of $\hat{\btheta}_{\PEL}$ in Theorem \ref{thm2} below will still hold under \eqref{eq:xi1} along with
more complicated notations to spell out the restrictions.
}

To allocate the parameter of interest $\btheta$ and those invalid moments in $\mathcal{A}^{\c}$,
we define an index set
$\mathcal{S}=\mathcal{P}\cup \mathcal{A}^{\c}$ and its cardinality $s=|\mathcal{S}|$.
For any $\bpsi \in\bPsi$, the set $\mathcal{S}$ picks out $\bpsi_{\calS}=(\btheta^\T,\bxi_{\cala^{\c}}^\T)^\T$.
Since  $\mathcal{A} = \mathcal{S}^{\c} = (\mathcal{P}\cup\mathcal{D}) \backslash \mathcal{S}$,
the corresponding auxiliary coefficients can be written as
 $\bpsi_{\calS^{\c}}=\bxi_{\cala}$ and furthermore
$\bpsi_{0,\calS^{\c}}=\bzero$  as $\mathcal{A}$ is the set of valid moments. Given some constant $C_*\in(0,1)$,
define $\mathcal{M}^*_{\bpsi}= \mathcal{I} \cup  \mathcal{D}^*_{\bpsi}$
with $\mathcal{D}^*_{\bpsi}=\{ j\in \mathcal{D}: |\bar{g}^{\calT}_{j}(\bpsi)|\ge C_*\nu\rho'_2(0^+)\}$.
We assume the existence of a sequence $\ell_n\to\infty$ such that
$$
\mathbb{P}\bigg(\max_{\bpsi\in\bPsi:\,|\bpsi_{\calS}-\bpsi_{0,\calS}|_\infty\le c_n,\,|\bpsi_{\calS^{\c}}|_1\leq \aleph_n }|\mathcal{M}^*_{\bpsi}| \le \ell_n\bigg) \to 1
$$
with some $c_n\gg\phi_n$.

Let $b_{1,n}=\max\{a_n,r_{1}\aleph_n^2\}$ and $b_{2,n}=\max\{b_{1,n},\nu^2 \}$. Then
$
\phi_n=\max\{pb_{1,n}^{1/2}, b_{2,n}^{1/2} \}$.
Define $\bPsi_*=\{\bpsi=(\bpsi_{\calS}^\T,\bpsi_{\calS^{\c}}^\T)^\T:|\bpsi_{\calS}-\bpsi_{0,\calS}|_\infty\le\varepsilon,|\bpsi_{\calS^{\c}}|_1\le \aleph_n \}$
for some fixed $\varepsilon>0$. Consider
\begin{align}\label{eq:locest}
\hat{\bpsi}=\arg\min_{\bpsi\in\bPsi_*}\max_{\blambda\in\hat{\Lambda}_{n}^{\calT}(\bpsi)}\bigg[\frac{1}{n}\sum_{i=1}^n\log\{1+\blambda^\T\bg^{\calT}_i(\bpsi)\} -\sum_{j\in \mathcal{D} } P_{2,\nu}(|\lambda_j|)
 +\sum_{k\in \mathcal{D} }P_{1,\pi}(|\xi_k|)\bigg]\,.
\end{align}
Proposition \ref{thm1} shows that such a $\hat{\bpsi}$ is a sparse local minimizer for \eqref{eq:est11}.

\begin{proposition} \label{thm1}
Let $P_{1,\pi}(\cdot),P_{2,\nu}(\cdot)\in\mathscr{P}$ for $\mathscr{P}$ defined in {\rm(\ref{eq:classp})}, and $P_{2,\nu}(\cdot)$ be convex with bounded second derivative around $0$. 
For $\hat\bpsi$ defined as {\rm\eqref{eq:locest}}, assume there exists a constant $\tilde{c}\in(C_*,1)$ such that $\mathbb{P}[\cup_{j\in\mathcal{T}}\{|\bar{g}_j^\calT(\hat\bpsi)|\in[\tilde{c}\nu\rho'_2(0^+),\nu\rho'_2(0^+))\}]\rightarrow0$.  Under Conditions {\rm \ref{A.1}--\ref{A.3}} and \eqref{eq:signal}, if $\log r=o(n^{1/3})$, $s^2\ell_n\phi_n^{2}=o(1)$, $b_{2,n}=o(n^{-2/\gamma})$, $\ell_n\aleph_n=o(\nu)$ and $\ell_n^{1/2}\aleph_n=o(\pi)$,
 then with probability approaching one this $\hat\bpsi=(\hat\btheta^\T,\hat\bxi^\T_{\cala^{\c}},\hat\bxi^\T_\cala)^\T$  provides a sparse local minimizer for the nonconvex optimization \eqref{eq:est11} such that {\rm(i)} $|\hat{\btheta}-\btheta_0|_\infty=O_\p(b_{1,n}^{1/2})$, {\rm(ii)} $|\hat{\bxi}_{\cala^{\c}}-\bxi_{0,\cala^{\c}}|_\infty=O_\p(\phi_n) $, and {\rm(iii)} $\mathbb{P}(\hat{\bxi}_{\cala}=\bzero)\to 1$ as $n\to\infty$.
\end{proposition}

 In the rest of this section, we focus on the sparse local minimizer $\hat{\bpsi}_{\PEL}=(\hat{\btheta}_{\PEL}^\T,\hat{\bxi}_{\PEL}^\T)^\T$ specified in \eqref{eq:locest}. We proceed with additional regularity conditions.

\begin{as}\label{A.7}
There exists a universal constant $K_5>1$ such that $K_5^{-1}<\lambda_{\rm min}\{\bQ_{\cali\cup\calB_1,\calB_2}^{\calT}\}\le\lambda_{\rm max}\{\bQ_{\cali\cup\calB_1,\calB_2}^{\calT}\}<K_5$ for any $\mathcal{B}_2\subset \mathcal{B}_1 \subset \mathcal{D}$ with $p+|\mathcal{B}_2| \le r_1+|\mathcal{B}_1| \le \ell_n$,
where $\bQ_{\cali\cup\calB_1,\calB_2}^{\calT}=([\mathbb{E}\{\nabla_{\bpsi_{\calp\cup\calB_2}}\bg_{i,\cali\cup\calB_1}^{\calT}(\bpsi_0)\}]^\T)^{\otimes2}$.
\end{as}
Condition \ref{A.7} is the sparse Riesz condition (\citealp{zhang2008sparsity}, \citealp{chen2008extended}) in our setting
to deal with the high-dimensional $\bpsi$ when $p+r_2>n$. Let $\hat\blambda(\hat\bpsi_\PEL)$ be the $r\times 1$ vector of
Lagrange multiplier defined at $\hat\bpsi_\PEL$:
\begin{align*}
\hat{\blambda}(\hat{\bpsi}_\PEL)=\arg\max_{\blambda\in\hat{\Lambda}_{n}^{\calT}(\hat{\bpsi}_{\PEL})}\bigg[\frac{1}{n}\sum_{i=1}^n\log\{1+\blambda^\T\bg^{\calT}_i(\hat{\bpsi}_\PEL)\}-\sum_{j\in\mathcal{D}}P_{2,\nu}(|\lambda_j|)\bigg]\,.
\end{align*}
Write $\hat{\blambda}:=\hat\blambda(\hat\bpsi_\PEL)=(\hat{\lambda}_1,\ldots,\hat{\lambda}_r)^\T$ and $\rho_2(t;\nu)=\nu^{-1}P_{2,\nu}(t)$. Let
$\mathcal{R}_n= \mathcal{I} \cup \{ j\in \mathcal{D}: \hat{\lambda}_j \neq 0 \} $
be the set of the estimated binding moments.
Since only those $\lambda_j$'s associated with $\mathcal{D}$ are penalized, it follows that
\[
\frac{1}{n}\sum_{i=1}^n\frac{g_{i,j}^{\calT}(\hat{\bpsi}_{\PEL})}{1+\hat{\blambda}_{\calR_n}^\T\bg_{i,\calR_n}^{\calT}(\hat{\bpsi}_{\PEL})}=\left\{\begin{aligned}
0\,,~~~~~~~~~~~~~~&\textrm{if}~j\in \mathcal{I} \,,\\
\nu\rho_2'(|\hat{\lambda}_j|;\nu)\sgn(\hat\lambda_j) \,,~~~&\textrm{if}~ j\in\mathcal{D} ~\textrm{with}~\hat{\lambda}_j\neq 0\,.
\end{aligned} \right.
\]
For any unbinding moment $j\in\mathcal{R}_n^{\c}$,
define
\begin{equation}\label{eq:bod}
\hat{\eta}_j := \frac{1}{n}\sum_{i=1}^n\frac{g_{i,j}^{\calT}(\hat{\bpsi}_{\PEL})}{1+\hat{\blambda}_{\calR_n}^\T\bg_{i,\calR_n}^{\calT}(\hat{\bpsi}_{\PEL})}\,.
\end{equation}
If $\hat{\lambda}_j=0$ for some $j\in\mathcal{D}$, the subdifferential at $\hat{\lambda}_j$ has to include the zero element \citep{bertsekas1997nonlinear}. That is, $\hat{\eta}_j\in[-\nu\rho_2'(0^+),\nu\rho_2'(0^+)]$ for any $j\in\mathcal{R}_n^{\c}$. In our theoretical analysis, we impose the next condition.
\begin{as}\label{AA1}
$
\mathbb{P}[\cup_{j\in\mathcal{R}_n^{\c}}\{|\hat{\eta}_j|=\nu\rho_2'(0^+)\}]\rightarrow0
$
as $n\rightarrow\infty$.
\end{as}

\begin{remark}
Condition \ref{AA1} requires that $\hat{\eta}_j$ $(j\in\mathcal{R}_n^{\c})$ does not lie on the boundary with probability approaching one, which is realistic in practice. If the distribution function of $\hat{\eta}_j$ is continuous at $\pm\nu\rho_2'(0^+)$,  we then have $\mathbb{P}\{|\hat{\eta}_j|=\nu\rho_2'(0^+)\}=0$. Condition \ref{AA1} makes sure that $\hat{\blambda}(\bpsi)$ is continuously differentiable at $\hat{\bpsi}_{\PEL}$ with probability approaching one. See Lemma A.5 in the supplementary materials.
\end{remark}

Define $\mathcal{A}_*=\{ j \in \mathcal{A}:\hat{\lambda}_j\neq 0\}$ and
$\mathcal{A}_{*,\c}=\{ j\in \mathcal{A}^{\c}:\hat{\lambda}_j\neq 0\}$.
Let $\mathcal{I}^*=\mathcal{I} \cup\mathcal{A}_*$,
and then  $\mathcal{R}_n$ can be decomposed into two disjoint parts $\mathcal{I}^*$
and $\mathcal{A}_{*,\c}$.
Furthermore, we define $\mathcal{S}_*= \mathcal{P} \cup  \mathcal{A}_{*,\c}$.
The fact $\mathcal{S}_*\subset\mathcal{S}$ implies that $|\mathcal{S}_*|=p+|\mathcal{A}_{*,\c}|\le |\mathcal{S}|=s$.
For any $\bpsi\in\bPsi$, we have $\bpsi_{\calS_*}=(\btheta^\T,\bxi^\T_{{\cala}_{*,\c}})^\T$. Define
\begin{align*}
\bJ^{\calT}_{\cali^*}&=\big([\mathbb{E}\{\nabla_{\btheta}\bg^{\calT}_{i,\cali^*}(\btheta_0)\}]^\T\{\bV^{\calT}_{\cali^*}(\btheta_0)\}^{-1/2}\big)^{\otimes2}
\end{align*}
with $\bV^{\calT}_{\cali^*}(\btheta_0)=\mathbb{E}\{\bg^\calT_{i,\cali^*}(\btheta_0)^{\otimes2}\}$,
and
$\hat\bzeta_{\calR_n}=\{\bJ^{\calT}_{\calR_n}\}^{-1}[\mathbb{E}\{\nabla_{\bpsi_{\calS_*}}\bg^{\calT}_{i,\calR_n}(\bpsi_0)\}]^\T\{\bV^{\calT}_{\calR_n}(\bpsi_0)\}^{-1}\hat\bfeta_{\calR_n}$, 
where $\bJ^{\calT}_{\calR_n}=([\mathbb{E}\{\nabla_{\bpsi_{\calS_*}}\bg^{\calT}_{i,\calR_n}(\bpsi_0)\}]^\T\{\bV^{\calT}_{\calR_n}(\bpsi_0)\}^{-1/2})^{\otimes2}$,
$\hat\bfeta=(\hat{\eta}_j)_{j\in \mathcal{T}}$ with $\hat\eta_j=0$
for $j\in \mathcal{I}$,
$\hat\eta_j=\nu\rho'_2(|\hat\lambda_j|;\nu){\rm sgn}(\hat\lambda_j)$ for $j\in \mathcal{D}$ with $\hat\lambda_j\neq0$, and $\hat\eta_j$ specified in \eqref{eq:bod} for $j\in\mathcal{R}_n^{\c}$. 
The limiting distribution of $\hat{\btheta}_{\PEL}$ is stated in Theorem \ref{thm2}.

\begin{theorem}\label{thm2}
Assume the conditions of Proposition {\rm\ref{thm1}} hold. Under Conditions {\rm \ref{A.7}} and {\rm\ref{AA1}}, if 
$\ell_n^{3/2}\log r=o(n^{1/2-1/\gamma})$ and
$\ell_nn^{1/2}s^{3/2} \phi_n \nu=o(1)$,
then
$
n^{1/2}\balpha^\T\{\bJ^{\calT}_{\cali^*}\}^{1/2}\{\hat\btheta_{{\PEL}}-\btheta_{0}
-\hat\bzeta_{\calR_n,(1)}\}\xrightarrow{d}\mathcal{N}(0,1)$
as $n\to\infty$ for any $\balpha\in\mathbb{R}^p$ with $|\balpha|_2=1$,  where $\hat\bzeta_{\calR_n,(1)}$ is the first $p$ components of $\hat\bzeta_{\calR_n}$ and $(a_n,\phi_n)$ are given in \eqref{eq:an}.
\end{theorem}

\begin{remark}\label{rek:3.4}
Notice that $a_n\lesssim s\pi$. To satisfy the restrictions in Theorem \ref{thm2},
it is sufficient for $(n,r,p,\ell_n,s)$ to have:  $r = o\{\exp(n^{1/3})\}$,
$\ell_n=o\{n^{(\gamma-2)/(3\gamma)}(\log r)^{-2/3}\}$,
$\ell_n^{5/2}s^{3/2}(\log r)\max\{p,\ell_n^{1/2}\}=o(n^{1/2})$,
the tuning parameters $\nu$ and $\pi$ satisfying
$
\ell_n^{1/2}\aleph_n \ll\pi\ll \min\{s^{-1}n^{-2/\gamma},\ell_n^{-4}s^{-4}p^{-2}(\log r)^{-1}\}$, $
\ell_n\aleph_n  \ll \nu \ll \min\{(\ell_n^3s^3p^2\log r)^{-1/2},n^{-1/\gamma},(\ell_n^2ns^3)^{-1/4} \}$ 
and $\nu^2\pi\ll(\ell_n^2ns^4p^2)^{-1}$.
\end{remark}

Thanks to the extra valid moments in $\bg^{\calD}$, the asymptotic covariance of $\hat\btheta_{\PEL}$ is smaller than that of $\hat{\btheta}_{\EL}^\calI$ which utilizes the information in $\bg^{\calI}$ only.
On the other hand,
given that $\hat{\btheta}_{\EL}^{\calI}$ provides asymptotic normality under
the set of correctly specified moments $\mathcal{I}$,
one may question whether it is worthwhile to consider all the available moments among which some may risk misspecification.
This is a trade-off between robustness and efficiency,
a recurrent theme in modern econometrics.
Although it inevitably depends on the data generating process (DGP),
in big data environments the efficiency gain may be substantial.
Benefits are demonstrated in the simulation exercises and the empirical application via very simple econometric models.\footnote{
See the RMSEs in Table \ref{tab5}, the length of confidence intervals in Figure \ref{fig2}, and the
standard deviations in Table \ref{tab:r1}.
}

\begin{remark}
The setting with the sets $\mathcal{I}$ and $\mathcal{D}= \mathcal{A}\cup \mathcal{A}^{\c}$ is the same as \citet{liao2013adaptive}.
When estimating (\ref{eq:esteq2}) with low-dimensional moments based on GMM, \cite{ChengLiao2015} require the number of estimating functions $r=o(n^{1/3})$,
and the signal of invalid estimating functions
$\min_{k\in\mathcal{A}^{\c}}|\mathbb{E}\{g_{i,k}^{\calD}(\btheta_0)\}|\gg (rn^{-1})^{1/2}$.
Our conditions on the relative size of the dimensions are more general. A sufficient condition for \eqref{eq:signal} is $\min_{k\in\mathcal{A}^{\c}}|\mathbb{E}\{g_{i,k}^{\calD}(\btheta_0)\}|\gg \max\{\nu,p(s\pi)^{1/2},pr_1^{1/2}\aleph_n\}$,
in view of $a_n\lesssim s\pi$.
Under the restrictions discussed in Remark \ref{rek:3.4},
if we select $\nu$ and $\pi$ sufficiently close to $\ell_n\aleph_n$ and $ \ell_n^{1/2}\aleph_n$,
respectively, then \eqref{eq:signal} holds provided that $\min_{k\in\mathcal{A}^{\c}}|\mathbb{E}\{g_{i,k}^{\calD}(\btheta_0)\}|\gg ps^{1/2}\ell_n^{1/4}\aleph_n^{1/2}$. This is similar to the ``beta-min condition'' which is necessary for consistent variable selection
by shrinkage methods \citep{buhlmann2013statistical}.
\end{remark}

\begin{remark}\label{rmk_thm2}
Under low-dimensional moments the asymptotic normality involves no bias;
see \citet[Theorem 3.5]{liao2013adaptive} and \citet[Theorem 3.3]{ChengLiao2015}.
In contrast, Theorem \ref{thm2} here makes clear that
the high-dimensional moments incur an
additional asymptotic bias $\hat \bzeta_{\calR_n,(1)}$.
To conduct hypothesis testing or construct confidence region
about  $\btheta$,
in principle we can estimate and correct the bias $\hat{\bzeta}_{\calR_n,(1)}$.
Such a direct bias correction approach, nevertheless, is undesirable in practice and in theory.
The bias term involves multiplication and inverse of large matrices,
which are difficult to estimate with precision in finite samples.
Illustrated in our simulation experiments in Section \ref{se:numst}, we are unsatisfied with the asymptotic normality approximation after estimating and correcting the bias.
Thus, we view Theorem \ref{thm2} as a characterization of the asymptotic behavior of
$\hat\btheta_{{\PEL}}$, but we do not encourage carrying out statistical inference based on it.
Instead, Section \ref{sec:infer} recommends PPEL for inference.
\end{remark}

While the augmented parameter $\bxi$ essentially validates all moments in $\mathcal{D}$,
the efficiency gain comes
from the penalty on $\bxi$ that shrinks some $\hat{\xi}_{k}$, $k \in \mathcal{A}$, to zero,  thereby confirming the validity of these associated moments.
To discuss moment selection, we write $\hat\bxi_{\PEL}=(\hat{\xi}_k)_{k\in \mathcal{D}}$.
If all valid estimating functions in $\mathcal{A}$ are selected by the optimization \eqref{eq:est11}, i.e., $\hat{\xi}_k = 0$ for all $k\in \mathcal{A}$,
then the asymptotic covariance of $\hat{\btheta}_{\PEL}$ is $\{\bJ^{\calH}\}^{-1}$.

Remind that $\{\bJ^{\calH}\}^{-1}$ in Proposition \ref{cy1} is the semiparametric efficiency bound for the estimation of $\btheta_0$ under the oracle, and
Proposition \ref{thm1} shows that $\mathbb{P}(\hat{\bxi}_{\PEL,\cala}=\bzero)\rightarrow1$ as $n\rightarrow\infty$, which provides a natural moment selection criterion
\begin{align}\label{eq:A}
\mathcal{\hat A}=\{k \in \mathcal{D}:\hat\xi_k=0\}\,
\end{align}
to identify the valid estimating functions in $\mathcal{A}$.
Notice that Proposition \ref{thm1} also indicates that $|\hat{\bxi}_{\PEL,\cala^{\c}}-\bxi_{0,\cala^{\c}}|_\infty=O_\p(\phi_n)$. Together with \eqref{eq:signal}, we have $\mathbb{P}[\cup_{k\in\mathcal{A}^{\c}}\{\hat{\xi}_k=0\}]\rightarrow0$ as $n\rightarrow\infty$.
Based on these arguments, Theorem \ref{thmoracle} supports our proposed moments selection criterion.

\begin{theorem}\label{thmoracle}
Under the conditions of Proposition {\rm\ref{thm1}}, it holds that $\mathbb{P}(\mathcal{\hat A}=\mathcal{A})\to1$ as $n\rightarrow\infty$.
\end{theorem}

Up to now, we have established the asymptotic properties of the PEL estimator for the moment-defined model (\ref{eq:esteq2}) with high-dimensional moments and low-dimensional $\btheta$.
The next section further extends the estimation to a high-dimensional structural parameter.

\section{High-dimensional moments with high-dimensional $\btheta$}\label{se:hpel}

A high-dimensional parameter $\btheta$ is present when many control variables are included in the structural model.
For example, the leading case of the linear IV model takes only one endogenous variable and it is accompanied by a few corresponding IVs. Nevertheless, such a simple setting may still involve many exogenous control variables in the main equation,
and these control variables are natural instruments for themselves.
An empirical example can be found in \citet{blundell1993we}. 
\citet{fan2014endogeneity} propose the \emph{focused GMM} for such a linear IV model, which is a special case of the moment-defined model.

In this section,  $p$ and $r_2$ are both allowed to be much larger than $n$,
whereas $r_1$ is fixed or diverges at some slow polynomial rate of $n$.
We assume $\btheta_0$ sparse in the sense that most of its components are zero.
Write $\btheta_0=(\theta_{0,l})_{l \in \mathcal{P}} $
and define the active set $\mathcal{P}_{\sharp}=\{l \in \mathcal{P}:\theta_{0,l}\neq0\}$ with cardinality $p_{\sharp}=|\mathcal{P}_{\sharp}|$.
To obtain a sparse estimate of $\btheta_0$, we add to \eqref{eq:est11} a penalty on $\btheta$ to produce the following optimization problem:
\begin{equation}\label{eq:est22}
\begin{split}
    (\hat{\btheta}_{{\PEL}}^\T,\hat{\bxi}_{{\PEL}}^\T)^\T = \arg\min_{\bpsi\in\bPsi}\max_{\blambda\in\hat{\Lambda}^{\calT}_{n}(\bpsi)}\bigg[\frac{1}{n}\sum_{i=1}^n\log\{1+\blambda^\T\bg^{\calT}_i(\bpsi)\}
    & -\sum_{j \in \mathcal{D} }  P_{2,\nu}(|\lambda_j|)\\
    +~  \sum_{k \in \mathcal{D} } P_{1,\pi}(|\xi_k|)
    & +\sum_{l \in \mathcal{P} } P_{1,\pi}(|\theta_l|)\bigg]\,.
\end{split}
    \end{equation}
With slight abuse of notation, we keep using $(\hat{\btheta}_{{\PEL}}^\T,\hat{\bxi}_{{\PEL}}^\T)^\T$ to denote the solution to \eqref{eq:est22}.

As $p$ can be bigger than $r_1$ in high dimension, 
here we update Condition \ref{A.1} for the identification of $\btheta_0$. 
In view of \citet{ChangTangWu2018}, we impose Condition \ref{A.1h} below for the identification of the nonzero components of $\btheta_0$.

\addtocounter{as}{-6}
\renewcommand{\theas}{\arabic{as}$'$}
\begin{as}\label{A.1h}
(i) There exists a universal constant $K'_1>0$ such that
$$\inf_{\btheta=(\btheta_{\calp_\sharp}^{\T},\btheta_{\calp_\sharp^{\c}}^{\T})^{\T}\in\bTheta:\,|\btheta_{\calp_\sharp}-\btheta_{0,\calp_\sharp}|_\infty>\varepsilon, \,\btheta_{\calp_\sharp^{\c}}=\bzero}|\mathbb{E}\{\bg^{\calI}_i(\btheta)\}|_\infty\ge K'_1 \varepsilon$$ for any $\varepsilon>0$. (ii) There exists a universal constant $K_3'>0$ such that $\sup_{\btheta\in\bTheta}|\mathbb{E}\{\nabla_{\btheta_{\calp_\sharp^{\c}}}\bg_i^{\calI}(\btheta)\}|_\infty\leq K_3'$. 
\end{as}
\renewcommand{\theas}{\arabic{as}}

As we have shown in Section \ref{se:pel}, the index set $\mathcal{S}$ and the quantities $a_n$ and $\phi_n$ play key roles in the theoretical analysis
of the low-dimensional $\btheta$ in \eqref{eq:est11}. Under the current high-dimensional setting, we define
\begin{align}
a_n=\sum_{k\in\mathcal{D}}P_{1,\pi}(|\xi_{0,k}|)+\sum_{l\in\mathcal{P}}P_{1,\pi}(|\theta_{0,l}|)~~\textrm{and}~~\mathcal{S}=\mathcal{P}_{\sharp}\cup\mathcal{A}^{\c}~~\textrm{with}~~s=|\mathcal{S}|\,.\label{eq:annew}\end{align}
This $\mathcal{S}$ indexes all nonzero components in the true augmented parameter $\bpsi_0$.
Compared to its counterpart $a_n$ in Section \ref{se:pel},
the newly defined $a_n$ here is amended with an extra term $\sum_{l\in\mathcal{P}}P_{1,\pi}(|\theta_{0,l}|)$ due to the penalty imposed on $\btheta$.
In the current high-dimensional setting, we also redefine
\begin{align}\label{eq:phinnew}
\phi_n=\max\{p_{\sharp} a_n^{1/2},p_{\sharp}r_1^{1/2}\aleph_n,\nu\}
\end{align}
with $a_n$ in \eqref{eq:annew}.
To control the bias introduced by $P_{1,\pi}(\cdot)$ on $\hat{\btheta}_{\PEL}$ and $\hat{\bxi}_{\PEL}$, similar to \eqref{eq:signal}, we assume that among the elements in $\bpsi_0=(\psi_{0,1},\ldots,\psi_{0,p+r_2})^\T$, the minimal active signal is
\begin{align}\label{eq:signal:2}
\min_{k\in\mathcal{S}}|\psi_{0,k}|\gg\max\{\phi_n,\pi\}\end{align}
with $\phi_n$ in \eqref{eq:phinnew} and $\max_{{k\in \mathcal{S}}}\sup_{c|\psi_{0,k}|<t<c^{-1}|\psi_{0,k}|}P'_{1,\pi}(t)=0$
for some constant $c\in(0,1)$.\footnote{See the arguments below \eqref{eq:xi} for the validity of these assumptions.}  
Given the newly defined $\mathcal{S}$ and $\phi_n$, we can further update $\ell_n$,
 $\mathcal{R}_n$, $\mathcal{A}_*$, $\mathcal{A}_{*,\c}$ and $\mathcal{I}^*$ in the same manner as
their counterparts in Section \ref{se:pel}.
Moreover, we also define $\bJ^{\calT}_{\calR_n}$
and $\hat\bzeta_{\calR_n}$ as specified in Section \ref{se:pel}
with $\mathcal{S}_*=\mathcal{P}_{\sharp}\cup \mathcal{A}_{*,\c}$. For any $\bpsi\in\bPsi$, we have $\bpsi_{\calS_*}=(\btheta_{\calp_{\sharp}}^\T,\bxi^\T_{{\cala}_{*,\c}})^\T$. 

Proposition A.2 in the supplementary materials shows that there exists a sparse local minimizer $(\hat\btheta_{\PEL}^\T,\hat\bxi_{\PEL}^\T)^\T\in\bPsi$ for the nonconvex optimization \eqref{eq:est22} such that $\mathbb{P}(\hat{\btheta}_{\PEL,\calp^{\c}_{\sharp}}=\bzero)\to1$ as $n\to\infty$, which means all zero components of $\btheta_0$ can be estimated  exactly as zero with high probability. The limiting distribution of such $\hat{\btheta}_{\PEL,\calp_{\sharp}}$ and the moment selection outcomes are stated in Theorem \ref{thm3} below.

\begin{theorem}\label{thm3}
Let $P_{1,\pi}(\cdot), P_{2,\nu}(\cdot)\in\mathscr{P}$ for $\mathscr{P}$ defined in \eqref{eq:classp},
and $P_{2,\nu}(\cdot)$ be convex with bounded second derivative around $0$. For the sparse local minimizer $\hat\bpsi_{{\PEL}}=(\hat{\btheta}_{{\PEL}}^\T,\hat{\bxi}_{{\PEL}}^\T)^\T$ for {\rm\eqref{eq:est22}} specified in Proposition {\rm A.2} in the supplementary materials, assume there exists a constant $\tilde{c}\in(C_*,1)$ such that $\mathbb{P}[\cup_{j\in\mathcal{T}}\{|\bar{g}_j^\calT(\hat\bpsi_{{\PEL}})|\in[\tilde{c}\nu\rho'_2(0^+),\nu\rho'_2(0^+))\}]\rightarrow0$.  
Suppose Conditions {\rm\ref{A.1h}}, {\rm \ref{A.2}--\ref{A.3}} and \eqref{eq:signal:2}  hold.
Furthermore, assume Condition {\rm\ref{A.7}} holds with the newly defined $\ell_n$,
replacing $\mathcal{P}$ by
$\mathcal{P}_{\sharp}$ and replacing $p$ by $p_{\sharp}$, and Condition {\rm\ref{AA1}} holds with the newly defined $\mathcal{R}_n$.
If $\log r=o(n^{1/3})$,  $\max\{a_n,\nu^2\}=o(n^{-2/\gamma})$, $\ell_n^{3/2}\log r=o(n^{1/2-1/\gamma})$, $\ell_nn^{1/2}s^{3/2} \phi_n \nu=o(1)$ and $\ell_n\aleph_n=o(\min\{\nu,\pi\})$, then we have
$$
    n^{1/2}\balpha^\T\{\bW^{\calT}_{\cali^*}\}^{1/2}\{\hat\btheta_{{\PEL},\calp_{\sharp}}-\btheta_{0,\calp_{\sharp}}-\hat\bzeta_{\calR_n,(1)}\}\xrightarrow{d}\mathcal{N}(0,1)
$$
for any $\balpha\in\mathbb{R}^{p_\sharp}$ with $|\balpha|_2=1$ as $n\to\infty$, where $\hat\bzeta_{\calR_n,(1)}$ is the first $p_{\sharp}$ components of $\hat\bzeta_{\calR_n}$,  $(a_n,\phi_n)$ are given by \eqref{eq:annew} and \eqref{eq:phinnew}, respectively, and $
    \bW^{\calT}_{\cali^*}=([\mathbb{E}\{\nabla_{\btheta_{\calp_{\sharp}}}\bg^{\calT}_{i,\cali^*}(\btheta_0)\}]^\T\{\bV^{\calT}_{\cali^*}(\btheta_0)\}^{-1/2})^{\otimes2}$ with $\bV^{\calT}_{\cali^*}(\btheta_0)=\mathbb{E}\{\bg^\calT_{i,\cali^*}(\btheta_0)^{\otimes2}\}$.
Moreover, under the conditions of Proposition {\rm A.2} in the supplementary materials, it holds that $\mathbb{P}(\mathcal{\hat A}=\mathcal{A})\to1$ as $n\rightarrow\infty$, where $\hat{\mathcal{A}}$ is specified in \eqref{eq:A}.
\end{theorem}

\begin{remark}
Instead of assuming $ \ell_n^{1/2} \aleph_n =o(\pi)$ as in Theorem \ref{thm2}, here Theorem \ref{thm3} strengthens $\ell_n  \aleph_n=o(\pi)$.
As shown in Section A.5 of the supplementary materials, this stronger condition guarantees that $\bpsi_{0,\calS^{\c}}$, the zero components of $\bpsi_0$, can be shrunk to zero with probability approaching one in the high-dimensional $\btheta$ setting.

\end{remark}

Compared to \eqref{eq:est11}, a penalty on $\btheta$ is added to \eqref{eq:est22}.
To appreciate the benefit from this additional penalty on the sparse parameter,
without loss of generality, we write $\btheta_{0}=(\btheta_{0,\calp_{\sharp}}^\T,\bzero^\T)^\T$ and
the inverse of $\bJ^{\calT}_{\cali^*}$ in Theorem \ref{thm2} in the following compatible partitioned matrix
\begin{align*}
\{\bJ^{\calT}_{\cali^*}\}^{-1}=\left(\begin{array}{cc}
~ [\{\bJ_{\cali^*}^{\calT}\}^{-1}]_{11}    &  [\{\bJ_{\cali^*}^{\calT}\}^{-1}]_{12} \\  ~[\{\bJ_{\cali^*}^{\calT}\}^{-1}]_{21}   &  [\{\bJ_{\cali^*}^{\calT}\}^{-1}]_{22}
\end{array}\right)\,,
\end{align*}
where $[\{\bJ_{\cali^*}^{\calT}\}^{-1}]_{11}$ is a $p_{\sharp}\times p_{\sharp}$ matrix.
The asymptotic covariance of $\hat{\btheta}_{\PEL,\calp_{\sharp}}$ by \eqref{eq:est11} is $[\{\bJ_{\cali^*}^{\calT}\}^{-1}]_{11}$,
and that in \eqref{eq:est22} is $\{\bW^{\calT}_{\cali^*}\}^{-1}$ according to Theorem \ref{thm3}.
Notice that
$\{\bW^{\calT}_{\cali^*}\}^{-1}=[\{\bJ_{\cali^*}^{\calT}\}^{-1}]_{11} -[\{\bJ_{\cali^*}^{\calT}\}^{-1}]_{12}[\{\bJ_{\cali^*}^{\calT}\}^{-1}]_{22}^{-1}[\{\bJ_{\cali^*}^{\calT}\}^{-1}]_{21} \le [\{\bJ_{\cali^*}^{\calT}\}^{-1}]_{11}$
by the inverse of a partitioned matrix.
Implicitly restricting the parameter space, the penalty on  $\btheta$ gains efficiency for the estimation of $\btheta_{0,\calp_{\sharp}}$, the nonzero components of $\btheta_0$.

\begin{remark}

Theorem \ref{thm3} spells out the limiting distribution and efficiency gain for the estimator of the nonzero components in $\btheta_0$.
If one is interested in some coefficient in $\mathcal{P}_{\sharp}^{\c}$,
we have consistency in that
$\mathbb{P}(\hat{\btheta}_{\PEL,\calp^{\c}_{\sharp}}=\bzero)\to1$ as $n\to\infty$, but the limiting distribution
is irregular due to the shrinkage.\footnote{This is a generic property shared by procedures of the oracle properties, for example 
SCAD \citep{FanLi2001} and the adaptive Lasso \citep{zou2006adaptive}.
}
Furthermore, parallel to Theorem \ref{thm2} the asymptotic bias is again present.
Similar to Remark \ref{rmk_thm2}, we do not suggest
conducting statistical inference predicated on this characterization of the asymptotic behavior of normality.
\end{remark}

Statistical inference is important in applied econometrics when researchers intend to assess
whether the estimated result supports or rejects a hypothesized value of the parameter.
The next section proposes an inferential procedure based on a projection of estimating functions, which is free of asymptotic biases.

\section{Confidence regions for a subset of $\bpsi$}
\label{sec:infer}

Suppose we are interested in the inference for a subset of parameters $\bpsi_{\calM}$ for some generic small subset
$\mathcal{M} \subset (\mathcal{P} \cup \mathcal{D}) $ with $|\mathcal{M}|=m$.
Here we allow $m$ to be fixed or diverge slowly with the sample size $n$.
What is novel here is that $\bpsi_{\calM}$ is allowed to contain part of the auxiliary parameter $\bxi$,
for which our method will provide a formal statistical inference for the validity of a subset of the
high-dimensional moment restrictions.
In contrast, \citet{liao2013adaptive} offers asymptotic normality for the structural parameter $\btheta$ under low-dimensional moments but does not characterize the asymptotic distribution of the auxiliary parameter $\bxi$.

A key technical issue is how to deal with the high-dimensional nuisance parameter $\bpsi_{\calM^\c}$,
where $\mathcal{M}^{\c}= (\mathcal{P} \cup \mathcal{D}) \backslash \mathcal{M}$.
Our approach follows \citet{Chang2020} to project out the influence of the
nuisance parameter. Such idea was also used in \citet{ning2017general} and \citet{neykov2018unified}.
Given an initial estimate $\bpsi^*$ for $\bpsi_0$, we first determine a linear transformation matrix $\bA_n = (\ba_k^n)_{k\in\mathcal{M}}^{\T} \in \mathbb{R}^{m\times r}$ with each row defined as
\begin{align}\label{eq:optim}
\ba_k^n = \arg\min_{\bu\in\mathbb{R}^r}|\bu|_1\quad \mathrm{s.t.} \quad |\{\nabla_{\bpsi}\bar\bg^{\calT}(\bpsi^*)\}^\T \bu - \bgamma_k|_\infty \le \varsigma 
\end{align}
for $k \in \mathcal{M}$,
where $\varsigma\rightarrow0$ as $n\rightarrow\infty$ is a tuning parameter,
and $\{\bgamma_k\}_{k\in \mathcal{M}}$ is a basis of the linear space
$\{
\bb=(b_j)_{j\in \mathcal{P}\cup\mathcal{D}  }:
\bb_{\mathcal{M}^{\c}} = \mathbf{0}
\}$.
In practice, we can specify the $(p+r_2)$-dimensional vector
$\bgamma_k = ( 1(j=k))_{j \in \mathcal{P} \cup \mathcal{D} }$
 with all zero elements except one unit entry.
As we will discuss in Remark \ref{initial_psi}, 
the solution from \eqref{eq:est11} or \eqref{eq:est22} can serve as the initial estimator $\bpsi^*$.

Based on $\bA_n$ as in (\ref{eq:optim}), we then obtain the new $m$-dimensional estimating functions
 $\bff^{\bA_n}(\cdot;\cdot)$ by projecting $\bg^{\calT}(\cdot;\cdot)$ on $\bA_n$:
\begin{align*}
\bff^{\bA_n}(\bX;\bpsi)  = \bA_n \bg^{\calT}(\bX;\bpsi) \,.
\end{align*}
Write $\bgamma_k=(\gamma_{k,j})_{j\in \mathcal{P}\cup\mathcal{D}}$ and
define an $m\times(p+r_2)$ matrix $\bGamma=(\gamma_{k,j})_{k\in \calM,
\, j \in \mathcal{P} \cup \mathcal{D}}$.
The definition of $\bA_n$ implies that $|\nabla_{\bpsi}\bar{\bff}^{\bA_n}(\bpsi^*)-\bGamma|_\infty\leq \varsigma$.
Since all the components in the $j$-th column of $\bGamma$ are zero
for $j\in\mathcal{M}^{\c}$, the newly defined estimating functions $\bff^{\bA_n}$ is uninformative of the nuisance parameter
 $\bpsi_{\calM^{\c}}$.  Informative is  $\bff^{\bA_n}$ of the parameter $\bpsi_{\calM}$ of interest,
as for $j\in\mathcal{M}$ some components in the $j$-th column of $\bGamma$ must be nonzero.

Given the projected estimating functions $\bff^{\bA_n}$, 
it is possible to directly borrow from \citet{Chang2020} to construct the confidence region of $\bpsi_{\calM}$ by the EL ratio
\[
w_n(\bpsi_{\calM})=2\max_{\blambda\in\tilde{\Lambda}_n(\bpsi_{\calM})}\sum_{i=1}^n\log\{1+\blambda^\T\bff_i^{\bA_n}(\bpsi_{\calM},\bpsi_{\calM^{\c}}^*)\}
\]
with respect to $\bpsi_{\calM}$, where $\tilde{\Lambda}_n(\bpsi_{\calM})=\{\blambda  \in\mathbb{R}^m : \blambda^\T \bff^{\bA_n}_i(\bpsi_{\calM},\bpsi_{\calM^{\c}}^*) \in \mathcal{V}~\textrm{for any}~i\in[n]\}$. 
Since $w_n(\bpsi_{0,\calM}) \xrightarrow{d}\chi_m^2$ as $n\rightarrow\infty$ for a fixed $m$,
the set $\{\bpsi_{\calM}\in\mathbb{R}^m:w_n(\bpsi_{\calM})\leq \chi_{m,1-\alpha}^2\}$ provides a $100(1-\alpha)\%$ confidence region  for $\bpsi_{\calM}$, where $\chi_{m,1-\alpha}^2$ is the $(1-\alpha)$-quantile of $\chi_m^2$ distribution. 
Nonetheless, the finite-sample performance  of such an asymptotically valid confidence region depends crucially on the convexity of $w_n(\bpsi_{\calM})$ near $\bpsi_{0,\calM}$. 
If convexity fails in a finite sample, this EL-ratio-based confidence region will be 
voluminous in magnitude due to numerical instability.
Verifying the convexity condition is onerous, in particular when $m$ is large, for $\bff_i^{\bA_n}(\bpsi_{\calM},\bpsi_{\calM^{\c}}^*)$ may be a nonlinear function of $\bpsi_{\calM}$. 

To secure a stable confidence region for $\bpsi_{\calM}$, in this paper we deviate from \citet{Chang2020} and instead
recommend a re-estimation procedure for a confidence region based on the asymptotic normality of the estimator. 
Let
\begin{align*}
\tilde{\bpsi}_{\calM} = \arg\min_{\bpsi_{\calM}\in\bPsi^*_{\calM}}\max_{\blambda\in\tilde\Lambda_n(\bpsi_{\calM})}\sum_{i=1}^n\log\{1+\blambda^\T \bff^{\bA_n}_i(\bpsi_{\calM},\bpsi_{\calM^{\c}}^*)\}\,,
\end{align*}
where $\bPsi^*_\calM = \{\bpsi_\calM: |\bpsi_\calM - \bpsi^*_\calM |_1 \le O_\p(\varpi_{1,n}) \}$ for some $\varpi_{1,n}\rightarrow0$ such that $|\bpsi^*_\calM-\bpsi_{0,\calM}|_1 = O_\p(\varpi_{1,n})$.
Since $\bpsi_\calM^*$ is an initial consistent estimator of $\bpsi_{0,\calM}$,
it is reasonable to search for $\tilde{\bpsi}_{\calM}$
in a small neighborhood of $\bpsi^*_\calM$.
We will specify $\varpi_{1,n}$ for some specific choices of $\bpsi_{\calM}^*$ in Remark \ref{initial_psi}.

To derive the limiting distribution of 
$\tilde\bpsi_\calM$, we assume the following condition.

\addtocounter{as}{5}
\begin{as}\label{A.10}
For each $k\in \mathcal{M}$, there is a non-random $\ba_k$ satisfying $[\mathbb{E}\{\nabla_{\bpsi}\bg^{\calT}_i(\bpsi_0)\}]^\T \ba_k = \bgamma_k$, $| \ba_k |_1\le K_6$  for some universal constant $K_6 > 0$, and $ \max_{k\in \calM }|\ba_k^n - \ba_k|_1 = O_\p(\omega_n)$ for some $\omega_n \to 0$. Let $\bA=(\ba_k)_{k\in\mathcal{M}}^{\T} \in \mathbb{R}^{m\times r}$. The eigenvalues of $\bA^{\otimes2}$ are uniformly bounded away from zero and infinity.
\end{as}

\begin{remark}
Let $\bXi=\mathbb{E}\{\nabla_{\bpsi}\bg^{\calT}_i(\bpsi_0)\}$ and $\widehat{\bXi}=\nabla_{\bpsi}\bar{\bg}^{\calT}(\bpsi^*)$. It follows from the existence of $\ba_k$ that
$\bgamma_k=\widehat{\bXi}^\T\ba_k+(\bXi-\widehat{\bXi})^\T\ba_k=\widehat{\bXi}^\T\ba_k+\bepsilonb_k$,
where $\bepsilonb_k = (\bXi-\widehat{\bXi})^\T\ba_k$.
Some mild conditions ensure $|\widehat{\bXi}-\bXi|_\infty=o_\p(1)$. This, together with the assumption $|\ba_k|_1\leq K_6$, implies that $\bepsilonb_k$ is stochastically small uniformly over all the components such that $|\bepsilonb_k|_\infty=o_\p(1)$,
which  can be viewed as an attempt to recover a nonrandom $\ba_k$ with no noise asymptotically (\citealp{CandesTao2007}, \citealp{Bickeletal2009}).
It follows that $|\ba_k^n-\ba_k|_1=o_\p(1)$ 
if $\widehat{\bXi}$ satisfies the routine conditions for sparse signal recovering.
Furthermore, the constant $K_6$ in Condition \ref{A.10} may be replaced by some diverging $\varphi_n$ and our main results remain valid.
\end{remark}

Theorem \ref{thm:clt} gives the limiting distribution of $\tilde{\bpsi}_{\calM}$
with a generic initial estimator $\bpsi^*$.

\begin{theorem}\label{thm:clt}
Let $|\bpsi_{\calM}^* - \bpsi_{0,\calM}|_1 = O_\p(\varpi_{1,n})$, $|\bpsi_{\calM^{\c}}^* - \bpsi_{0,\calM^{\c}}|_1 = O_\p(\varpi_{2,n})$ for some $\varpi_{1,n}\to0$ and $\varpi_{2,n}\to0$. Under Conditions {\rm\ref{A.2}}--{\rm\ref{A.3}} and {\rm\ref{A.10}}, if $m=o(n^{\min\{1/5,(\gamma-2)/(3\gamma)\}})$,
$m\omega_n^2(m^2+\log r)=o(1)$,
$m\varpi_{1,n}=o(1)$ and $n m \varpi_{2,n}^2 (\varsigma^2 + \varpi_{1,n}^2 + \varpi_{2,n}^2) = o(1)$, then 
$
n^{1/2}\balpha^\T(\hat\bJ^*)^{1/2}(\tilde{\bpsi}_{\calM} - \bpsi_{0,\calM})\xrightarrow{d} \mathcal{N}(0,1)$ as $n\to\infty$ 
for any $\balpha \in \mathbb{R}^m$ with $| \balpha |_2 = 1$,  where
$
\hat\bJ^* = [ \{ \nabla_{\bpsi_{\calM}} \bar{\bff}^{\bA_n}(\tilde\bpsi_{\calM},\bpsi_{\calM^{\c}}^*) \}^\T \{ \widehat{\bV}_{\bff^{\bA_n}}(\tilde\bpsi_{\calM},\bpsi_{\calM^{\c}}^*) \}^{-1/2} ]^{\otimes2}
$
with $\widehat{\bV}_{\bff^{\bA_n}}(\tilde\bpsi_{\calM},\bpsi_{\calM^{\c}}^*)
= \mathbb{E}_n\{\bff^{\bA_n}_i(\tilde\bpsi_{\calM},\bpsi_{\calM^{\c}}^*)^{\otimes2}\}$.
\end{theorem}

The above theorem is stated for $\tilde{\bpsi}_{\calM} $. It includes the estimation of $\btheta_{\calM}$ as a special case if one's research interest falls on the main parameter for economic interpretation, and makes it possible to infer the validity of a subset of moments in view of selecting $\bpsi_{\calM}=\bxi_{\calM}$.

\begin{remark}\label{initial_psi}
We verify that the PEL estimator is qualified to serve as $\bpsi^*$ in Theorem \ref{thm:clt}.
Define $\bar{s}=|\mathcal{S}\cap\mathcal{M}|$, and thus $|\mathcal{S}\cap\mathcal{M}^{\c}|=s-\bar{s}$.
If $\btheta_0$ is low-dimensional and we choose $\bpsi^*=\hat{\bpsi}_{\PEL}$ in \eqref{eq:est11},
we have $\varpi_{1,n}=\bar{s}\phi_n$ and $\varpi_{2,n}=(s-\bar{s})\phi_n$
due to $|\hat{\bpsi}_{\PEL,\calS}-\bpsi_{0,\calS}|_\infty=O_\p(\phi_n)$ and $\mathbb{P}(\hat{\bpsi}_{\PEL,\calS^{\c}}=\bzero)\rightarrow1$.
Then $ms\phi_n=o(1)$ and $nms^2\phi_n^{2}(\varsigma^2+s^2\phi_n^{2})=o(1)$ are sufficient for the restrictions imposed on $\varpi_{1,n}$ and $\varpi_{2,n}$.
Given $s$ and $\phi_n$ in Theorem \ref{thm2}, if $m$, $\omega_n$ and $\varsigma$ satisfy  $m=o(n^{\min\{1/5,(\gamma-2)/(3\gamma)\}})$, $m\omega_n^2(m^2+\log r)=o(1)$, $ms\phi_n=o(1)$ and  $nms^2\phi_n^{2}(\varsigma^2+s^2\phi_n^{2})=o(1)$, the asymptotic normality of Theorem \ref{thm:clt} holds
under this choice of the initial value $\bpsi^* = \hat{\bpsi}_{\PEL}$.
Analogously, if  $\bpsi^*=\hat{\bpsi}_{\PEL}$ in \eqref{eq:est22} when $\btheta_0$ is of high dimension, Theorem \ref{thm:clt} also holds provided that $m$, $\omega_n$ and $\varsigma$ satisfy the same restrictions with the newly defined $s$ and $\phi_n$  in Section \ref{se:hpel}.
\end{remark}


So far, we have established statistical inference results based on the PPEL estimator $\tilde{\bpsi}_{\calM}$ when we are interested in a small subset
$\mathcal{M}$ of $\bpsi$.
In the next section, we check the finite sample performance via simulations.

\section{Numerical studies}\label{se:numst}

One of the most important models in econometrics is the linear IV regression.
We design a linear IV model here to mimic the empirical application in Section \ref{emp_app}, whereas
simulation results of a nonlinear  panel regression with time-varying individual heterogeneity are presented in the supplementary materials.

Suppose that the researcher has at hand a dataset
of $n$ independent observations of a vector $\left(y_{i},x_{i},\bz_{i},w_{1i},\bw_{2i},\bw_{3i}\right)$, and is interested in estimating the main structural equation
\begin{equation}
y_{i}=\beta_{x}x_{i}+\bbeta_{\bz}^\T\bz_{i}+\epsilon_{i}\,,
\label{eq:main}
\end{equation}
where $x_{i}$ is a scalar endogenous variable, and
$\bz_{i}$ is a $d_{z}\times1$ vector of exogenous variables (including the intercept).
Such an equation with a scalar endogenous variable is the leading case of IV regressions \citep{andrews2019weak}.
Due to space limitations, we report the results when we specify
$\bz_i=(1,z_{1,i},z_{2,i})^\T\in\mathbb{R}^3$ with $( z_{1,i}, z_{2,i} )^\T \sim \mathcal{N}(\bzero,\rm{\bI}_2)$, and
the coefficients $ (\beta_x, \bbeta^\T_\bz)^\T=(0.5,0.5,0.5,0.5)^\T$ in a plausible setting.
The numerical performance are robust across our experiments when the parameters are varied.

The corresponding reduced-form equation is
$
x_{i}=\gamma_{w1}w_{1i}+\bgamma^\T_{\bw2}\bw_{2i}+\bgamma_{\bz}^\T\bz_{i}+u_{i}$, 
where $w_{1i}\in\mathbb{R}$ and $\bw_{2i}\in\mathbb{R}^{d_{\bw2}}$ are excluded instruments.
We specify
$ (w_{1i}, \bw_{2i}^\T)^\T \sim \mathcal{N}(\bzero,{\rm\bf I}_{d_{\bw2}+1})$,
$ \gamma_{w1}=0.8$, $\bgamma_{\bw2}=(\gamma_{\bw2,j})_{j\in[d_{\bw2}]}$ with $\gamma_{\bw2,j}=0.4-0.3(j-1)/(d_{\bw2}-1)$,
and $\bgamma_\bz=(0.8,0.8,0.8)^\T$.
The two error terms $(\epsilon_i,u_i)^{\T}$ are generated from the two-dimensional normal distribution with mean $0$, covariance $1$ and correlation $0.5$ which are independent of $\left(\bz_{i},w_{1i},\bw_{2i}\right)$.
An additional vector $\bw_{3i}=(w_{3i,j})_{j\in[s]}\in\mathbb{R}^{s}$, which serves as the invalid IV, follows
$
w_{3i,j}=\delta_j\epsilon_{i}+v_{i}$, 
where $v_{i}\sim \mathcal{N}\left(0,1\right)$ is
independent of $\left(\epsilon_{i},u_{i}\right)$, and $\delta_j \neq 0$ controls the strength of correlation.
Write $\bdelta=(\delta_j)_{j\in[s]}$. To emulate the scenarios of \emph{weak}, \emph{moderate} and \emph{strong} correlations between $\epsilon_i$ and $\bw_{3i}$, we set $\delta_j=0.3+0.2(j-1)/(s-1)$, $0.5+0.2(j-1)/(s-1)$ and $0.7+0.2(j-1)/(s-1)$, respectively.
It is expected that the smaller is the coefficient, the less severe is misspecification, so estimation is more prone to selection mistakes
in finite samples.

While the researcher is confident about the validity of
$w_{1i}$ in that
$
\bg_i^{\calI} (\btheta)=
(w_{1i},  \bz_{i}^\T)^\T
\times \left(y_{i}-\beta_{x}x_{i}-\bbeta_{\bz}^\T\bz_{i}\right),
$
where $\bz_i$ is self-instrumented,
she is uncertain about the validity of $(\bw_{2i}, \bw_{3i})$ and
therefore must detect the invalid IVs.
The two classes of undetermined (to the researcher) IVs consist of 
$
\bg_i^{\calD} (\btheta)=
(\bw_{2i}^\T, \bw_{3i}^\T)^\T
\times \left(y_{i}-\beta_{x}x_{i}-\bbeta_{\bz}^\T\bz_{i}\right).
$
According to our DGP design, $\bw_{2i}$ is the valid IV vector so the moments
involving  $\bw_{2i}$ equal to zero under the true parameter;
those moments involving $\bw_{3i}$ are invalid.

Let $d_w =d_{\bw2}+s+1$ be the number of all the IVs, among which
$s$ IVs are invalid. In the low-dimensional setting, we fix $s=6$ and
consider $(n,d_w)=(100,50)$ and $(200,100)$.
In the high-dimensional setting, we vary the number of invalid IVs to be
$s\in\{6,8,13\}$ for $(n,d_w)=(100,120)$, and $s\in\{6,12,17\}$ for $(n,d_w)=(200,240)$, where $s$ is specified by rounding, in addition, $2\log n$ and $3 n^{1/3}$.

In terms of the numerical implementation, we compute the PEL estimates by  the
\emph{modified two-layer coordinate descent algorithm} \citep{ChangTangWu2018}.
The SCAD penalty is used for both $P_{1,\pi}(\cdot)$ and $P_{2,\nu}(\cdot)$ in \eqref{eq:est11} for all the numerical experiments in this paper with local quadratic approximation \citep{FanLi2001}, and the tuning parameters $\nu$ and $\pi$ are chosen by the Bayesian information criterion (BIC).
Specifically, we use the following BIC type function:
\begin{align}\label{eq:BIC}
\mathrm{BIC} = \ell(\hat{\bpsi}_{\PEL}) + (\log n) \cdot \mathrm{df}(\hat{\bpsi}_{\PEL})\,,
\end{align}
where $\mathrm{df}(\hat{\bpsi}_{\PEL})$ denotes the number of nonzero elements in $\hat{\bpsi}_{\PEL}$, and the log likelihood term is the EL ratio
$
\ell(\hat{\bpsi}_{\PEL})=2\sum_{i=1}^n\log\{1+\hat\blambda(\hat{\bpsi}_{\PEL})^\T\bg^{\calT}_i(\hat{\bpsi}_{\PEL})\}$.\footnote{\cite{leng2012penalized} employ this BIC to choose the tuning parameter for the high-dimensional parameter $\bpsi$, and \cite{ChangTangWu2018} apply it to select multiple tuning parameters. We follow their practice as \eqref{eq:BIC} remains valid in our setting where two tuning parameters are included. }



\begin{table}[ht]
 \small
\centering \caption{ PEL's performance in moment selection} \label{tab1}
\begin{spacing}{1.4}
\begin{tabular}{llllllllll}  \hline
               & Correlation: & \multicolumn{2}{c}{weak}                        &                      & \multicolumn{2}{c}{moderate}                    &                      & \multicolumn{2}{c}{strong}                      \\ 
$(n, d_w, s)$     & Method                   & \multicolumn{1}{c}{FP} & \multicolumn{1}{c}{FN} & \multicolumn{1}{c}{} & \multicolumn{1}{c}{FP} & \multicolumn{1}{c}{FN} & \multicolumn{1}{c}{} & \multicolumn{1}{c}{FP} & \multicolumn{1}{c}{FN} \\ \hline
\multicolumn{10}{c}{Panel A: low-dimensional setting}                                                                                                                                                                                         \\
(100, 50, 6)   & PEL                      & 0.1995                 & 0.0267                 &                      & 0.1933                 & 0.0003                 &                      & 0.2359                 & 0.0000                 \\
               & DB-PEL                   & 0.1983                 & 0.0273                 &                      & 0.1923                 & 0.0003                 &                      & 0.2343                 & 0.0000                 \\
(200, 100, 6)  & PEL                      & 0.0942                 & 0.0033                 &                      & 0.1012                 & 0.0000                 &                      & 0.0971                 & 0.0000                 \\
         & DB-PEL                   & 0.0930                 & 0.0033                 &                      & 0.1003                 & 0.0000                 &                      & 0.0962                 & 0.0000                 \\ \hline
\multicolumn{10}{c}{Panel B: high-dimensional setting}                                                                                                                                                                                        \\ 
(100, 120, 6)  & PEL                      & 0.1067                 & 0.0550                 &                      & 0.0850                 & 0.0047                 &                      & 0.1141                 & 0.0013                 \\
               & DB-PEL                   & 0.1061                 & 0.0580                 &                      & 0.0845                 & 0.0050                 &                      & 0.1131                 & 0.0013                 \\
(200, 240, 6)  & PEL                      & 0.0478                 & 0.0073                 &                      & 0.0476                 & 0.0007                 &                      & 0.0444                 & 0.0007                 \\ 
         & DB-PEL                   & 0.0468                 & 0.0093                 &                      & 0.0464                 & 0.0007                 &                      & 0.0435                 & 0.0007                 \\ \hline
(100, 120, 8)  & PEL                      & 0.1009                 & 0.0675                 &                      & 0.0872                 & 0.0063                 &                      & 0.1087                 & 0.0010                 \\
               & DB-PEL                   & 0.1004                 & 0.0693                 &                      & 0.0863                 & 0.0065                 &                      & 0.1074                 & 0.0015                 \\
(200, 240, 12) & PEL                      & 0.0498                 & 0.0072                 &                      & 0.0497                 & 0.0003                 &                      & 0.0494                 & 0.0003                 \\ 
               & DB-PEL                   & 0.0486                 & 0.0087                 &                      & 0.0488                 & 0.0003                 &                      & 0.0485                 & 0.0003                 \\ \hline
(100, 120, 13) & PEL                      & 0.0709                 & 0.0892                 &                      & 0.0558                 & 0.0168                 &                      & 0.1288                 & 0.0042                 \\
               & DB-PEL                   & 0.0706                 & 0.0905                 &                      & 0.0554                 & 0.0175                 &                      & 0.1268                 & 0.0043                 \\
(200, 240, 17) & PEL                      & 0.0402                 & 0.0086                 &                      & 0.0469                 & 0.0002                 &                      & 0.0456                 & 0.0000                 \\         & DB-PEL                   & 0.0394                 & 0.0116                 &                      & 0.0458                 & 0.0002                 &                      & 0.0447                 & 0.0000                \\ \hline
\end{tabular}
\end{spacing}
\end{table}

We first report results of moment selection by our selection criterion in \eqref{eq:A}. Let ``FP'' (false positive) denote the frequency that the valid moments being not selected, and ``FN'' (false negative) denote the frequency that the invalid moments being selected.
In Table \ref{tab1}, PEL and DB-PEL denote the moment selection criterion based on the PEL estimator and its de-biased version, respectively.
As expected, the strength of correlations between $\epsilon_i$ and $\bw_{3i}$ does not affect FP,
whereas FN quickly vanishes as the correlations get stronger.
In all cases, larger sample sizes help reduce the chance of moment selection error.

\begin{table}[htbp]
\setlength\tabcolsep{3pt}
\small
\centering
\caption{Point estimations for $\beta_x$}\label{tab5}
\begin{spacing}{1.2}
\begin{tabular}{llrrrrrrrrrrr} \hline
               &Correlation: & \multicolumn{3}{c}{weak}                                                      &                      & \multicolumn{3}{c}{moderate}                                                  &                      & \multicolumn{3}{c}{strong}                                                    \\
$(n, d_w, s)$     & Method                   & \multicolumn{1}{c}{RMSE} & \multicolumn{1}{c}{BIAS} & \multicolumn{1}{c}{STD} & \multicolumn{1}{c}{} & \multicolumn{1}{c}{RMSE} & \multicolumn{1}{c}{BIAS} & \multicolumn{1}{c}{STD} & \multicolumn{1}{c}{} & \multicolumn{1}{c}{RMSE} & \multicolumn{1}{c}{BIAS} & \multicolumn{1}{c}{STD} \\ \hline
\multicolumn{13}{c}{Panel A: low-dimensional setting}                                                                                                                                                                                                                                                                                   \\
(100, 50, 6)              & Oracle                   & 0.074                    & 0.020                    & 0.071                   &                      & 0.074                    & 0.020                    & 0.071                   &                      & 0.074                    & 0.020                    & 0.071                   \\
               & PEL                      & 0.088                    & 0.030                    & 0.082                   &                      & 0.092                    & 0.032                    & 0.086                   &                      & 0.081                    & 0.019                    & 0.079                   \\
               & DB-PEL                   & 0.080                    & 0.032                    & 0.073                   &                      & 0.082                    & 0.035                    & 0.075                   &                      & 0.077                    & 0.026                    & 0.073                   \\
               & 2SLS                     & 0.145                    & -0.004                   & 0.145                   &                      & 0.145                    & -0.004                   & 0.145                   &                      & 0.145                    & -0.004                   & 0.145                   \\
(200, 100, 6)  & Oracle                   & 0.039                    & 0.010                    & 0.038                   &                      & 0.039                    & 0.010                    & 0.038                   &                      & 0.039                    & 0.010                    & 0.038                   \\ & PEL                      & 0.044                    & 0.013                    & 0.042                   &                      & 0.047                    & 0.013                    & 0.045                   &                      & 0.045                    & 0.013                    & 0.044                   \\
               & DB-PEL                   & 0.042                    & 0.021                    & 0.036                   &                      & 0.042                    & 0.020                    & 0.037                   &                      & 0.042                    & 0.020                    & 0.037                   \\
         & 2SLS                     & 0.101                    & -0.003                   & 0.101                   &                      & 0.101                    & -0.003                   & 0.101                   &                      & 0.101                    & -0.003                   & 0.101                   \\ \hline
\multicolumn{13}{c}{Panel B: high-dimensional setting}                                                                                                                                                                                                                                                                                  \\
(100, 120, 6)  & Oracle                   & 0.067                    & 0.018                    & 0.064                   &                      & 0.067                    & 0.018                    & 0.064                   &                      & 0.067                    & 0.018                    & 0.064                   \\& PEL                      & 0.090                    & 0.031                    & 0.084                   &                      & 0.098                    & 0.040                    & 0.090                   &                      & 0.092                    & 0.033                    & 0.087                   \\
               & DB-PEL                   & 0.083                    & 0.029                    & 0.078                   &                      & 0.082                    & 0.031                    & 0.076                   &                      & 0.077                    & 0.026                    & 0.073                   \\
               & 2SLS                     & 0.181                    & -0.016                   & 0.180                   &                      & 0.181                    & -0.016                   & 0.180                   &                      & 0.181                    & -0.016                   & 0.180                   \\
(200, 240, 6)  & Oracle                   & 0.035                    & 0.012                    & 0.033                   &                      & 0.035                    & 0.012                    & 0.033                   &                      & 0.035                    & 0.012                    & 0.033                   \\& PEL                      & 0.045                    & 0.014                    & 0.043                   &                      & 0.044                    & 0.014                    & 0.042                   &                      & 0.037                    & 0.014                    & 0.034                   \\
               & DB-PEL                   & 0.043                    & 0.017                    & 0.040                   &                      & 0.042                    & 0.017                    & 0.039                   &                      & 0.035                    & 0.016                    & 0.031                   \\
        & 2SLS                     & 0.128                    & 0.002                    & 0.128                   &                      & 0.128                    & 0.002                    & 0.128                   &                      & 0.128                    & 0.002                    & 0.128                   \\ \hline
(100, 120, 8)  & Oracle                   & 0.063                    & 0.015                    & 0.061                   &                      & 0.063                    & 0.015                    & 0.061                   &                      & 0.063                    & 0.015                    & 0.061                   \\& PEL                      & 0.110                    & 0.036                    & 0.104                   &                      & 0.116                    & 0.045                    & 0.107                   &                      & 0.112                    & 0.039                    & 0.105                   \\
               & DB-PEL                   & 0.100                    & 0.033                    & 0.094                   &                      & 0.098                    & 0.038                    & 0.090                   &                      & 0.092                    & 0.032                    & 0.087                   \\
               & 2SLS                     & 0.211                    & -0.011                   & 0.211                   &                      & 0.211                    & -0.011                   & 0.211                   &                      & 0.211                    & -0.011                   & 0.211                   \\
(200, 240, 12) & Oracle                   & 0.034                    & 0.010                    & 0.033                   &                      & 0.034                    & 0.010                    & 0.033                   &                      & 0.034                    & 0.010                    & 0.033                   \\& PEL                      & 0.038                    & 0.013                    & 0.036                   &                      & 0.036                    & 0.013                    & 0.033                   &                      & 0.036                    & 0.013                    & 0.033                   \\
               & DB-PEL                   & 0.039                    & 0.017                    & 0.035                   &                      & 0.038                    & 0.016                    & 0.034                   &                      & 0.038                    & 0.016                    & 0.034                   \\
          & 2SLS                     & 0.126                    & -0.005                   & 0.126                   &                      & 0.126                    & -0.005                   & 0.126                   &                      & 0.126                    & -0.005                   & 0.126                   \\ \hline
(100, 120, 13) & Oracle                   & 0.065                    & 0.020                    & 0.062                   &                      & 0.065                    & 0.020                    & 0.062                   &                      & 0.065                    & 0.020                    & 0.062                   \\& PEL                      & 0.117                    & 0.051                    & 0.105                   &                      & 0.139                    & 0.069                    & 0.121                   &                      & 0.110                    & 0.036                    & 0.104                   \\
               & DB-PEL                   & 0.103                    & 0.045                    & 0.092                   &                      & 0.113                    & 0.054                    & 0.099                   &                      & 0.093                    & 0.033                    & 0.087                   \\
               & 2SLS                     & 0.200                    & -0.008                   & 0.200                   &                      & 0.200                    & -0.008                   & 0.200                   &                      & 0.200                    & -0.008                   & 0.200                   \\
(200, 240, 17) & Oracle                   & 0.036                    & 0.012                    & 0.034                   &                      & 0.036                    & 0.012                    & 0.034                   &                      & 0.036                    & 0.012                    & 0.034                  \\ & PEL                      & 0.047                    & 0.013                    & 0.045                   &                      & 0.052                    & 0.011                    & 0.051                   &                      & 0.041                    & 0.011                    & 0.039                   \\
               & DB-PEL                   & 0.044                    & 0.016                    & 0.041                   &                      & 0.053                    & 0.015                    & 0.051                   &                      & 0.039                    & 0.015                    & 0.036                   \\
        & 2SLS                     & 0.133                    & -0.014                   & 0.132                   &                      & 0.133                    & -0.014                   & 0.132                   &                      & 0.133                    & -0.014                   & 0.132         \\ \hline         
\end{tabular}

\end{spacing}
\end{table}

The parameter of interest in the linear IV model is $\beta_{x}$ in \eqref{eq:main} as it characterizes the ``causal effect'' of the endogenous variable, which bears economic interpretation. The root-mean-square error (RMSE), bias (BIAS) and standard deviation (STD) are calculated for
PEL, DB-PEL and the classical two-stage least squares (2SLS)  for \eqref{eq:main}, 
and an oracle estimator is added for comparison.\footnote{
The oracle EL in Section \ref{sec:add} is for low-dimensional parameters and moments. 
To handle the large pool of orthogonal IVs, 
the oracle estimator here is a 2SLS estimator taking advantage of a few most relevant IVs---those with the top $0.1n$ big coefficients $\gamma_{\mathbf{w}2,j}$ in the reduced-form equation.
}
The results are summarized in Table \ref{tab5}.
The performance of PEL and the oracle is comparable,
and the gaps are narrowed when the sample size is increased from $n=100$ to $n=200$, suggesting the capacity for PEL to mimic the oracle by absorbing the signal from the valid moments and in the 
meantime keeping the invalid ones at bay. 
The RMSE of 2SLS is significantly larger than those of PEL and DB-PEL.

\begin{figure}[htbp]
\small
  \centering
  \includegraphics[width=0.7\textwidth]{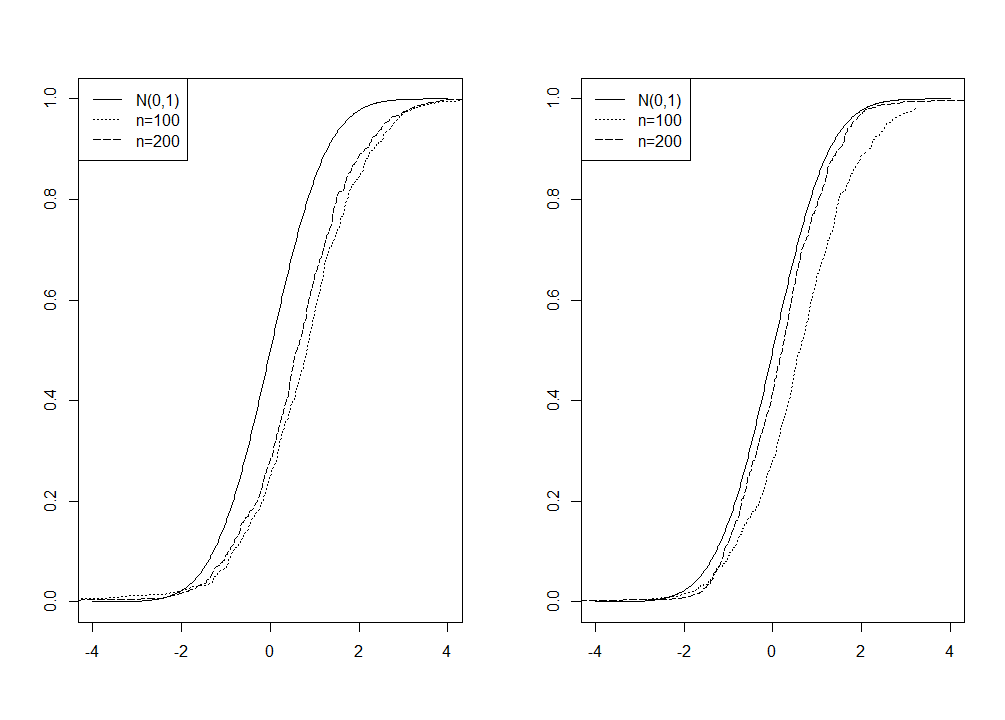}
  \caption{ ECDF of DB-PEL (left) and PPEL (right) of $\beta_x$ with moderate correlation between $\epsilon_i$ and $\bw_{3i}$}
  \label{fig1}

  \includegraphics[width=0.7\textwidth]{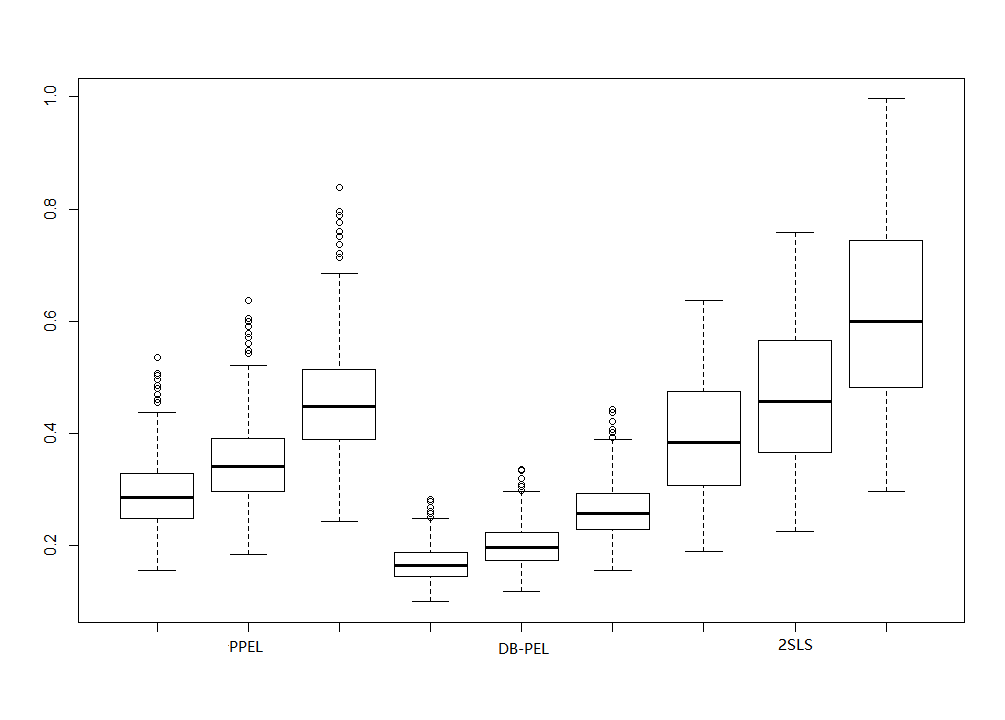}
  \caption{Width of $90\%$ (left), $95\%$ (middle), and $99\%$ (right) CI for $\beta_x$ estimated by PPEL, DB-PEL and 2SLS under moderate correlation between $\epsilon_i$ and $\bw_{3i}$ with $(n,d_w,s)=(100,120,8)$}
  \label{fig2}
\end{figure}

The left panel of Figure \ref{fig1} plots the empirical cumulative distribution functions (ECDF) of the DB-PEL estimates. The dotted curve corresponds to the case of $(n,d_w,s)=(100, 120, 8)$, the dashed curve to
that of $(n,d_w,s)=(200,240,12)$, and the solid curve is the cumulative distribution function (CDF) of $\mathcal{N}(0,1)$
for comparison.
A better normal approximation can be obtained by the PPEL estimate, as shown in the right panel of
Figure \ref{fig1}, with its tuning parameter $\varsigma=0.08(n^{-1}\log p)^{1/2}$.
We also present the confidence intervals (CI) according to PPEL, DB-PEL and 2SLS, as reported in Table \ref{tab9},
for the 90\%, 95\% and 99\% levels, where the CIs for DB-PEL and PPEL are predicated on the asymptotic normality from Theorems \ref{thm2} and \ref{thm:clt}, respectively, and the CIs for 2SLS are based on its asymptotic normality as in standard textbooks.
PPEL's coverage probability to the nominal counterpart
is the best among the three estimators, and is much better than PEL.  2SLS's
coverage probability seems too high in the high-dimensional setting, though it is acceptable in the low-dimensional case.
Figure \ref{fig2} shows that the width of 2SLS's CI is much wider than that of PPEL, which is caused by efficiency loss
from abandoning the potentially valid estimating equations.

\begin{table}[htbp]
\setlength\tabcolsep{3pt}
\small
\centering
\caption{Coverage probabilities for the CIs of $\beta_x$}
\label{tab9}
\begin{spacing}{1.4}

\begin{tabular}{lllllllllllll} \hline
               & Correlation: & \multicolumn{3}{c}{weak}                                                 &                      & \multicolumn{3}{c}{moderate}                                             &                      & \multicolumn{3}{c}{strong}                                               \\
$(n, d_w, s)$     & Method                   & \multicolumn{1}{c}{90} & \multicolumn{1}{c}{95} & \multicolumn{1}{c}{99} & \multicolumn{1}{c}{} & \multicolumn{1}{c}{90} & \multicolumn{1}{c}{95} & \multicolumn{1}{c}{99} & \multicolumn{1}{c}{} & \multicolumn{1}{c}{90} & \multicolumn{1}{c}{95} & \multicolumn{1}{c}{99} \\ \hline
\multicolumn{13}{c}{Panel A: low-dimensional setting}                                                                                                                                                                                                                                                                    \\
(100, 50, 6)   & PPEL                     & 0.894                  & 0.940                  & 0.988                  &                      & 0.890                  & 0.940                  & 0.986                  &                      & 0.888                  & 0.942                  & 0.986                  \\
               & DB-PEL                      & 0.736                  & 0.830                  & 0.926                  &                      & 0.726                  & 0.794                  & 0.916                  &                      & 0.744                  & 0.818                  & 0.902                  \\
               & 2SLS                     & 0.912                  & 0.964                  & 0.994                  &                      & 0.912                  & 0.964                  & 0.994                  &                      & 0.912                  & 0.964                  & 0.994                  \\
(200, 100, 6)  & PPEL                     & 0.902                  & 0.950                  & 0.994                  &                      & 0.902                  & 0.948                  & 0.994                  &                      & 0.902                  & 0.946                  & 0.994                  \\
               & DB-PEL                      & 0.762                  & 0.838                  & 0.934                  &                      & 0.734                  & 0.842                  & 0.940                  &                      & 0.752                  & 0.852                  & 0.928                  \\
               & 2SLS                     & 0.920                  & 0.956                  & 1.000                  &                      & 0.920                  & 0.956                  & 1.000                  &                      & 0.920                  & 0.956                  & 1.000                  \\ \hline
\multicolumn{13}{c}{Panel B: high-dimensional setting}                                                                                                                                                                                                                                                                   \\
(100, 120, 6)  & PPEL                     & 0.908                  & 0.956                  & 0.996                  &                      & 0.892                  & 0.948                  & 0.994                  &                      & 0.888                  & 0.940                  & 0.992                  \\
               & DB-PEL                      & 0.746                  & 0.832                  & 0.922                  &                      & 0.784                  & 0.850                  & 0.926                  &                      & 0.768                  & 0.836                  & 0.936                  \\
               & 2SLS                     & 0.956                  & 0.982                  & 1.000                  &                      & 0.956                  & 0.982                  & 1.000                  &                      & 0.956                  & 0.982                  & 1.000                  \\
(200, 240, 6)  & PPEL                     & 0.914                  & 0.964                  & 0.988                  &                      & 0.916                  & 0.960                  & 0.988                  &                      & 0.906                  & 0.954                  & 0.988                  \\
               & DB-PEL                      & 0.754                  & 0.846                  & 0.966                  &                      & 0.778                  & 0.854                  & 0.958                  &                      & 0.776                  & 0.866                  & 0.966                  \\
         & 2SLS                     & 0.966                  & 0.992                  & 1.000                  &                      & 0.966                  & 0.992                  & 1.000                  &                      & 0.966                  & 0.992                  & 1.000                  \\ \hline
(100, 120, 8)  & PPEL                     & 0.888                  & 0.942                  & 0.992                  &                      & 0.870                  & 0.930                  & 0.986                  &                      & 0.852                  & 0.924                  & 0.980                  \\
               & DB-PEL                      & 0.778                  & 0.836                  & 0.904                  &                      & 0.738                  & 0.818                  & 0.920                  &                      & 0.730                  & 0.806                  & 0.910                  \\
               & 2SLS                     & 0.960                  & 0.988                  & 1.000                  &                      & 0.960                  & 0.988                  & 1.000                  &                      & 0.960                  & 0.988                  & 1.000                  \\
(200, 240, 12) & PPEL                     & 0.924                  & 0.962                  & 0.990                  &                      & 0.914                  & 0.958                  & 0.986                  &                      & 0.892                  & 0.950                  & 0.982                  \\
               & DB-PEL                      & 0.794                  & 0.872                  & 0.954                  &                      & 0.788                  & 0.864                  & 0.946                  &                      & 0.764                  & 0.860                  & 0.938                  \\
         & 2SLS                     & 0.972                  & 0.998                  & 1.000                  &                      & 0.972                  & 0.998                  & 1.000                  &                      & 0.972                  & 0.998                  & 1.000                  \\ \hline
(100, 120, 13) & PPEL                     & 0.884                  & 0.940                  & 0.992                  &                      & 0.852                  & 0.914                  & 0.976                  &                      & 0.862                  & 0.912                  & 0.974                  \\
               & DB-PEL                      & 0.788                  & 0.862                  & 0.938                  &                      & 0.744                  & 0.798                  & 0.904                  &                      & 0.688                  & 0.768                  & 0.882                  \\
               & 2SLS                     & 0.942                  & 0.986                  & 0.996                  &                      & 0.942                  & 0.986                  & 0.996                  &                      & 0.942                  & 0.986                  & 0.996                  \\
(200, 240, 17) & PPEL                     & 0.916                  & 0.956                  & 0.984                  &                      & 0.914                  & 0.958                  & 0.982                  &                      & 0.886                  & 0.938                  & 0.982                  \\
               & DB-PEL                      & 0.792                  & 0.884                  & 0.968                  &                      & 0.780                  & 0.866                  & 0.954                  &                      & 0.798                  & 0.864                  & 0.952                  \\
         & 2SLS                     & 0.974                  & 0.988                  & 1.000                  &                      & 0.974                  & 0.988                  & 1.000                  &                      & 0.974                  & 0.988                  & 1.000         \\ \hline        
\end{tabular}

\end{spacing}
\end{table}

\section{Empirical application}\label{emp_app}

Colonialism was widespread over the globe prior to the Second World War.
While European institutions which protected private properties and checked
government powers
were successfully replicated in a few colonies,
many others suffered expropriation of natural and human resources.
In an influential study, \cite{Acemoglu2001} (AJR, henceforth) systematically explored the relationship
between Europeans' mortality rates and the modes of institutions.

We revisit AJR's open-access dataset.
AJR's main structural equation is 
$
y_i = \gamma_y + \beta_x x_i+ \beta_z z_i + \epsilon_i$, 
where the dependent variable $y_i$ is {\tt logarithm GDP per capita in 1995},
and the key explanatory variable of interest $x_i$ is {\tt average protection against expropriation risk}, a continuous variable of a scale between 0 and 10. The {\tt latitude} of a colony, denoted as $z_i$, is an additional control variable.
For substantial measurement errors in the institutional index $x_i$ can bias the OLS estimator,
the credibility of the empirical evidence counts on IVs.

AJR compiled from historical documents a seminal variable {\tt logarithm of European settler mortality},
and argued forcefully that it qualifies as a valid IV for the endogenous variable $x_i$.
Their empirical evidence was based on the 2SLS regression in cross-sectional data of countries.
The baseline result in AJR's Column (2) of Table 4 (p.1386) corresponds to the three estimating functions
$
\bg^{\calI}_i(\btheta)= (w_{1i}, z_i, 1)^{\T} \times (y_i -\gamma_y - \beta_x x_i - \beta_z z_i)$ by the notations in our simulation,  
and the 2SLS reports $\hat{\beta}_x = 1.00$ with STD 0.22.

There are another 11 potential health variables and institutional variables,
which serve as potential IVs. AJR were uncertain about their validity,
and they experimented in their Table 7 (p.1392) and Table 8 (p.1394) the empirical results under various IV configurations.
These variables, denoted as $\bw_2$, are
(i) {\tt Malaria in 1994},
(ii) {\tt Yellow fever},
(iii) {\tt Life expectancy},
(iv) {\tt Infant mortality},
(v) {\tt Mean temperature},
(vi) {\tt Distance from coast},
(vii) {\tt European settlements in 1990},
(viii) {\tt Democracy (1st year of independence)},
(ix) {\tt Constraint on executive (1st year of independence)},
(x) {\tt Democracy in 1900}, and
(xi) {\tt Constraint on executive in 1900}.
(i)--(iv) are health variables, (v)--(vi) are geographic variables, and (vii)--(xi) are institutional variables.
All these extra IVs are associated with the estimating functions
$
\bg^{\calD}_i (\btheta)=\bw_{2i} \times (y_i-\gamma_y - \beta_x x_i - \beta_z z_i) \, .
$
Given the moderate sample size of 56 countries,
a total of 12 IVs is non-trivial.


Our PPEL estimates the main equation as
\[
\widehat{\mbox{log GDP per capita}} = \underset{(1.194)}{2.048} + \underset{(0.126)}{0.945} \times \mathrm{institution}
 - \underset{(0.977)}{0.785} \times  \mbox {lattitude}\,.
\]
For the estimation and inference of
the key parameter of interest $\beta_x$, we compare the results of  PEL, DB-PEL, PPEL and the conventional 2SLS.
Its point estimation (PE), STD, and CIs are presented in Table \ref{tab:r1}.
Enjoying the efficiency gain from the extra IVs, the STD of  PPEL is 0.126
when we select the tuning parameter $\varsigma=\zeta_c(n^{-1}\log p)^{1/2}$ with $\zeta_c=0.08$ (the boldface row), as suggested by our simulations.
This STD achieves a 37\% reduction relative to that of the plain 2SLS with the sole valid IV.
Moreover,
when we vary the constant $\zeta_c=0.08$ in $\varsigma$ as 0.04, 0.06, 0.12 and 0.16, PPEL is rather robust over the wide range of tuning parameters.

\begin{table}[htbp]
\centering
\small

\caption{ Estimation of the effect of institution ($\beta_x$)  }\label{tab:r1}
\begin{spacing}{1.4}

\begin{tabular}{lc|ccc}\hline
                      & $\zeta_c$       & PE    & STD   & 95\% CI       \\ \hline
PEL                   & NA            & 0.937 & 0.078 & (0.786, 1.090) \\
DB-PEL                & NA            & 0.938 & 0.078 & (0.786, 1.090) \\
2SLS                  & NA            & 0.945 & 0.200 & (0.553, 1.338) \\\hline
\multirow{5}{*}{PPEL} & 0.04          & 0.942 & 0.159 & (0.631, 1.254) \\
                      & 0.06          & 0.941 & 0.136 & (0.675, 1.207) \\
                      & \textbf{0.08} & \textbf{0.945} & \textbf{0.126} & \textbf{(0.698, 1.193)} \\
                      & 0.12          & 0.964 & 0.150 & (0.669, 1.259) \\
                      & 0.16          & 0.967 & 0.152 & (0.669, 1.266) \\ \hline
\end{tabular}

\begin{flushleft}
\footnotesize{Note: Our sample size is 56, after removing countries with missing variables from the original sample of 64 countries,
so the 2SLS point estimate is 0.945, slightly different from AJR's 1.00.}
\end{flushleft}

\end{spacing}
\end{table}

Our method is particularly important in unifying the IV selection in AJR's Tables 7 and 8 into a single set of
automatically selected IVs.
Among the 11 potential IVs, our moment selection criterion \eqref{eq:A} invalidates
all institutional variables. Five IVs survive the testing: the climate variable {\tt mean temperature}, the geographic variable {\tt distance from coast}, and
three health variables {\tt dummy of yellow fever}, {\tt infant mortality} and {\tt life expectancy}.
All the other health and institutional variables are assessed as endogenous and are unsuitable for IVs in this study.
The justification for {\tt mean temperature} and {\tt distance from coast} is straightforward because humans were unable to interfere with these natural conditions in the era of colonialism.
Furthermore, AJR argued for the validity of {\tt dummy of yellow fever}, though due to concerns of lack of variation they did not employ it as the main IV (AJR's p.1393, Paragraph 2). Our variable selection result provides supportive evidence
to AJR's heuristics.

\section{Conclusion}
\label{s6}
This paper considers a general setting of an economic structural model
with many potential moments,
some of which may be invalid. These invalid moments must be disciplined in order to estimate the
structural parameter consistently.
We propose a PEL approach to estimate the parameter of interest while coping with the invalid moments.
We show that the PEL estimator is normally distributed asymptotically, and invalid moments can be consistently
detected thanks to the oracle property.
To overcome the difficulty of estimating the bias in the limiting distribution of the PEL estimator, we further devise
the PPEL approach for statistical inference of a low-dimensional object of interest,
which is useful for hypothesis testing and confidence region construction.
Simulation exercises are carried out to demonstrate excellent finite sample performance of our methods.
We revisit an empirical application concerning economic development and shed new insight about
its candidate instruments.

\section*{Supplementary materials}
The supplementary materials consist of three parts. Part A provides the proofs and technical details about the methods developed in the present article. Part B reports additional simulation results concerning the liner IV model in the main text and an additional dynamic panel data model, respectively. Part C checks the robustness of PEL in the empirical application.



\singlespacing
\small
\bibliography{bib}


\newpage
\onehalfspacing
\normalsize

\begin{center}
	{\bf  \LARGE
		Supplement to ``Culling the Herd of Moments
with Penalized Empirical Likelihood''}  \\~\\
 {\large Jinyuan Chang, Zhentao Shi, and Jia Zhang}
\end{center}

\setcounter{page}{1}

\bigskip

\setcounter{equation}{0}
\setcounter{section}{0}
\setcounter{table}{0}
\setcounter{figure}{0}

\renewcommand{\thefigure}{S\arabic{figure}} 
\renewcommand{\thetable}{S\arabic{table}} 

\renewcommand{\thesection}{\Alph{section}}

This supplement consists of three parts. Part A provides the proofs and technical details
about the method developed in the present article. Part B reports additional simulation results
concerning the liner IV model in the main text and an additional dynamic panel data model,
respectively. Part C checks the robustness of PEL in the empirical application.

We use ``$C$'' to denote a generic positive finite constant that may be different in different uses.

\section{Theoretical details and technical proofs}


\subsection{Proposition \ref{pn:0}}\label{se:p1}

Let $\bV^{\calI}(\btheta)=\mathbb{E}\{\bg^{\calI}_i(\btheta)^{\otimes2}\}$ and define
$
\bJ^{\calI}=([\mathbb{E}\{\nabla_{\btheta}\bg^{\calI}_i(\btheta_0)\}]^\T\{\bV^{\calI}(\btheta_0)\}^{-1/2})^{\otimes2}.
$
Proposition \ref{pn:0} gives the asymptotic normality of the standard EL estimator $\hat{\btheta}_{\EL}^{\calI}$.

\begin{proposition}\label{pn:0}
Assume that: {\rm(i)} There exists a universal constant $C_1>0$ such that
\begin{equation}\label{eq:ginde}
\inf_{\btheta\in\{\btheta\in\bTheta:\,|\btheta-\btheta_{0}|_\infty>\varepsilon\}}|\mathbb{E}\{\bg^{\calI}_i(\btheta)\}|_\infty\geq C_1 \varepsilon
\end{equation}
for any $\varepsilon>0$.
{\rm(ii)} There exist universal constants $C_2>0$, $C_3>1$ and $\gamma>4$ such that
    \begin{align}
    &~~~~~~~~~~~~~~~~~~~~\max_{j\in\mathcal{I}}\mathbb{E}\bigg\{\sup_{\btheta\in\bTheta}|g^{\calI}_{i,j}(\btheta)|^{\gamma}\bigg\}\leq C_2\,,\label{eq:c1}    \\
    &\mathbb{P}\bigg[C_3^{-1}\leq\inf_{\btheta\in\bTheta}\lambda_{\min}\{\widehat{\bV}^{\calI}(\btheta)\}\leq\sup_{\btheta\in\bTheta}\lambda_{\max}\{\widehat{\bV}^{\calI}(\btheta)\}\leq C_3\bigg]\rightarrow1\label{eq:c2}
    \end{align}
    with $\widehat{\bV}^{\calI}(\btheta)=\mathbb{E}_n\{\bg^{\calI}_i(\btheta)^{\otimes2}\}$.
{\rm(iii)}
 Each element of $\bg^{\calI}(\bX;\btheta)$ is twice continuously differentiable with respect to $\btheta$ for any $\bX$, and
\begin{align}\label{eq:pn0_reg}
\sup_{\btheta\in\bTheta}\bigg(|
    \mathbb{E}_n[\{\nabla_{\btheta}\bg^{\calI}_{i}(\btheta)\}^{\circ2}]|_\infty + \max_{j\in \mathcal{I}}
 | \mathbb{E}_n[\{\nabla^2_{\btheta}g^{\calI}_{i,j}(\btheta)\}^{\circ2}]|_\infty +
|\mathbb{E}_n[\{\bg_i^{\calI}(\btheta)\}^{\circ\gamma}]|_{\infty}\bigg)
 =O_\p(1)\,,
\end{align}
where $\gamma$ is specified in \eqref{eq:c1}.
{\rm(iv)} There exists a universal constant $C_4>1$ such that
\begin{equation}\label{eq:parcov}
C_4^{-1}<\lambda_{\min}\{\bQ^{\calI}\}\leq\lambda_{\max}\{\bQ^{\calI}\}<C_4\,,
\end{equation}
where $\bQ^{\calI}=([\mathbb{E}\{\nabla_{\btheta}\bg^{\calI}_i(\btheta_0)\}]^\T)^{\otimes2}.$
{\rm(v)} $r_1^{3}n^{-1+2/\gamma}=o(1)$ and $r_1^{3}p^2n^{-1}=o(1)$.
Then
\[
\sqrt{n}\balpha^\T\{\bJ^{\calI}\}^{1/2}\{\hat{\btheta}_{{\EL}}^{\calI}-\btheta_0\}\xrightarrow{d}\mathcal{N}(0,1)
\]
as $n\rightarrow\infty$ for any $\balpha\in\mathbb{R}^p$ with $|\balpha|_2=1$.
\end{proposition}

\begin{remark}

To understand the relative magnitude of $r_1$, $p$ and $n$,
consider the special case with $\gamma = \infty$ and fixed $p$, under which
$r_1 = o(n^{1/3})$ satisfies the condition. This is consistent with the literature
of GMM involving a diverging number of moments under a fixed $p$ \citep{koenker1999gmm};
when $p$ diverges, $p\leq r_1 =o(n^{1/5})$ is sufficient for the asymptotic normality in Proposition \ref{pn:0}.

\end{remark}

\begin{remark}
Strong identification of the parameter of interest is assumed in \eqref{eq:ginde}. It is possible
to generalize $\varepsilon$ on the right-hand side of (\ref{eq:ginde}) to $\varepsilon^{\beta}$
for some universal constant $\beta >0$ 
at the cost of much complicated expressions for the admissible range of $r_1$ and $p$. \eqref{eq:c1} restricts the population moments
uniformly over the parameter space. \eqref{eq:c2} and \eqref{eq:parcov} bound away from zero
and infinity the eigenvalues of $\mathbb{E}_n\{\bg^{\calI}_i(\btheta)^{\otimes2}\}$ and $([\mathbb{E}\{\nabla_{\btheta}\bg^{\calI}_i(\btheta_0)\}]^\T)^{\otimes2}$, respectively.
The first and second derivatives of the estimating functions are further regularized by \eqref{eq:pn0_reg}.

\end{remark}

\begin{remark}

If there are some envelope functions $\{B_{n,j}(\cdot)\}_{j\in \mathcal{I}}$ such that $\sup_{\btheta\in\bTheta}|g_{j}^{\calI}(\bX;\btheta)|^\gamma\leq B_{n,j}(\bX)$ for any $j$, and $\max_{j\in\mathcal{I}}\mathbb{E}\{B_{n,j}^m(\bX_i)\}\leq Km!H^{m-2}$ for any integer $m\geq 2$, where $K$ and $H$ are two universal positive constants independent of $j$,
\cite{Petrov1995}'s Theorem 2.8 implies $\sup_{\btheta\in\bTheta}|\mathbb{E}_n[\{\bg_i^{\calI}(\btheta)\}^{\circ\gamma}]|_{\infty}=O_\p(1)$ provided  $\log r_1=o(n)$; the other two requirements in \eqref{eq:pn0_reg} can be satisfied in the same manner.
The stochastic order $O_\p(1)$ in \eqref{eq:pn0_reg} can be replaced by $O_\p(\varphi_n)$ for some diverging $\varphi_n$. Our theoretical results still hold in this broader situation at the expense of more complicated restrictions among $r_1$, $p$ and $n$. 
\end{remark}

\noindent{\it Proof}.
Define $A_n(\btheta,\blambda)=n^{-1}\sum_{i=1}^n\log\{1+\blambda^\T\bg_{i}^{\calI}(\btheta)\}$ for any $\btheta\in\bTheta$ and $\blambda\in\hat{\Lambda}_{n}^{\calI}(\btheta)$. Then $\hat{\btheta}_{\EL}^{\calI}$ and its associated Lagrange multiplier $\hat{\blambda}$ satisfy the score equation $\nabla_{\blambda}{A}_n\{\hat{\btheta}_{\EL}^{\calI},\hat{\blambda}\}=\bzero$, i.e.
 \[
 \bzero=\frac{1}{n}\sum_{i=1}^n\frac{\bg_{i}^{\calI}\{\hat{\btheta}_{\EL}^{\calI}\}}{1+\hat{\blambda}^\T\bg_{i}^{\calI}\{\hat{\btheta}_{\EL}^{\calI}\}}\,.
 \]
 By the Taylor expansion, we have
\begin{equation*}\label{eq:expan1}
\bzero=\frac{1}{n}\sum_{i=1}^n\bg_{i}^{\calI}\{\hat{\btheta}_{\EL}^{\calI}\}-\bigg(\frac{1}{n}\sum_{i=1}^n\frac{\bg_{i}^{\calI}\{\hat{\btheta}_{\EL}^{\calI}\}^{\otimes2}}{[1+c\hat{\blambda}^\T\bg_{i}^{\calI}\{\hat{\btheta}_{\EL}^{\calI}\}]^2}\bigg)\hat{\blambda}
\end{equation*}
for some $|c|<1$, which implies
\[
\hat{\blambda}=\bigg(\frac{1}{n}\sum_{i=1}^n\frac{\bg_{i}^{\calI}\{\hat{\btheta}_{\EL}^{\calI}\}^{\otimes2}}{[1+c\hat{\blambda}^\T\bg_{i}^{\calI}\{\hat{\btheta}_{\EL}^{\calI}\}]^2}\bigg)^{-1}\bar{\bg}^{\calI}\{\hat{\btheta}_{\EL}^{\calI}\}\,.
\]
 By the implicit function theorem [Theorem 9.28 of \cite{Rudin1976}], for all $\btheta$ in a $|\cdot|_2$-neighborhood of $\hat{\btheta}_{\EL}^{\calI}$, there is a $\hat{\blambda}(\btheta)$ such that $\nabla_{\blambda}{A}_n\{\btheta,\hat{\blambda}(\btheta)\}=\bzero$ and $\hat{\blambda}(\btheta)$ is continuously differentiable in $\btheta$. By the concavity of ${A}_n(\btheta,\blambda)$ with respect to (w.r.t) $\blambda$, ${A}_n\{\btheta,\hat{\blambda}(\btheta)\}=\max_{\blambda\in\hat{\Lambda}_{n}^{\calI}(\btheta)}{A}_n(\btheta,\blambda)$. It follows from the envelope theorem that
\begin{equation*}
\bzero=\nabla_{\btheta}{A}_n\{\btheta,\hat{\blambda}(\btheta)\}\big|_{\btheta=\hat{\btheta}_{\EL}^{\calI}}=\bigg[\frac{1}{n}\sum_{i=1}^n\frac{\nabla_{\btheta}\bg_{i}^{\calI}\{\hat{\btheta}_{\EL}^{\calI}\}}{1+\hat{\blambda}^\T\bg_{i}^{\calI}\{\hat{\btheta}_{\EL}^{\calI}\}}\bigg]^\T\hat{\blambda}\,.
\end{equation*}
Therefore, we have
\begin{equation}\label{eq:exp1}
\bzero=\bigg[\frac{1}{n}\sum_{i=1}^n\frac{\nabla_{\btheta}\bg_{i}^{\calI}\{\hat{\btheta}_{\EL}^{\calI}\} }{1+\hat{\blambda}^\T\bg_{i}^{\calI}\{\hat{\btheta}_{\EL}^{\calI}\}}\bigg]^\T\bigg(\frac{1}{n}\sum_{i=1}^n\frac{\bg_{i}^{\calI}\{\hat{\btheta}_{\EL}^{\calI}\}^{\otimes2}}{[1+c\hat{\blambda}^\T\bg_{i}^{\calI}\{\hat{\btheta}_{\EL}^{\calI}\}]^2}\bigg)^{-1}\bar{\bg}^{\calI}\{\hat{\btheta}_{\EL}^{\calI}\}\,.
\end{equation}

Define $F_n(\btheta)=\max_{\blambda\in\hat{\Lambda}_{n}^{\calI}(\btheta)}A_n(\btheta,\blambda)$ and let $b_n=r_1n^{-1}$. As shown in the proof of Proposition 1 of \cite{ChangTangWu2018}, we have $\max_{\blambda\in\hat{\Lambda}_{n}^{\calI}(\btheta_0)}A_n(\btheta_0,\blambda)=O_\p(r_1n^{-1})$ which implies $F_n(\btheta_0)=O_\p(b_n)$.
As ${F}_n\{\hat{\btheta}_{\EL}^{\calI}\}\leq {F}_n(\btheta_0)$, we have $F_n\{\hat{\btheta}_{\EL}^{\calI}\}=O_\p(b_n)$.
We will first show that for any $\epsilon_n\rightarrow\infty$ satisfying $b_n\epsilon_n^{2}n^{2/\gamma}=o(1)$,  there exists a universal constant $K>0$ independent of $\btheta$ such that $\mathbb{P}\{{F}_n(\btheta)>Kb_n\epsilon_n^{2}\}\rightarrow1$ as $n\rightarrow\infty$ for any
$\btheta\in\bTheta$ satisfying $|\btheta-\btheta_{0}|_\infty> \epsilon_nb_n^{1/2}$. Thus $|\hat{\btheta}_{\EL}^{\calI}-\btheta_{0}|_\infty=O_\p(\epsilon_nb_n^{1/2})$.
Notice that we can select an arbitrary slowly diverging $\epsilon_n$,
to ensure $|\hat{\btheta}_{\EL}^{\calI}-\btheta_{0}|_\infty=O_\p(b_n^{1/2})$,
 following a standard result from probability theory.
To do this, we will use the technique developed for the proof of Theorem 1 in \cite{ChangTangWu2013}. For any ${\btheta}\in\bTheta$ satisfying $|{\btheta}-\btheta_{0}|_\infty> \epsilon_nb_n^{1/2}$, let $j_0=\arg\max_{j\in\mathcal{I}}|\mathbb{E}\{g_{i,j}^{\calI}({\btheta})\}|$. Define $\mu_{j_0}=\mathbb{E}\{g_{i,j_0}^{\calI}({\btheta})\}$ and $\tilde{\blambda}=\delta b_n^{1/2}\epsilon_n\bfe_{j_0}$ where $\delta>0$ is a constant to be determined later, and $\bfe_{j_0}$ is an $r_1$-dimensional vector with the $j_0$-th component being $1$ and other components being $0$. Without loss of generality, we assume $\mu_{j_0}>0$. (\ref{eq:c1}) and the Markov inequality yield that $\max_{i\in[n]}|g_{i,j_0}^{\calI}(\btheta)|=O_\p(n^{1/\gamma})$, which implies $\max_{i\in[n]}|\tilde{\blambda}^\T\bg_{i}^{\calI}(\btheta)|=O_\p(b_n^{1/2}\epsilon_n n^{1/\gamma})=o_\p(1)$. Then $\tilde{\blambda}\in\hat{\Lambda}_{n}^{\calI}(\btheta)$ with probability approaching one (w.p.a.1).
Write $\tilde{\blambda}=(\tilde{\lambda}_1,\ldots,\tilde{\lambda}_r)^\T$. By the definition of ${F}_n(\btheta)$, it
holds w.p.a.1 that
\begin{equation*}\label{eq:equ2}
\begin{split}
{F}_n(\btheta)\geq\frac{1}{n}\sum_{i=1}^n\log\{1+\tilde{\blambda}^\T\bg_{i}^{\calI}({\btheta})\}
=&~\frac{1}{n}\sum_{i=1}^n\tilde{\lambda}_{j_0}g_{i,j_0}^{\calI}({\btheta})-\frac{1}{2n}\sum_{i=1}^n\frac{\{\tilde{\lambda}_{j_0}g_{i,j_0}^{\calI}({\btheta})\}^2}{\{1+c\tilde{\lambda}_{j_0}g_{i,j_0}^{\calI}({\btheta})\}^2}\\
\geq&~\frac{1}{n}\sum_{i=1}^n\tilde{\lambda}_{j_0}g_{i,j_0}^{\calI}({\btheta})-\frac{1}{n}\sum_{i=1}^n\{\tilde{\lambda}_{j_0}g_{i,j_0}^{\calI}({\btheta})\}^2\\
\end{split}
\end{equation*}
for some $|c|<1$ and $\tilde{\lambda}_{j_0}=\delta b_n^{1/2}\epsilon_n$. Therefore, it holds that
\[
\begin{split}
\mathbb{P}\big\{{F}_n({\btheta})\leq Kb_n\epsilon_n^{2}\big\}\leq\mathbb{P} \bigg[\frac{1}{n}\sum_{i=1}^n\{g_{i,j_0}^{\calI}({\btheta})-\mu_{j_0}\}\leq b_n^{1/2}\epsilon_n \bigg\{\frac{K}{\delta}+\frac{\delta}{n}\sum_{i=1}^n|g_{i,j_0}^{\calI}(\btheta)|^2\bigg\}-\mu_{j_0}\bigg]+o(1)\,.
\end{split}
\]
From (\ref{eq:c1}) and the Markov inequality, there exists a universal positive constant $L$ independent of $\btheta$ such that $\mathbb{P}\{n^{-1}\sum_{i=1}^n|g_{i,j_0}^{\calI}(\btheta)|^2>L\}\rightarrow0$ as $n\rightarrow\infty$. Thus, with $\delta=(K/L)^{1/2}$, we have
\[
\mathbb{P}\big\{{F}_n({\btheta})\leq Kb_n\epsilon_n^{2}\big\}\leq\mathbb{P} \bigg[\frac{1}{n}\sum_{i=1}^n\{g_{i,j_0}^{\calI}({\btheta})-\mu_{j_0}\}\leq2b_n^{1/2}\epsilon_n(KL)^{1/2}-\mu_{j_0}\bigg]+o(1)\,.
\]
From (\ref{eq:ginde}), we know that $\mu_{j_0}\geq C_1\epsilon_n  b_n^{1/2}$ with $C_1$ specified in (\ref{eq:ginde}).
 For sufficiently small $K$ independent of $\btheta$, we have
$
2b_n^{1/2}\epsilon_n (KL)^{1/2}-\mu_{j_0}\leq-c\mu_{j_0}
$
for some $c\in(0,1)$, which implies that
$
n^{1/2}\{2b_n^{1/2}\epsilon_n (KL)^{1/2}-\mu_{j_0}\}\leq -cn^{1/2}\mu_{j_0}\lesssim -\epsilon_n b_n^{1/2}n^{1/2}\rightarrow-\infty.
$
Since $n^{-1/2}\sum_{i=1}^n\{g_{i,j_0}^{\calI}({\btheta})-\mu_{j_0}\}\stackrel{d}{\rightarrow}\mathcal{N}(0,\sigma^2)$ for some $\sigma>0$, it holds that $\mathbb{P}\{{F}_n({\btheta})\leq Kb_n\epsilon_n^{2}\}\rightarrow0$. Therefore, we have $|\hat{\btheta}_{\EL}^{\calI}-\btheta_{0}|_\infty=O_\p(b_n^{1/2})$.

Under (\ref{eq:c1}) and (\ref{eq:c2}), Proposition 1 of \cite{ChangTangWu2018} implies that $|\bar{\bg}^{\calI}\{\hat{\btheta}_{\EL}^{\calI}\}|_2=O_\p(r_1^{1/2}n^{-1/2})$. It follows from the Taylor expansion that $\bar{\bg}^{\calI}\{\hat{\btheta}_{\EL}^{\calI}\}-\bar{\bg}^{\calI}(\btheta_0)=\{\nabla_{\btheta}\bar{\bg}^{\calI}(\dot{\btheta})\}\{\hat{\btheta}_{\EL}^{\calI}-\btheta_0\}$ for some $\dot{\btheta}$ between $\hat{\btheta}_{\EL}^{\calI}$ and $\btheta_0$. Similar to Lemma 3 of \cite{ChangTangWu2018}, we know  $|[\nabla_{\btheta}\bar{\bg}^{\calI}(\dot{\btheta})-\mathbb{E}\{\nabla_{\btheta}\bg_{i}^{\calI}(\btheta_0)\}]\bz|_2=|\bz|_2\cdot[O_\p(r_1^{1/2}p^{3/2}b_n^{1/2})+O_\p\{(n^{-1}r_1p\log r_1)^{1/2}\}]$ holds uniformly over $\bz\in\mathbb{R}^{p}$. If $r_1p^{3}b_n=o(1)$ and $n^{-1}r_1p\log r_1=o(1)$, (\ref{eq:parcov}) implies that $\lambda_{\min}([\{\nabla_{\btheta}\bar{\bg}^{\calI}(\dot{\btheta})\}^\T]^{\otimes2})$ is uniformly bounded away from zero w.p.a.1. Recall $|\bar{\bg}^{\calI}(\btheta_0)|_2=O_\p(r_1^{1/2}n^{-1/2})$. Then $O_\p(r_1n^{-1})\geq \lambda_{\min}([\{\nabla_{\btheta}\bar{\bg}^{\calI}(\dot{\btheta})\}^\T]^{\otimes2})|\hat{\btheta}_{\EL}^{\calI}-\btheta_0|_2^2$, which implies $|\hat{\btheta}_{\EL}^{\calI}-\btheta_0|_2=O_\p(r_1^{1/2}n^{-1/2})$. Repeating the proof of Lemma 3 of \cite{ChangTangWu2018}, we can improve the convergence rate of $|[\nabla_{\btheta}\bar{\bg}^{\calI}(\dot{\btheta})-\mathbb{E}\{\nabla_{\btheta}\bg_{i}^{\calI}(\btheta_0)\}]\bz|_2$. More specifically, $|[\nabla_{\btheta}\bar{\bg}^{\calI}(\dot{\btheta})-\mathbb{E}\{\nabla_{\btheta}\bg_{i}^{\calI}(\btheta_0)\}]\bz|_2=|\bz|_2\cdot O_\p(r_1pn^{-1/2})$ holds uniformly over $\bz\in\mathbb{R}^{p}$. Identical to the proof of Proposition 1 of \cite{ChangTangWu2018}, we have $|\hat{\blambda}|_2=O_\p(r_1^{1/2}n^{-1/2})$. Under (\ref{eq:c2}) and (\ref{eq:pn0_reg}), similar to Lemmas 1--3 of \cite{ChangTangWu2018}, we have the following two results:
\[
\begin{split}
&\bigg\|\frac{1}{n}\sum_{i=1}^n\frac{\bg_{i}^{\calI}\{\hat{\btheta}_{\EL}^{\calI}\}^{\otimes2}}{[1+c\hat{\blambda}^\T\bg_{i}^{\calI}\{\hat{\btheta}_{\EL}^{\calI}\}]^2}-\bV^{\calI}(\btheta_0)\bigg\|_2\\
&~~~~~~~~~~~~~~~~~~~~~~=O_\p(r_1n^{-1/2+1/\gamma})+O_\p(r_1p^{1/2}n^{-1/2})+O_\p\{r_1(n^{-1}\log r_1)^{1/2}\}
\end{split}
\]
and
\begin{equation*}\label{eq:b2}
\begin{split}
&\bigg|\bigg[\frac{1}{n}\sum_{i=1}^n\frac{\nabla_{\btheta}\bg_{i}^{\calI}\{\hat{\btheta}_{\EL}^{\calI}\}}{1+\hat{\blambda}^\T\bg_{i}^{\calI}\{\hat{\btheta}_{\EL}^{\calI}\}}-\mathbb{E}\{\nabla_{\btheta}\bg_{i}^{\calI}(\btheta_0)\}\bigg]\bz\bigg|_2=|\bz|_2\cdot O_\p(r_1pn^{-1/2})\\
\end{split}
\end{equation*}
holds uniformly over $\bz\in\mathbb{R}^{p}$. Therefore, by (\ref{eq:exp1}), for any $\bdelta\in\mathbb{R}^p$ with finite $L_2$-norm, we have
\begin{equation}\label{eq:asy1}
\begin{split}
&~n^{1/2}\bdelta^\T[\mathbb{E}\{\nabla_{\btheta}\bg_{i}^{\calI}(\btheta_0)\}]^\T\{\bV^{\calI}(\btheta_0)\}^{-1}[\bar{\bg}^{\calI}\{\hat{\btheta}_{\EL}^{\calI}\}-\bar{\bg}^{\calI}({\btheta}_0)]\\
=&-n^{1/2}\bdelta^\T[\mathbb{E}\{\nabla_{\btheta}\bg_{i}^{\calI}(\btheta_0)\}]^\T\{\bV^{\calI}(\btheta_0)\}^{-1}\bar{\bg}^{\calI}({\btheta}_0)+O_\p(r_1^{3/2}n^{-1/2}\log^{1/2}r_1)\\
&+O_\p(r_1^{3/2}n^{-1/2+1/\gamma})+O_\p(r_1^{3/2}pn^{-1/2})\,.
\end{split}
\end{equation}
Recall $\bar{\bg}^{\calI}\{\hat{\btheta}_{\EL}^{\calI}\}-\bar{\bg}^{\calI}(\btheta_0)=\{\nabla_{\btheta}\bar{\bg}^{\calI}(\dot{\btheta})\}\{\hat{\btheta}_{\EL}^{\calI}-\btheta_0\}$,
$\bJ^{\calI}=([\mathbb{E}\{\nabla_{\btheta}\bg_{i}^{\calI}(\btheta_0)\}]^\T\{\bV^{\calI}(\btheta_0)\}^{-1/2})^{\otimes2}$
and $|[\nabla_{\btheta}\bar{\bg}^{\calI}(\dot{\btheta})-\mathbb{E}\{\nabla_{\btheta}\bg_{i}^{\calI}(\btheta_0)\}]\bz|_2=|\bz|_2\cdot O_\p(r_1pn^{-1/2})$ holds uniformly over $\bz\in\mathbb{R}^{p}$. Thus, (\ref{eq:asy1}) implies
\begin{equation}\label{eq:asy2}
\begin{split}
n^{1/2}\bdelta^\T\bJ^{\calI}\{\hat{\btheta}_{\EL}^{\calI}-\btheta_0\}=&-n^{1/2}\bdelta^\T[\mathbb{E}\{\nabla_{\btheta}\bg_{i}^{\calI}(\btheta_0)\}]^\T\{\bV^{\calI}(\btheta_0)\}^{-1}\bar{\bg}^{\calI}({\btheta}_0)\\
&+O_\p(r_1^{3/2}n^{-1/2}\log^{1/2}r_1)+O_\p(r_1^{3/2}n^{-1/2+1/\gamma})+O_\p(r_1^{3/2}pn^{-1/2})\,.
\end{split}
\end{equation}
For any $\balpha\in\mathbb{R}^p$ with unit $L_2$-norm, let $\bdelta=\{\bJ^{\calI}\}^{-1/2}\balpha$. Write $\bU=\{\bV^{\calI}(\btheta_0)\}^{-1/2}\mathbb{E}\{\nabla_{\btheta}\bg_{i}^{\calI}(\btheta_0)\}$ and $\bJ^{\calI}=\bU^\T\bU$. Notice that $\bU^\T\bV^{\calI}(\btheta_0)\bU=([\mathbb{E}\{\nabla_{\btheta}\bg_{i}^{\calI}(\btheta_0)\}]^\T)^{\otimes2}$. Then,
\begin{align}
    |\mathbb{E}\{\nabla_{\btheta}\bg_{i}^{\calI}(\btheta_0)\}\bdelta|_2^2=&~\balpha^\T(\bU^\T\bU)^{-1/2}\bU^\T\bV^{\calI}(\btheta_0)\bU(\bU^\T\bU)^{-1/2}\balpha \notag\\
    \le&~\lambda_{\rm max}\{\bV^{\calI}(\btheta_0)\}|\bU(\bU^\T\bU)^{-1/2}\balpha|_2^2=\lambda_{\rm max}\{\bV^{\calI}(\btheta_0)\}\,.\notag
\end{align}
Then, it follows from \eqref{eq:c2} and \eqref{eq:parcov} that $
|\bdelta|_2^2\le\lambda_{\rm max}\{\bV^{\calI}(\btheta_0)\}\lambda^{-1}_{\rm min}\{([\mathbb{E}\{\nabla_{\btheta}\bg_{i}^{\calI}(\btheta_0)\}]^\T)^{\otimes2}\}=O(1)$. From (\ref{eq:asy2}), the Central Limit Theorem implies that
\[
\begin{split}
n^{1/2}\balpha^\T\{\bJ^{\calI}\}^{1/2}\{\hat{\btheta}_{\EL}^{\calI}-\btheta_0\}=&-n^{1/2}\balpha^\T\{\bJ^{\calI}\}^{-1/2}[\mathbb{E}\{\nabla_{\btheta}\bg_{i}^{\calI}(\btheta_0)\}]^\T\{\bV^{\calI}(\btheta_0)\}^{-1}\bar{\bg}^{\calI}({\btheta}_0)\\
&+O_\p(r_1^{3/2}n^{-1/2}\log^{1/2}r_1)+O_\p(r_1^{3/2}n^{-1/2+1/\gamma})+O_\p(r_1^{3/2}pn^{-1/2})\\
\xrightarrow{d}&~\mathcal{N}
(0,1)
\end{split}
\]
provided that $r_1^{3}n^{-1+2/\gamma}=o(1)$ and $r_1^{3}p^2n^{-1}=o(1)$. $\hfill\Box$

\subsection{Proof of Proposition \ref{prop11}}

Define
$L^{\calI}(\btheta)=\max\big\{\prod_{i=1}^n\pi_i:\pi_i>0\,,\sum_{i=1}^n\pi_i=1\,,\sum_{i=1}^n\pi_i\bg^{\calI}_i(\btheta)=\bzero\big\}$ and
$L^{\calT}(\btheta,\bxi)=\max\big\{\prod_{i=1}^n\pi_i:\pi_i>0\,,\sum_{i=1}^n\pi_i=1\,,\sum_{i=1}^n\pi_i\bg^{\calI}_i(\btheta)=\bzero\,,\sum_{i=1}^n\pi_i\bg^{\calD}_i(\btheta)-\bxi=\bzero\big\}
$.
Let $\hat{\bpi}^{\calI}=\{\hat{\pi}^{\calI}_1,\ldots,\hat{\pi}^{\calI}_n\}$ and $\hat{\bpi}^{\calT}=\{\hat{\pi}^{\calT}_1,\ldots,\hat{\pi}^{\calT}_n\}$ be the associated $\bpi=(\pi_1,\ldots,\pi_n)$'s such that $L^{\calI}\{\hat\btheta_{\EL}^{\calI}\}=\prod_{i=1}^n\hat{\pi}_i^{\calI}$
and $L^{\calT}\{\hat\btheta_{\EL}^{\calT},\hat\bxi_{\EL}^{\calT}\}=\prod_{i=1}^n\hat{\pi}_i^{\calT}$.
Due to $\hat{\pi}_i^{\calT}>0$, $\sum_{i=1}^n \hat{\pi}_i^{\calT} = 1 $ and $\sum_{i=1}^n \hat{\pi}_i^{\calT} \bg_i^{\calI}\{\hat\btheta_{\EL}^{\calT}\} = \bzero$, we have
$
L^{\calI}\{\hat\btheta_{\EL}^{\calT}\} \ge \prod_{i=1}^n \hat{\pi}_i^{\calT}$.
Due to $\hat{\pi}_i^{\calI}>0$, $\sum_{i=1}^n \hat{\pi}_i^{\calI} = 1 $ and $\sum_{i=1}^n \hat{\pi}_i^{\calI} \bg_i^{\calI}\{\hat\btheta_{\EL}^{\calI}\} = \bzero$, letting $\hat\bxi = \sum_{i=1}^n \hat{\pi}_i^{\calI} \bg_i^{\calD}\{\hat\btheta_{\EL}^{\calI}\}$, we have
$
L^{\calT}\{\hat\btheta_{\EL}^{\calI}, \hat\bxi\} \ge \prod_{i=1}^n \hat{\pi}_i^{\calI}$. Since $\{\hat\btheta_{\EL}^{\calT,\T}, \hat\bxi_{\EL}^{\calT,\T}\}^\T = \arg\max_{(\btheta^\T,\bxi^\T)^\T\in\bPsi} L^{\calT}(\btheta,\bxi)$, then
$\prod_{i=1}^n \hat{\pi}_i^{\calT} =L^{\calT}\{\hat\btheta_{\EL}^{\calT}, \hat\bxi_{\EL}^{\calT}\}\ge L^{\calT}\{\hat\btheta_{\EL}^{\calI}, \hat\bxi\} \ge \prod_{i=1}^n \hat{\pi}_i^{\calI}$.
Hence, $L^{\calI}\{\hat\btheta_{\EL}^{\calT}\} \ge \prod_{i=1}^n \hat{\pi}_i^{\calT} \ge \prod_{i=1}^n \hat{\pi}_i^{\calI} = L^{\calI}\{\hat\btheta_{\EL}^{\calI}\} $. Notice that $\hat\btheta_{\EL}^{\calI} = \arg\max_{\btheta\in\bTheta} L^{\calI}(\btheta)$. Then $L^{\calI}\{\hat\btheta_{\EL}^{\calI}\}= L^{\calI}\{\hat\btheta_{\EL}^{\calT}\}$.
Since \eqref{eq:elestk} has a unique solution, we have $\hat\btheta_{\EL}^{\calI}=\hat\btheta_{\EL}^{\calT}$. $\hfill\Box$


\subsection{Proof of Proposition \ref{thm1}}\label{se:b1}
 As we have defined in Section \ref{se:pel}, $\mathcal{M}^*_{\bpsi}=\mathcal{I}\cup\mathcal{D}^*_{\bpsi}$ with $\mathcal{D}^*_{\bpsi}=\{j\in\mathcal{D}:|\bar{g}^{\calT}_{j}(\bpsi)|\ge C_*\nu\rho'_2(0^+)\}$ for any $\bpsi\in\bPsi$, where $C_*\in(0,1)$ is a prescribed constant. Define $\mathcal{D}_{\bpsi}({c})=\{j\in\mathcal{D}:|\bar g_j^{\calT}(\bpsi)|\ge {c}\nu\rho'_2(0^+)\}$ for any $c\in(C_*,1)$ and $\mathcal{M}_{\bpsi}({c})=\mathcal{I}\cup\mathcal{D}_{\bpsi}({c})$. For any index set $\mathcal{F}\subset\mathcal{T}$ and $\bpsi\in\bPsi$, we write $\widehat{\bV}_{\calF}^{\calT}(\bpsi)=n^{-1}\sum_{i=1}^n\bg_{i,\calF}^{\calT}(\bpsi)^{\otimes2}$. For any $\blambda=(\lambda_1,\ldots,\lambda_r)^\T$ and $\bpsi=(\theta_1,\ldots,\theta_p,\xi_1,\ldots,\xi_{r_2})^\T\in\bPsi$, we define
\begin{align}\label{eq:flampsi}
&f(\blambda;\bpsi)=\frac{1}{n}\sum_{i=1}^n\log\{1+\blambda^\T\bg_i^{\calT}(\bpsi)\}-\sum_{j\in\mathcal{D}}P_{2,\nu}(|\lambda_j|)\,,\\
&~~~~~~~~~S_n(\bpsi)=\max_{\blambda\in\hat\Lambda_{n}^{\calT}(\bpsi)}f(\blambda;\bpsi)+\sum_{k\in\mathcal{D}}P_{1,\pi}(|\xi_k|)\,.\notag
\end{align}
Write $\bxi_0=(\xi_{0,1},\ldots,\xi_{0,r_2})^\T$. Recall that $\aleph_n=(n^{-1}\log r)^{1/2}$, $\mathcal{S}=\mathcal{P}\cup\mathcal{A}^{\c}$ with $s=|\mathcal{S}|$, $\bpsi_{0,\calS^{\c}}=\bzero$ and $\bPsi_*=\{\bpsi=(\bpsi_{\calS}^\T,\bpsi_{\calS^{\c}}^\T)^\T:|\bpsi_{\calS}-\bpsi_{0,\calS}|_\infty\le\varepsilon,|\bpsi_{\calS^{\c}}|_1\le \aleph_n \}$
for some fixed $\varepsilon>0$. Then
\begin{align*}
\hat{\bpsi}=\arg\min_{\bpsi\in\bPsi_*}S_n(\bpsi)\,.
\end{align*}


The proof of Proposition \ref{thm1} requires the following lemmas. The proof of Lemma \ref{lem1} is similar to that of Lemma 1 in \cite{ChangTangWu2018} and we omit it here. Lemma \ref{prop3} presents general properties of the Lagrange multiplier $\hat{\blambda}(\bpsi)$ when $\bpsi$ {is} in a small neighborhood of $\bpsi_0$, whose proof is given in Section \ref{sec:pflem1}. If we just focus on the properties of $\hat{\blambda}(\bpsi_0)$, Lemma \ref{lem11} states a refined version of Lemma \ref{prop3} with proof given in Section \ref{sec:pflem2}.

\begin{lemma}\label{lem1}
Let $\mathscr{F}=\{\mathcal{F}\subset\mathcal{T}:|\mathcal{F}|\le\ell_n\}$ and $\bPsi_n=\{\bpsi\in\bPsi:|\bpsi_{\calS}-\bpsi_{0,\calS}|_\infty=O_\p(\zeta_{1,n}), |\bpsi_{\calS^{\c}}|_1\le\zeta_{2,n}\}$ for some $\zeta_{1,n},\, \zeta _{2,n}\rightarrow0$ as $n\rightarrow\infty$. If Conditions {\rm\ref{A.4}} and {\rm\ref{A.3}} hold, $\log r=o(n^{1/3})$, $\ell_n(s^2\zeta_{1,n}^2+\zeta_{2,n}^2)=o(1)$ and $\ell_n\aleph_n=o(1)$, then
$
\sup_{\bpsi\in\bPsi_n}\sup_{\calF\in\mathscr{F}}\|\widehat\bV_{\calF}^{\calT}(\bpsi)-\bV_\calF^{\calT}(\bpsi_0)\|_2=O_\p\{\ell_n^{1/2}(s\zeta_{1,n}+\zeta_{2,n})\}+O_\p(\ell_n\aleph_n)$.
\end{lemma}

\begin{lemma}\label{prop3}
Let $\{\bpsi_n\}$ be a sequence in $\bPsi$ and $P_{2,\nu}(\cdot)\in\mathscr{P}$ be a convex function for $\mathscr{P}$ defined as {\rm(\ref{eq:classp})}. For some ${c}\in(C_*,1)$, assume that all the eigenvalues of $\widehat\bV_{\calM_{\bpsi_n}({c})}^{\calT}(\bpsi_n)$ are uniformly bounded away form zero and infinity w.p.a.1. Let $|\bar\bg^{\calI}(\bpsi_n)|_2^2+|\bar{\bg}^{\calT}_{\cald_{\bpsi_n}({c})}(\bpsi_n)-\nu\rho'_2(0^+)\sgn\{\bar\bg_{\cald_{\bpsi_n}({c})}^{\calT}(\bpsi_n)\}|_2^2=O_\p(u_n^2)$ for some $u_n\to 0$, and $\max_{j\in\mathcal{T}}{n}^{-1}\sum_{i=1}^n|g^{\calT}_{i,j}(\bpsi_n)|^\gamma=O_\p(1)$. For some non-random sequence $\{m_n\}$ such that $\mathbb{P}(|\mathcal{M}_{\bpsi_n}^*|\leq m_n)\rightarrow1$ as $n\rightarrow\infty$,
if $m_n^{1/2}u_n=o(\nu)$ and $m_n^{1/2}u_nn^{1/\gamma}=o(1)$, then w.p.a.1 there is a sparse {global} maximizer $\hat\blambda(\bpsi_n)$ for
$
f(\blambda;\bpsi_n)
$
satisfying the following three results: {\rm (i)} $|\hat\blambda(\bpsi_n)|_2=O_\p(u_n)$, {\rm (ii)} $\supp\{\hat\blambda_{\cald}(\bpsi_n)\}\subset\mathcal{D}_{\bpsi_n}({c})$, and {\rm (iii)} $\sgn(\hat\lambda_{n,j})=\sgn\{\bar g^{\calT}_{j}(\bpsi_n)\}$ for any $j\in\mathcal{D}_{\bpsi_n}({c})$ with $\hat\lambda_{n,j}\neq 0$, where $\hat\blambda(\bpsi_n)=(\hat\lambda_{n,1},\ldots,\hat\lambda_{n,r})^\T$.
\end{lemma}

\begin{lemma}\label{lem11}
Let $P_{2,\nu}(\cdot) \in \mathscr{P}$ be a convex function for $\mathscr{P}$ defined as \eqref{eq:classp}. Assume that all the eigenvalues of $\widehat\bV_{\calM_{\bpsi_0}({c})}^{\calT}(\bpsi_0)$ are uniformly bounded away from zero and infinity w.p.a.1 for some $c\in(C_*,1)$, and $\max_{j\in\mathcal{T}}n^{-1}\sum_{i=1}^n|g_{i,j}^{\calT}(\bpsi_0)|^\gamma=O_\p(1)$. If $\log r=o(n^{1/3})$ and $r_1 \aleph_n=o[\min\{n^{-1/\gamma},\nu\}]$, then w.p.a.1 there is a sparse global maximizer $\hat{\blambda}(\bpsi_0)$ for $f(\blambda;\bpsi_0)$ satisfying ${\rm supp}\{\hat{\blambda}_{\cald}(\bpsi_{0})\}\subset \mathcal{D}_{\bpsi_0}({c})$.
\end{lemma}

Now we begin to prove Proposition \ref{thm1}. Recall 
$
S_n(\bpsi)=\max_{\blambda\in\hat\Lambda_{n}^{\calT}(\bpsi)}f(\blambda;\bpsi)+\sum_{k\in\mathcal{D}}P_{1,\pi}(|\xi_k|)$ and $a_n=\sum_{k\in\mathcal{D}}P_{1,\pi}(|\xi_{0,k}|)$. Then $S_n(\bpsi_0)=f\{\hat\blambda(\bpsi_0);\bpsi_0\}+a_n$, where $\hat\blambda(\bpsi_0)=\arg\max_{\blambda\in\hat\Lambda_{n}^{\calT}(\bpsi_0)}f(\blambda;\bpsi_0)$. Let $\mathcal{G}=\mathcal{I}\cup\supp\{\hat\blambda_{\cald}(\bpsi_0)\}$ and write $\hat\blambda(\bpsi_0)=(\hat\lambda_1,\ldots,\hat\lambda_r)^\T$. It holds that
\begin{align}
f\{\hat\blambda(\bpsi_0);\bpsi_0\}&=\frac{1}{n}\sum_{i=1}^n\log\{1+\hat\blambda_\calG(\bpsi_0)^\T\bg_{i,\calG}^{\calT}(\bpsi_0)\}-\sum_{j\in\mathcal{D}:\,\hat{\lambda}_j\neq0}P_{2,\nu}(|\hat\lambda_j|)\notag\\
&\le\max_{\bfeta\in\hat\Lambda_{n}^{\dag}(\bpsi_0)}\frac{1}{n}\sum_{i=1}^n\log\{1+\bfeta^\T\bg_{i,\calG}^{\calT}(\bpsi_0)\}\,,\label{7.00}
\end{align}
where $\hat{\Lambda}_{n}^{\dag}(\bpsi_0)=\{\bfeta=(\eta_1,\ldots,\eta_{|\calG|})^\T\in\mathbb{R}^{|\mathcal{G}|}:\bfeta^\T\bg_{i,\calG}^{\calT}(\bpsi_0)\in\mathcal{V}~\textrm{for any}~i\in[n]\}$ for some open interval $\mathcal{V}$ containing zero. We will first prove that
\be\label{7.0}
\max_{\bfeta\in\hat\Lambda_{n}^{\dag}(\bpsi_0)}\frac{1}{n}\sum_{i=1}^n\log\{1+\bfeta^\T\bg_{i,\calG}^{\calT}(\bpsi_0)\}=O_\p(r_1\aleph_n^2)\,.
\ee
Based on (\ref{7.0}), we have $f\{\hat\blambda(\bpsi_0);\bpsi_0\}=O_\p(r_1\aleph_n^2)$. Let $A_n(\bpsi,\bfeta)=n^{-1}\sum_{i=1}^n\log\{1+\bfeta^\T\bg_{i,\calG}^{\calT}(\bpsi)\}$ and $\tilde\bfeta=\arg\max_{\bfeta\in\hat\Lambda_{n}^{\dag}(\bpsi_0)}A_n(\bpsi_0,\bfeta)$. As we have shown in the proof of Lemma \ref{lem11} that $|\mathcal{M}_{\bpsi_0}(c)|\leq2r_1$ w.p.a.1, it then follows from Lemma \ref{lem11} that $|\mathcal{G}|\le |\mathcal{M}_{\bpsi_0}(c)| \leq2r_1$ w.p.a.1. Pick $\delta_n=o(r_1^{-1/2} n^{-1/\gamma})$ and $r_1^{1/2}\aleph_n=o(\delta_n)$, which can be guaranteed by $r_1\aleph_n=o(n^{-1/\gamma})$. Define $\Lambda_n=\{\bfeta\in\mathbb{R}^{|\mathcal{G}|}:|\bfeta|_2\le\delta_n\}$ and let $\bar\bfeta=\arg\max_{\bfeta\in\Lambda_n}A_n(\bpsi_0,\bfeta)$. It follows from the last requirement of Condition \ref{A.4} that $\max_{i\in[n]}|\bg_{i,\calG}^{\calT}(\bpsi_0)|_2=O_\p(r_1^{1/2} n^{1/\gamma})$, which implies that $\max_{i\in[n]}\sup_{\bfeta\in\Lambda_n}|\bfeta^\T\bg_{i,\calG}^{\calT}(\bpsi_0)|=o_\p(1)$. By the Taylor expansion, it holds w.p.a.1 that
\be\label{7.1}
\begin{split}
0=A_n(\bpsi_0,\bzero)\le A_n(\bpsi_0,\bar\bfeta)&=\bar\bfeta^\T\bar\bg_\calG^{\calT}(\bpsi_0)-\frac{1}{2n}\sum_{i=1}^n\frac{\bar\bfeta^\T\bg_{i,\calG}^{\calT}(\bpsi_0)^{\otimes2}\bar\bfeta}{\{1+\bar{c}\bar\bfeta^\T\bg_{i,\calG}^{\calT}(\bpsi_0)\}^2} \\
&\le|\bar\bfeta|_2|\bar{\bg}_\calG^{\calT}(\bpsi_0)|_2-C|\bar\bfeta|_2^2\{1+o_\p(1)\}
\end{split}
\ee
for some $\bar{c}\in(0,1)$, where the last inequality is implied by Condition \ref{A.3} and Lemma \ref{lem1}. As we have shown in the proof of Lemma \ref{lem11} that $|\bar{\bg}^{\calT}(\bpsi_0)|_\infty=O_\p(\aleph_n)$, then $|\bar{\bg}_{\calG}^{\calT}(\bpsi_0)|_2=O_\p(r_1^{1/2}\aleph_n)$. It follows from \eqref{7.1} that $|\bar\bfeta|_2=O_\p(r_1^{1/2}\aleph_n)=o_\p(\delta_n)$. Hence, $\bar\bfeta\in{\rm int}(\Lambda_n)$ w.p.a.1. Since $\Lambda_n\subset\hat\Lambda_{n}^{\dag}(\bpsi_0)$ w.p.a.1, by the concavity of $A_n(\bpsi_0,\bfeta)$ and the convexity of $\hat\Lambda_{n}^{\dag}(\bpsi_0)$, we have $\tilde\bfeta=\bar\bfeta$ w.p.a.1. Then we can obtain \eqref{7.0} from \eqref{7.1}.

Recall that $b_{1,n}=\max\{a_n,r_1\aleph_n^2\}$. Then $S_n(\bpsi_0)=O_\p(r_1\aleph_n^2)+a_n=O_\p(b_{1,n})$. Notice that $\hat{\bpsi}=\arg\min_{\bpsi\in\bPsi_*}S_n(\bpsi)$ with $\bPsi_*=\{\bpsi=(\bpsi_{\calS}^\T,\bpsi_{\calS^{\c}}^\T)^\T:|\bpsi_{\calS}-\bpsi_{0,\calS}|_\infty\le\varepsilon,|\bpsi_{\calS^{\c}}|_1\le \aleph_n  \}$, and $\bpsi_{0,\calS^{\c}}=\bzero$. We then have $\bpsi_0\in\bPsi_*$ which implies that $S_n(\hat\bpsi)\le S_n(\bpsi_0)=O_\p(b_{1,n})$. We need to show $\hat\bpsi\in{\rm int}(\bPsi_*)$ w.p.a.1, which indicates that $\hat{\bpsi}$ is a local minimizer of $S_n(\bpsi)$. Our proof includes three parts: (i) to show that for any $\epsilon_n\to\infty$ satisfying $b_{1,n}\epsilon_n^{2}n^{2/\gamma}=o(1)$ and any $\bpsi=(\btheta^\T,\bxi^\T)^\T\in\bPsi_*$ satisfying $| \btheta-\btheta_0 |_\infty>\epsilon_nb_{1,n}^{1/2}$, there exists a universal constant $K>0$ independent of $\bpsi$ such that $\mathbb{P}\{S_n(\bpsi)>Kb_{1,n}\epsilon_n^{2}\}\to 1$ as $n\to\infty$. Due to $b_{1,n}=o(n^{-2/\gamma})$, we can select an arbitrary slowly diverging $\epsilon_n$ satisfying $b_{1,n}\epsilon_n^{2}n^{2/\gamma}=o(1)$. Thus, we have $|\hat\btheta-\btheta_0|_\infty=O_\p(b_{1,n}^{1/2})$; (ii) letting $b_{2,n}=\max\{b_{1,n},\nu^2\}$ and $\phi_n=\max\{p b_{1,n}^{1/2},b_{2,n}^{1/2}\}$, to show that for any $\varepsilon_n\to\infty$ satisfying $b_{2,n} \varepsilon_n^2 n^{2/\gamma}=o(1)$ and $\bpsi=(\btheta^\T,\bxi_\cala^\T,\bxi_{\cala^{\c}}^\T)^\T\in\bPsi_*$ satisfying $|\btheta-\btheta_0|_\infty \leq O(\varepsilon_n^{1/2}b_{1,n}^{1/2})$ and $|\bxi_{\cala^{\c}}-\bxi_{0,\cala^{\c}}|_\infty>\varepsilon_n\phi_n$, there exists a universal constant $M>0$ independent of $\bpsi$ such that $\mathbb{P}\{S_n(\bpsi)>M b_{2,n}\varepsilon_n^{2}\}\to 1$ as $n\to\infty$. Recall $\hat{\bpsi}=(\hat{\btheta}^\T,\hat{\bxi}_{\cala}^\T,\hat{\bxi}_{\cala^{\c}}^\T)^\T$. Due to $|\hat{\btheta}-\btheta|_\infty=O_\p(b_{1,n}^{1/2})$, we know $|\hat{\btheta}-\btheta_0|\leq O(\varepsilon_n^{1/2}b_{1,n}^{1/2})$ w.p.a.1. Since we can select an arbitrary slowly diverging $\varepsilon_n$ satisfying $b_{2,n} \varepsilon_n^2 n^{2/\gamma}=o(1)$, it holds that $|\hat\bxi_{\cala^{\c}}-\bxi_{0,\cala^{\c}}|_\infty=O_\p(\phi_n)$; (iii) to show that $\hat\bpsi_{\calS^{\c}}=\bzero$ w.p.a.1.

 \underline{{\it Proof of Part} (i).} The proof is similar to that for Part (i) of Proposition \ref{pn:0}. For any $\bpsi=(\btheta^\T,\bxi^\T)^\T\in\bPsi_*$ satisfying $|\btheta-\btheta_0|_\infty>\epsilon_nb_{1,n}^{1/2}$, let $j_0=\arg\max_{j\in\mathcal{I}}|\mathbb{E}\{g_{i,j}^{\calI}(\btheta)\}|$ and $\mu_{j_0}=\mathbb{E}\{g_{i,j_0}^{\calI}(\btheta)\}$. Select $\tilde\blambda=\delta b_{1,n}^{1/2}\epsilon_n\bfe_{j_0}$, where $\delta>0$ is a sufficiently small constant, and $\bfe_{j_0}$ is an $r$-dimensional vector with the $j_0$-th component being $1$ and other components being $0$. Then $\tilde\blambda\in\hat\Lambda_{n}^{\calT}(\bpsi)$ w.p.a.1. Without loss of generality, we assume that $\mu_{j_0}>0$. Write $\tilde\blambda=(\tilde\lambda_1,\ldots,\tilde\lambda_r)^\T$. Notice that $j_0\notin\mathcal{D}$. By the Taylor expansion, it holds w.p.a.1 that
$
    S_n(\bpsi)\ge n^{-1}\sum_{i=1}^n\log\{1+\tilde\blambda^\T\bg_{i}^{\calT}(\bpsi)\}-\sum_{j\in\mathcal{D}}P_{2,\nu}(|\tilde\lambda_j|)\geq{n}^{-1}\sum_{i=1}^n\tilde\lambda_{j_0}g_{i,j_0}^{\calI}(\btheta)-{n}^{-1}\sum_{i=1}^n\{\tilde\lambda_{j_0}g_{i,j_0}^{\calI}(\btheta)\}^2$.
Thus,
\begin{align*}
\mathbb{P}\{S_n(\bpsi)\le Kb_{1,n}\epsilon_n^{2}\}\le&~\mathbb{P}\bigg[\frac{1}{n}\sum_{i=1}^n\tilde\lambda_{j_0}g_{i,j_0}^{\calI}(\btheta)-\frac{1}{n}\sum_{i=1}^n\{\tilde\lambda_{j_0}g_{i,j_0}^{\calI}(\btheta)\}^2\le Kb_{1,n}\epsilon_n^{2}\bigg]+o(1)\\
\le&~\mathbb{P}\bigg[\bar{g}^{\calI}_{j_0}(\btheta)-\mu_{j_0}\le b_{1,n}^{1/2}\epsilon_n\bigg\{\frac{K}{\delta}+ \frac{\delta}{n}\sum_{i=1}^n |g_{i,j_0}^{\calI}(\btheta)|^2\bigg\}-\mu_{j_0}\bigg]+o(1)\,.
\end{align*}
Using the same arguments stated in the proof of Proposition \ref{pn:0}, we have $\mathbb{P}\{S_n(\bpsi)>Kb_{1,n}\epsilon_n^{2}\}\to1$ as $n\rightarrow\infty$. We complete the proof of Part (i).

\underline{{\it Proof of Part} (ii).} The proof is also similar to that for Part (i) of Proposition \ref{pn:0}. For any $\bpsi=(\btheta^\T,\bxi^\T)^\T\in\bPsi_*$ with $\bxi=(\bxi_\cala^\T,\bxi_{\cala^{\c}}^\T)^\T$ satisfying $|\btheta-\btheta_0|_\infty \leq O(\varepsilon_n^{1/2}b_{1,n}^{1/2})$ and $|\bxi_{\cala^{\c}}-\bxi_{0,\cala^{\c}}|_\infty>\varepsilon_n\phi_n$, let $j_0=\arg\max_{j\in\cala^{\c}}|\xi_j-\xi_{0,j}|$ and $\mu_{j_0}=\mathbb{E}\{g_{i,j_0}^{\calT}(\bpsi)\}$. Without loss of generality, we assume $\xi_{0,j_0}-\xi_{j_0}>0$. Select $\tilde\blambda=\delta b_{2,n}^{1/2}\varepsilon_n\bfe_{j_0}$, where $\delta>0$ is a sufficiently small constant, and $\bfe_{j_0}$ is similarly defined as that in the proof of Part (i). Then $\tilde\blambda\in\hat\Lambda_{n}^{\calT}(\bpsi)$ w.p.a.1. By the Taylor expansion, it holds w.p.a.1 that
$
    S_n(\bpsi)\ge{n}^{-1}\sum_{i=1}^n\tilde\lambda_{j_0}g_{i,j_0}^{\calT}(\bpsi)-{n}^{-1}\sum_{i=1}^n\{\tilde\lambda_{j_0}g_{i,j_0}^{\calT}(\bpsi)\}^2-P_{2,\nu}(|\tilde\lambda_{j_0}|)\ge{n}^{-1}\sum_{i=1}^n\tilde\lambda_{j_0}g_{i,j_0}^{\calT}(\bpsi)-{n}^{-1}\sum_{i=1}^n\{\tilde\lambda_{j_0}g_{i,j_0}^{\calT}(\bpsi)\}^2-C\nu\tilde\lambda_{j_0}$.
Thus,
\begin{align*}
\mathbb{P}\{S_n(\bpsi)\le M b_{2,n}\varepsilon_n^{2}\}\le&~\mathbb{P}\bigg[\frac{1}{n}\sum_{i=1}^n\tilde\lambda_{j_0}g_{i,j_0}^{\calT}(\bpsi)-\frac{1}{n}\sum_{i=1}^n\{\tilde\lambda_{j_0}g_{i,j_0}^{\calT}(\bpsi)\}^2-C\nu\tilde\lambda_{j_0}\le M b_{2,n}\varepsilon_n^{2}\bigg]+o(1)\\
\le&~\mathbb{P}\bigg[\bar{g}^{\calT}_{j_0}(\bpsi)-\mu_{j_0}\le b_{2,n}^{1/2}\varepsilon_n\bigg\{\frac{M}{\delta}+ \frac{\delta}{n}\sum_{i=1}^n |g_{i,j_0}^{\calT}(\bpsi)|^2\bigg\}+C\nu-\mu_{j_0}\bigg]+o(1)\,.
\end{align*}
By the Taylor expansion and Condition \ref{A.4}, $|\mathbb{E}\{g_{i,j_0}^{\calD}(\btheta)\}-\mathbb{E}\{g_{i,j_0}^{\calD}(\btheta_0)\}|\leq |\mathbb{E}\{\nabla_{\btheta}g_{i,j_0}^{\calD}(\dot{\btheta})\}|_\infty|\btheta-\btheta_0|_1\leq O(\varepsilon_n^{1/2}pb_{1,n}^{1/2})$. Recall $b_{2,n}=\max\{b_{1,n},\nu^2 \}$ and $\phi_n=\max\{p b_{1,n}^{1/2},b_{2,n}^{1/2}\}$. Then
\begin{align}
\mu_{j_0}&=\mathbb{E}\{g_{i,j_0}^{\calD}(\btheta)\}-\mathbb{E}\{g_{i,j_0}^{\calD}(\btheta_0)\}+\xi_{0,j_0}-\xi_{j_0} \ge \varepsilon_n\phi_n - O(\varepsilon_n^{1/2}p b_{1,n}^{1/2}) \ge \varepsilon_n b_{2,n}^{1/2}/2 \label{eq:muj0}
\end{align}
when $n$ is sufficiently large.
Using the same arguments stated in the proof of Proposition \ref{pn:0}, we have $\mathbb{P}\{S_n(\bpsi)>M b_{2,n}\varepsilon_n^{2}\}\to1$ as $n\rightarrow\infty$. We complete the proof of Part (ii).

\underline{{\it Proof of Part} (iii).} If $\hat\bpsi_{\calS^{\c}}\neq \bzero$, we define $\hat\bpsi^*=(\hat\bpsi_{\calS}^\T,\bzero^\T)^\T$ and will show $S_n(\hat{\bpsi}^*)<S_n(\hat\bpsi)$ w.p.a.1. This contradicts the definition of $\hat\bpsi$. Then we have $\hat\bpsi_{\calS^{\c}}=\bzero$ w.p.a.1. Write $\hat{\bpsi}=(\hat{\psi}_1,\ldots,\hat{\psi}_{p+r_2})^\T=(\hat{\theta}_1,\ldots,\hat{\theta}_p,\hat{\xi}_1,\ldots,\hat{\xi}_{r_2})^\T$ and $\bpsi_0=(\theta_{0,1},\ldots,\theta_{0,p},\xi_{0,1},\ldots,\xi_{0,r_2})^\T$. Recall $\xi_{0,k}=0$ for any $k\in\mathcal{A}$ and $\xi_{0,k}\neq 0$ for any $k\in\mathcal{A}^{\c}$. As shown in Part (ii) that $\max_{k\in\mathcal{A}^{\c}}|\hat{\xi}_k-\xi_{0,k}|=O_\p(\phi_n)$, due to $\phi_n=o(\min_{k\in\mathcal{A}^{\c}}|\xi_{0,k}|)$ and \eqref{eq:xi}, \eqref{7.00} and \eqref{7.0} imply that
\begin{align}
\max_{\blambda\in\hat\Lambda_{n}^{\calT}(\hat\bpsi)}f(\blambda;\hat\bpsi)&\le \max_{\blambda\in\hat\Lambda_{n}^{\calT}(\bpsi_0)}f(\blambda;\bpsi_0) +\sum_{k\in\mathcal{D}}P_{1,\pi}(|\xi_{0,k}|)-\sum_{k\in\mathcal{D}}P_{1,\pi}(|\hat\xi_k|)\notag \\
&\le O_\p(r_1\aleph_n^2)+\sum_{k\in\mathcal{A}^c}P_{1,\pi}(|\xi_{0,k}|)-\sum_{k\in\mathcal{A}^c}P_{1,\pi}(|\hat\xi_k|)\label{eq:diff}\\
&= O_\p(r_1\aleph_n^2)\,.\notag
\end{align}
Notice that $\ell_n\aleph_n=o(n^{-1/\gamma})$.
We pick $\delta_n=o(\ell_n^{-1/2}n^{-1/\gamma})$ and $\ell_n^{1/2}\aleph_n=o(\delta_n)$. Recall $\mathcal{M}_{\bpsi}({c})=\mathcal{I}\cup\mathcal{D}_{\bpsi}({c})$ with $\mathcal{D}_{\bpsi}({c})=\{j\in\mathcal{D}:|\bar g_j^{\calT}(\bpsi)|\ge {c}\nu\rho'_2(0^+)\}$ for any ${c}\in(C_*,1)$. Define
\begin{equation*}
\bbeta_{\calM_{\hat\bpsi}(\tilde{c}_1)}(\hat\bpsi):=\Bigg(\begin{array}{c}\bar\bg^{\calI}(\hat\bpsi)\\\bar\bg_{\cald_{\hat\bpsi}(\tilde{c}_1)}^{\calT}(\hat\bpsi)-\nu\rho'_2(0^+)\sgn\{\bar\bg_{\cald_{\hat\bpsi}(\tilde{c}_1)}^{\calT}(\hat\bpsi)\}
\end{array}\Bigg)
\end{equation*}
for some $\tilde{c}_1\in(C_*,1)$. Select $\blambda^*$ satisfying $\blambda^*_{\calM_{\hat\bpsi}(\tilde{c}_1)}=\delta_n\bbeta_{\calM_{\hat\bpsi}(\tilde{c}_1)}(\hat\bpsi)/|\bbeta_{\calM_{\hat\bpsi}(\tilde{c}_1)}(\hat\bpsi)|_2$ and $\blambda^*_{\calM^{\c}_{\hat\bpsi}(\tilde{c}_1)}=\bzero$. Since $|\mathcal{M}_{\hat{\bpsi}}(\tilde{c}_1)|\leq \ell_n$ w.p.a.1, it holds that $\max_{i\in[n]}|\blambda^{*,\T}\bg_i^{\calT}(\hat{\bpsi})|\leq |\blambda_{\calM_{\hat{\bpsi}}(\tilde{c}_1)}^*|_2\max_{i\in[n]}|\bg_{i,\calM_{\hat{\bpsi}}(\tilde{c}_1)}^{\calT}(\hat{\bpsi})|_2=o(\ell_n^{-1/2}n^{-1/\gamma})\cdot O_\p(\ell_n^{1/2}n^{1/\gamma})=o_\p(1)$, which indicates that $\blambda^*\in\hat\Lambda_{n}^{\calT}(\hat\bpsi)$ w.p.a.1. Write $\blambda^*=(\lambda_1^*,\ldots,\lambda_r^*)^\T$.

Recall $P_{2,\nu}(t)=\nu\rho_2(t;\nu)$ for any $t\geq0$. Notice that $\mathbb{P}[\cup_{j\in\mathcal{T}}\{|\bar{g}_j^\calT(\hat\bpsi)|\in[\tilde{c}\nu\rho'_2(0^+),\nu\rho'_2(0^+))\}]\rightarrow0$ for some constant $\tilde{c}\in(C_*,1)$. Then $\{j\in\mathcal{T}:\,\tilde{c}_1\nu\rho_2'(0^+)\leq|\bar{g}_j^{\calT}(\hat{\bpsi})|<\nu\rho_2'(0^+)\} = \emptyset$ w.p.a.1 by letting $\tilde{c}_1 = \tilde{c}$. Notice that $r_1\lesssim \ell_n$. By the Taylor expansion, it holds w.p.a.1 that
\begin{align*}
o_\p(\delta_n^2)&= \max_{\blambda\in\hat\Lambda_{n}^{\calT}(\hat\bpsi)}f(\blambda;\hat\bpsi)\\
&\ge \frac{1}{n}\sum_{i=1}^n\log\{1+\blambda^{*,\T}_{\calM_{\hat\bpsi}(\tilde{c}_1)}\bg_{i,\calM_{\hat\bpsi}(\tilde{c}_1)}^{\calT}(\hat{\bpsi})\}-\sum_{j\in{\cald_{\hat\bpsi}(\tilde{c}_1)}}P_{2,\nu}(|\lambda_j^*|)\\
&=\blambda^{*,\T}_{\calM_{\hat\bpsi}(\tilde{c}_1)}\bar{\bg}_{\calM_{\hat\bpsi}(\tilde{c}_1)}^{\calT}(\hat\bpsi)-\frac{1}{2n}\sum_{i=1}^n\frac{\blambda^{*,\T}_{\calM_{\hat\bpsi}(\tilde{c}_1)}\bg_{i,\calM_{\hat\bpsi}(\tilde{c}_1)}^{\calT}(\hat\bpsi)^{\otimes2}\blambda^{*}_{\calM_{\hat\bpsi}(\tilde{c}_1)}}{\{1+c^*\blambda^{*,\T}_{\calM_{\hat\bpsi}(\tilde{c}_1)}\bg_{i,\calM_{\hat\bpsi}(\tilde{c}_1)}^{\calT}(\hat{\bpsi})\}^2}\\
&~~~~~~-\sum_{j\in\cald_{\hat{\bpsi}}(\tilde{c}_1)}\nu\rho_2'(0^+)|\lambda_j^*|-\frac{1}{2}\sum_{j\in\cald_{\hat{\bpsi}}(\tilde{c}_1)}\nu\rho_2''(c_j|\lambda_j^*|;\nu)|\lambda_j^*|^2\\
&\ge\blambda^{*,\T}_{\calM_{\hat\bpsi}(\tilde{c}_1)}\bbeta_{\calM_{\hat\bpsi}(\tilde{c}_1)}(\hat\bpsi)-C\delta_n^2\{1+o_\p(1)\}-2\nu\rho_2'(0^+)\sum_{j\in\mathcal{T}:\,\tilde{c}_1\nu\rho_2'(0^+)\leq|\bar{g}_j^{\calT}(\hat{\bpsi})|<\nu\rho_2'(0^+)}|\lambda_j^*|\\
&\ge\delta_n|\bbeta_{\calM_{\hat\bpsi}(\tilde{c}_1)}(\hat\bpsi)|_2-C\delta_n^2\{1+o_\p(1)\}
\end{align*}
for some $c^*, c_j\in(0,1)$. Thus, $|\bbeta_{\calM_{\hat\bpsi}(\tilde{c}_1)}(\hat\bpsi)|_2=O_\p(\delta_n)$. For any $\epsilon_n\to0$, choose $\blambda^{**}$ satisfying $\blambda^{**}_{\calM_{\hat\bpsi}(\tilde{c}_1)}=\epsilon_n\bbeta_{\calM_{\hat\bpsi}(\tilde{c}_1)}(\hat\bpsi)$ and $\blambda^{**}_{\calM^{\c}_{\hat\bpsi}(\tilde{c}_1)}=\bzero$. Then, $|\blambda^{**}|_2=o_\p(\delta_n)$. Using the same arguments given above, we can obtain
$
\epsilon_n|\bbeta_{\calM_{\hat\bpsi}(\tilde{c}_1)}(\hat\bpsi)|_2^2-C\epsilon_n^2|\bbeta_{\calM_{\hat\bpsi}(\tilde{c}_1)}(\hat\bpsi)|_2^2\{1+o_\p(1)\}=O_\p(r_1\aleph_n^2)
$,
which implies that $\epsilon_n|\bbeta_{\calM_{\hat\bpsi}(\tilde{c}_1)}(\hat\bpsi)|_2^2=O_\p(r_1\aleph_n^2)$. Since we can select an arbitrary slow $\epsilon_n\rightarrow0$,  we have $|\bbeta_{\calM_{\hat\bpsi}(\tilde{c}_1)}(\hat\bpsi)|_2^2=O_\p(r_1 \aleph_n^2)$ following a standard result from probability theory. Then Lemmas \ref{lem1} and \ref{prop3} imply that $|\hat\blambda(\hat\bpsi)|_2=O_\p(r_1^{1/2}\aleph_n)$. Recall $\hat\blambda(\bpsi)=\arg\max_{\blambda\in\hat\Lambda_{n}^{\calT}(\bpsi)}f(\blambda;\bpsi)$. Write $\hat\blambda=\hat\blambda(\hat\bpsi)$ and $\hat\blambda^*=\hat\blambda(\hat\bpsi^*)$. Notice that $
S_n(\bpsi)=\max_{\blambda\in\hat\Lambda_{n}^{\calT}(\bpsi)}f(\blambda;\bpsi)+\sum_{k\in\mathcal{D}}P_{1,\pi}(|\xi_k|)$ and $\mathcal{S}=\mathcal{P}\cup\mathcal{A}^{\c}$. Then
\begin{align}
S_n(\hat\bpsi^*)-  S_n(\hat\bpsi)=&~f(\hat\blambda^*;\hat\bpsi^*)-f(\hat\blambda;\hat\bpsi)-\sum_{k\in \mathcal{A}}P_{1,\pi}(|\hat\xi_k|)\notag\\
 \le&~f(\hat\blambda^*;\hat\bpsi^*)- f(\hat\blambda^*;\hat\bpsi)-\sum_{k\in \mathcal{A}}P_{1,\pi}(|\hat\xi_k|)\label{eq:diff2}\\
 =&~\frac{1}{n}\sum_{i=1}^n\log\{1+\hat\blambda^{*,\T}\bg_{i}^{\calT}(\hat\bpsi^*)\}-\frac{1}{n}\sum_{i=1}^n\log\{1+\hat\blambda^{*,\T}\bg_{i}^{\calT}(\hat\bpsi)\}-\sum_{k\in \mathcal{A}}P_{1,\pi}(|\hat\xi_k|)\,.\notag
\end{align}
It follows from the Taylor expansion that
\begin{align}\label{eq:sn1}
    S_n(\hat\bpsi^*)\le&~S_n(\hat\bpsi)-\underbrace{\frac{1}{n}\sum_{i=1}^n\frac{\hat\blambda^{*,\T}\nabla_{\bpsi_{\calS^{\c}}}\bg_{i}^{\calT}(\check{\bpsi})}{1+\hat\blambda^{*,\T}\bg_{i}^{\calT}(\check{\bpsi})}\hat\bpsi_{\calS^c}}_{{\rm I}}
    -\underbrace{\sum_{k\in\mathcal{A}}P_{1,\pi}(|\hat\xi_k|)}_{{\rm II}}\,,
\end{align}
where $\check{\bpsi}$ is on the jointing line between $\hat\bpsi$ and $\hat\bpsi^*$. We need to show ${\rm I}+{\rm II}>0$ w.p.a.1.

To do this, we first use Lemma \ref{prop3} to bound $|\hat{\blambda}^*|_2$. Given some $\tilde{c}_2\in(\tilde{c}_1,1)$, we define
\begin{equation*}
\bbeta_{\calM_{\hat\bpsi^*}(\tilde{c}_2)}(\hat\bpsi^*):=\Bigg(\begin{array}{c}\bar\bg^{\calI}(\hat\bpsi^*)\\\bar\bg_{\cald_{\hat\bpsi^*}(\tilde{c}_2)}^{\calT}(\hat\bpsi^*)-\nu\rho'_2(0^+)\sgn\{\bar\bg_{\cald_{\hat\bpsi^*}(\tilde{c}_2)}^{\calT}(\hat\bpsi^*)\}
\end{array}\Bigg)\,.
\end{equation*}
It holds that
\begin{align}
|\bbeta_{\calM_{\hat\bpsi^*}(\tilde{c}_2)}(\hat\bpsi^*)|_2\le &~ |\bar\bg^{\calI}(\hat\bpsi^*)|_2 + \big|\bar\bg_{\cald_{\hat\bpsi^*}(\tilde{c}_2)}^{\calT}(\hat\bpsi^*)-\nu\rho'_2(0^+)\sgn\big\{\bar\bg_{\cald_{\hat\bpsi^*}(\tilde{c}_2)}^{\calT}(\hat\bpsi^*)\big\}\big|_2 \notag\\
\le &~ |\bar\bg^{\calI}(\hat\bpsi)|_2 + \underbrace{\big|\bar\bg_{\cald_{\hat\bpsi^*}(\tilde{c}_2) \cap \cald_{\hat\bpsi}(\tilde{c}_1)}^{\calT}(\hat\bpsi^*)-\nu\rho'_2(0^+)\sgn\big\{\bar\bg_{\cald_{\hat\bpsi^*}(\tilde{c}_2) \cap \cald  _{\hat\bpsi}(\tilde{c}_1)}^{\calT}(\hat\bpsi^*)\big\}\big|_2}_{T_1} \label{betastar}\\
&~  +  \underbrace{|\bar\bg^{\calI}(\hat\bpsi^*) - \bar\bg^{\calI}(\hat\bpsi)|_2}_{T_2} + \underbrace{\big|\bar\bg_{\cald_{\hat\bpsi^*}(\tilde{c}_2) \cap \cald^{\c}_{\hat\bpsi}(\tilde{c}_1)}^{\calT}(\hat\bpsi^*)-\nu\rho'_2(0^+)\sgn\big\{\bar\bg_{\cald_{\hat\bpsi^*}(\tilde{c}_2) \cap \cald^{\c}_{\hat\bpsi}(\tilde{c}_1)}^{\calT}(\hat\bpsi^*)\big\}\big|_2}_{T_3} \notag\,.
\end{align}
As we have shown that $|\bbeta_{\calM_{\hat\bpsi}(\tilde{c}_1)}(\hat\bpsi)|_2^2=|\bar\bg^{\calI}(\hat\bpsi)|_2^2+|\bar\bg_{\cald_{\hat\bpsi}(\tilde{c}_1)}^{\calT}(\hat\bpsi)-\nu\rho'_2(0^+)\sgn\{\bar\bg_{\cald_{\hat\bpsi}(\tilde{c}_1)}^{\calT}(\hat\bpsi)\}|_2^2=O_\p(r_1\aleph_n^2)$, then $|\bar\bg^{\calI}(\hat\bpsi)|_2=O_\p(r_1^{1/2}\aleph_n)=|\bar\bg_{\cald_{\hat\bpsi}(\tilde{c}_1)}^{\calT}(\hat\bpsi)-\nu\rho'_2(0^+)\sgn\{\bar\bg_{\cald_{\hat\bpsi}(\tilde{c}_1)}^{\calT}(\hat\bpsi)\}|_2$. For the term $T_1$ in \eqref{betastar}, we have
\begin{align*}
T_1\le&~ \big|\bar\bg_{\cald_{\hat\bpsi}(\tilde{c}_1)}^{\calT}(\hat\bpsi^*)-\nu\rho'_2(0^+)\sgn\big\{\bar\bg_{\cald_{\hat\bpsi}(\tilde{c}_1)}^{\calT}(\hat\bpsi^*)\big\}\big|_2 \\
\le&~ \big|\bar\bg_{\cald_{\hat\bpsi}(\tilde{c}_1)}^{\calT}(\hat\bpsi)-\nu\rho'_2(0^+)\sgn\big\{\bar\bg_{\cald_{\hat\bpsi}(\tilde{c}_1)}^{\calT}(\hat\bpsi)\big\}\big|_2 + \big| \bar\bg_{\cald_{\hat\bpsi}(\tilde{c}_1)}^{\calT}(\hat\bpsi^*) - \bar\bg_{\cald_{\hat\bpsi}(\tilde{c}_1)}^{\calT}(\hat\bpsi)\big|_2 \\
&~ + \nu\rho'_2(0^+)\big|\sgn\big\{\bar\bg_{\cald_{\hat\bpsi}(\tilde{c}_1)}^{\calT}(\hat\bpsi)\big\} - \sgn\big\{\bar\bg_{\cald_{\hat\bpsi}(\tilde{c}_1)}^{\calT}(\hat\bpsi^*)\big\}\big|_2\\
\le&~O_\p(r_1^{1/2}\aleph_n)+\big| \bar\bg_{\cald_{\hat\bpsi}(\tilde{c}_1)}^{\calT}(\hat\bpsi^*) - \bar\bg_{\cald_{\hat\bpsi}(\tilde{c}_1)}^{\calT}(\hat\bpsi)\big|_2+\nu\rho'_2(0^+)\big|\sgn\big\{\bar\bg_{\cald_{\hat\bpsi}(\tilde{c}_1)}^{\calT}(\hat\bpsi)\big\} - \sgn\big\{\bar\bg_{\cald_{\hat\bpsi}(\tilde{c}_1)}^{\calT}(\hat\bpsi^*)\big\}\big|_2\,.
\end{align*}
Notice that $\max_{j\in\mathcal{D}}|\bar{g}_j^{\calT}(\hat{\bpsi}^*)-\bar{g}_j^{\calT}(\hat{\bpsi})|\leq |\hat{\bpsi}_{\calS^{\c}}|_1\cdot O_\p(1)$. Then $| \bar\bg_{\cald_{\hat\bpsi}(\tilde{c}_1)}^{\calT}(\hat\bpsi^*) - \bar\bg_{\cald_{\hat\bpsi}(\tilde{c}_1)}^{\calT}(\hat\bpsi)|_2 \le \ell_n^{1/2} |\hat\bpsi_{\calS^{\c}}|_1\cdot O_\p(1) = O_\p(\ell_n^{1/2}\aleph_n)$. Recall $| \bar{g}_j^{\calT}(\hat{\bpsi}) | \ge \tilde{c}_1 \nu\rho'_2(0^+)$ for any $j \in \mathcal{D}_{\hat\bpsi}(\tilde{c}_1)$. Due to $|\hat{\bpsi}_{\calS^{\c}}|_1\leq \aleph_n=o(\nu)$, it holds that $\sgn\{\bar{g}_j^{\calT}(\hat{\bpsi}^*)\}=\sgn\{\bar{g}_j^{\calT}(\hat{\bpsi})\}$ for any $j\in\mathcal{D}_{\hat{\bpsi}}(\tilde{c}_1)$. Hence, $T_1=O_\p(\ell_n^{1/2}\aleph_n)$. Analogously, we also have
$
T_2\le r_1^{1/2}|\hat\bpsi_{\calS^{\c}}|_1\cdot O_\p(1)= O_\p(\ell_n^{1/2}\aleph_n)$.
For the term $T_3$, notice that for any $j\in\mathcal{D}_{\hat{\bpsi}^*}(\tilde{c}_2)\cap\mathcal{D}_{\hat{\bpsi}}^{\c}(\tilde{c}_1)$, we have $|\bar{g}_j^{\calT}(\hat{\bpsi}^*)|\geq \tilde{c}_2\nu\rho_2'(0^+)$ and $|\bar{g}_j^{\calT}(\hat{\bpsi})|<\tilde{c}_1\nu\rho_2'(0^+)$ for some $\tilde{c}_2>\tilde{c}_1$. Since  $\max_{j\in\mathcal{T}}|\bar{g}_j^{\calT}(\hat{\bpsi})-\bar{g}_j^{\calT}(\hat{\bpsi}^*)| \leq |\hat{\bpsi}_{\mathcal{S}^{\c}}|_1\cdot O_\p(1)=o_\p(\nu)$, it holds that $\mathcal{D}_{\hat{\bpsi}^*}(\tilde{c}_2)\cap\mathcal{D}_{\hat{\bpsi}}^{\c}(\tilde{c}_1)=\emptyset$ w.p.a.1, which implies that $T_3=0$ w.p.a.1. Therefore, we have $|\bbeta_{\calM_{\hat\bpsi^*}(\tilde{c}_2)}(\hat\bpsi^*)|_2=O_\p(\ell_n^{1/2}\aleph_n)$. Together with Lemma \ref{prop3}, we have $|\hat{\blambda}^*|_2=O_\p(\ell_n^{1/2}\aleph_n)$, which implies
$
\max_{i\in[n]}|\hat{\blambda}^{*,\T}\bg^{\calT}_i(\check{\bpsi})|=\max_{i\in[n]}|\hat{\blambda}_{\calM_{\hat{\bpsi}^*}(\tilde{c}_2)}^{*,\T}\bg^{\calT}_{i,{\calM_{\hat{\bpsi}^*}(\tilde{c}_2)}}(\check{\bpsi})|=o_\p(1)$
for $\check{\bpsi}$ specified in \eqref{eq:sn1}.

Recall $\bg_i^{\calT}(\bpsi)=\{\bg_i^{\calI}(\btheta)^\T,\bg_i^{\calD}(\btheta)^\T-\bxi^\T\}^\T$ and $\bpsi_{\calS^{\c}}=\bxi_{\cala}$. We have $\hat{\blambda}^{*,\T}\nabla_{\bpsi_{\calS^{\c}}}\bg_{i}^{\calT}(\check{\bpsi})=-\hat{\blambda}^{*,\T}_{\cala}$. For ${\rm I}$, since $\max_{i\in[n]}|\hat\blambda^{*,\T}\bg_{i}^{\calT}(\check{\bpsi})|=o_\p(1)$, we have
\be\label{I}
\begin{split}
    |{\rm I}|=\bigg|\frac{1}{n}\sum_{i=1}^n\frac{\hat{\blambda}^{*,\T}_{\cala}\hat\bpsi_{\calS^{\c}}}{1+\hat\blambda^{*,\T}\bg_{i}^{\calT}(\check{\bpsi})}\bigg|&\le|\hat\blambda^*|_\infty|\hat\bpsi_{\calS^{\c}}|_1\{1+o_\p(1)\}\leq |\hat\blambda^*|_2|\hat\bpsi_{\calS^{\c}}|_1\{1+o_\p(1)\}\,.\\
\end{split}
\ee
As we have shown $|\hat{\blambda}^*|_2=O_\p(\ell_n^{1/2}\aleph_n)$, 
it then holds $|{\rm I}|\le|\hat\bpsi_{\calS^{\c}}|_1\cdot O_\p(\ell_n^{1/2}\aleph_n)$. On the other hand, ${\rm II}$ in \eqref{eq:sn1} satisfies
$
{\rm II}=\sum_{k\in \mathcal{A}}\pi\rho'_1(c_k|\hat\xi_k|;\pi)|\hat\xi_k|\ge C\pi|\hat\bpsi_{\calS^{\c}}|_1
$
for some $c_k\in(0,1)$. Due to $\ell_n^{1/2}\aleph_n=o(\pi)$, we can obtain ${\rm I}+{\rm II}>0$ w.p.a.1, which implies that $S_n(\hat\bpsi^*)<S_n(\hat\bpsi)$ w.p.a.1. We complete the proof of Part (iii).$\hfill\Box$

\subsection{Proof of Theorem \ref{thm2}}\label{se:b2}

Select $\hat{\bpsi}_{\PEL}$ as the sparse local minimizer given in Proposition \ref{thm1}. Recall that the estimate $\hat{\blambda}(\bpsi)=\arg\max_{\blambda\in\hat{\Lambda}_n^{\calT}(\bpsi)}f(\blambda;\bpsi)$ is the Lagrange multiplier associated with $\bpsi$, where $
f(\blambda;\bpsi)=n^{-1}\sum_{i=1}^n\log\{1+\blambda^\T\bg^{\calT}_{i}(\bpsi)\}-\sum_{j\in\mathcal{D}}P_{2,\nu}(|\lambda_j|)
$
for any $\bpsi \in\bPsi$ and $\blambda = (\lambda_1, \ldots, \lambda_r)^\T$. Write $\hat{\blambda}=\hat\blambda(\hat\bpsi_{\PEL})=(\hat\lambda_1,\ldots,\hat\lambda_r)^\T$. Recall $\mathcal{R}_n=\mathcal{I}\cup \supp(\hat\blambda_{\mathcal{D}})$, $\mathcal{A}_*=\{ j \in \mathcal{A}:\hat{\lambda}_j\neq 0\}$ and $\mathcal{A}_{*,\c}=\{ j\in \mathcal{A}^{\c}:\hat{\lambda}_j\neq 0\}$. Then $\mathcal{R}_n$ can be decomposed into three disjoint sets $\mathcal{R}_n=\mathcal{I}\cup\mathcal{A}_*\cup\mathcal{A}_{*,\c}$. Write $\mathcal{I}^*=\mathcal{I}\cup\mathcal{A}_*$. Notice that $\mathcal{S}_*= \mathcal{P} \cup  \mathcal{A}_{*,\c} $ and $\mathcal{S}=\mathcal{P}\cup\mathcal{A}^{\c}$. Then $\mathcal{S}_*\subset\mathcal{S}$ and $s_*:=|\mathcal{S}_*|\le |\mathcal{S}|=s$.  For any $\bpsi\in\bPsi$, we have $\bpsi_{\calS_*}=(\btheta^\T,\bxi^\T_{{\cala}_{*,\c}})^\T$. To prove Theorem \ref{thm2}, we also need the following two lemmas. The proof of Lemma \ref{lem3} is similar to that of Lemma 3 in \cite{ChangTangWu2018} and we omit it here. The proof of Lemma \ref{lemenvlope} is given in Section \ref{sec:pflem5}.

\begin{lemma}\label{lem3}
Assume the conditions of Proposition {\rm\ref{thm1}} hold. Then $\sup_{\mathcal{F}\in\mathscr{F}}\big|[\nabla_{\bpsi_{\calS_*}}\bar\bg^{\calT}_{\calF}(\hat\bpsi_{\PEL})-\mathbb{E}\{\nabla_{\bpsi_{\calS_*}}\bg^{\calT}_{i,\calF}(\bpsi_0)\}]\bz\big|_2=|\bz|_2\cdot\{O_\p(s^{3/2}\ell_n^{1/2}\phi_n)+O_\p(s^{1/2}\ell_n^{1/2}\aleph_n)\}$ holds uniformly over $\bz\in\mathbb{R}^{s_*}$, where $\mathscr{F}$ is defined in Lemma {\rm\ref{lem1}}.
\end{lemma}

\begin{lemma}\label{lemenvlope}
Assume Condition {\rm\ref{AA1}} and the conditions of Proposition {\rm\ref{thm1}} hold. It then holds w.p.a.1 that $\hat{\blambda}(\bpsi)$ is continuously differentiable at $\hat{\bpsi}_{\PEL}$ and $\nabla_{\bpsi}\hat{\blambda}_{\calR_n^{\c}}(\hat{\bpsi}_{\PEL})=\bzero$.
\end{lemma}

Now we begin to prove Theorem \ref{thm2}. Define
 \begin{align}\label{eq:Hpsilam}
H_n(\bpsi,\blambda)=\frac{1}{n}\sum_{i=1}^n\log\{1+\blambda^\T\bg^{\calT}_{i}(\bpsi)\}+\sum_{k\in\mathcal{D}}P_{1,\pi}(|\xi_k|)-\sum_{j\in\mathcal{D}}P_{2,\nu}(|\lambda_j|)\,.
 \end{align}
By the definition of $\hat\blambda$, we have $\nabla_{\blambda} {H}_n(\hat\bpsi_{\PEL},\hat\blambda)=\bzero$, that is,
\[
\bzero=\frac{1}{n}\sum_{i=1}^n\frac{\bg_{i}^{\calT}(\hat\bpsi_{\PEL})}{1+\hat\blambda^\T\bg_{i}^{\calT}(\hat\bpsi_{\PEL})}-\hat\bfeta\,,
\]
where $\hat\bfeta=(\hat\eta_1,\ldots,\hat\eta_r)^\T$ with $\hat\eta_j=0$ for $j\in\mathcal{I}$, $\hat\eta_j=\nu\rho'_2(|\hat\lambda_j|;\nu){\rm sgn}(\hat\lambda_j)$ for $j\in\mathcal{D}$ and $\hat\lambda_j\neq0$, and $\hat\eta_j \in[-\nu\rho'_2(0^+),\nu\rho'_2(0^+)]$
for $j\in\mathcal{R}_n^{\c}$. It follows from the Taylor expansion that
\begin{align}
\bzero&=\frac{1}{n}\sum_{i=1}^n \bg_{i,\calR_n}^{\calT}(\hat\bpsi_{\PEL})-\frac{1}{n}\sum_{i=1}^n\frac{ \bg_{i,\calR_n}^{\calT}(\hat\bpsi_{\PEL})^{\otimes2}\hat\blambda_{\calR_n}}{\{1+c\hat\blambda_{\calR_n}^\T\bg_{i,\calR_n}^{\calT}(\hat\bpsi_{\PEL})\}^2}-\hat\bfeta_{\calR_n}\nonumber\\
&=:\bar\bg_{\calR_n}^{\calT}(\hat\bpsi_{\PEL})-\bC(\hat\bpsi_{\PEL})\hat\blambda_{\calR_n}-\hat\bfeta_{\calR_n}\label{optlambda}
\end{align}
for some $|c|<1$. Hence,
$\hat\blambda_{\calR_n}=\{\bC(\hat\bpsi_{\PEL})\}^{-1}\{\bar\bg_{\calR_n}^{\calT}(\hat\bpsi_{\PEL})-\hat\bfeta_{\calR_n}\}$. By the definition of $\hat\bpsi_\PEL$, we have $\bzero=\nabla_{\bpsi}{H}_n\{\bpsi,\hat\blambda(\bpsi)\}|_{\bpsi=\hat\bpsi_{\PEL}}$. Notice that
\begin{align*}
&~\nabla_{\bpsi}{H}_n\{\bpsi,\hat\blambda(\bpsi)\}|_{\bpsi=\hat\bpsi_{\PEL}}\\
=&~\frac{\partial{H}_n(\hat\bpsi_\PEL,\hat\blambda)}{\partial\bpsi} + \bigg\{\underbrace{\frac{\partial{H}_n(\hat\bpsi_\PEL,\hat\blambda)}{\partial\blambda_{\calR_n}^\T}}_{\rm I} \frac{\partial\hat\blambda_{\calR_n}(\hat{\bpsi}_{\PEL})}{\partial\bpsi} + \frac{\partial{H}_n(\hat\bpsi_\PEL,\hat\blambda)}{\partial\blambda_{\calR_n^{\c}}^\T} \underbrace{\frac{\partial\hat\blambda_{\calR_n^{\c}}(\hat{\bpsi}_{\PEL})}{\partial\bpsi}}_{\rm II}\bigg\}^\T\,.
\end{align*}
Due to $\hat{\blambda}(\hat{\bpsi}_{\PEL})=\arg\max_{\blambda\in\hat{\Lambda}_n^{\calT}(\hat{\bpsi}_{\PEL})}f(\blambda;\hat{\bpsi}_{\PEL})=\arg\max_{\blambda\in\hat{\Lambda}_n^{\calT}(\hat{\bpsi}_{\PEL})}H_n(\hat{\bpsi}_{\PEL},\blambda)$, then ${\rm I} = \bzero$. On the other hand, Lemma \ref{lemenvlope} implies that ${\rm II} = \bzero$. Thus, $\bzero=\partial{H}_n(\hat\bpsi_\PEL,\hat\blambda)/\partial\bpsi$. Together with \eqref{optlambda}, we have
\begin{align}
    \bzero&=\bigg\{\frac{1}{n}\sum_{i=1}^n\frac{\nabla_{\bpsi_{\calS_*}}\bg_{i,\calR_n}^{\calT}(\hat\bpsi_{\PEL})}{1+\hat\blambda_{\calR_n}^\T\bg_{i,\calR_n}^{\calT}(\hat\bpsi_{\PEL})}\bigg\}^\T\hat\blambda_{\calR_n}+\hat{\bfvarsigma}_{\calS_*} \nonumber\\
    &=\bigg\{\frac{1}{n}\sum_{i=1}^n\frac{\nabla_{\bpsi_{\calS_*}}\bg_{i,\calR_n}^{\calT}(\hat\bpsi_{\PEL})}{1+\hat\blambda_{\calR_n}^\T\bg_{i,\calR_n}^{\calT}(\hat\bpsi_{\PEL})}\bigg\}^\T\{\bC(\hat\bpsi_{\PEL})\}^{-1}\{\bar\bg_{\calR_n}^{\calT}(\hat\bpsi_{\PEL})-\hat\bfeta_{\calR_n}\}+\hat{\bfvarsigma}_{\calS_*}\label{7.7}\\
    &=:\{\bD(\hat\bpsi_{\PEL})\}^\T\{\bC(\hat\bpsi_{\PEL})\}^{-1}\{\bar\bg_{\calR_n}^{\calT}(\hat\bpsi_{\PEL})-\hat\bfeta_{\calR_n}\}+\hat{\bfvarsigma}_{\calS_*} \,, \nonumber
\end{align}
where $\hat\bfvarsigma_{\calS_*}=\{\sum_{k\in\mathcal{D}}\nabla_{\bpsi_{\calS_*}}P_{1,\pi}(|\xi_k|)\}|_{\bpsi=\hat\bpsi_{{\PEL}}}$. Recall that $\mathcal{S}_*=\mathcal{P}\cup\mathcal{A}_{*,\c}$. Proposition \ref{thm1} and \eqref{eq:xi} imply that $\hat\bfvarsigma_{\calS_*}=\bzero$ w.p.a.1. To construct the asymptotic normality, we need the following lemma. The proof of Lemma \ref{lem2} is similar to that of Lemma 2 in \cite{ChangTangWu2018} and we omit it here.
\begin{lemma}\label{lem2}
Assume the conditions of Proposition {\rm\ref{thm1}} hold. Then
$
\|\bC(\hat\bpsi_{\PEL})-\widehat\bV_{\calR_n}^{\calT}(\hat\bpsi_{\PEL})\|_2=O_\p(\ell_nn^{1/\gamma}\aleph_n)
$,
and
$
|\{\bD(\hat\bpsi_{\PEL})-\nabla_{\bpsi_{\calS_*}}\bar\bg_{\calR_n}^{\calT}(\hat\bpsi_{\PEL})\}\bz|_2=|\bz|_2\cdot O_\p(\ell_ns^{1/2}\aleph_n)
$
holds uniformly over $\bz\in\mathbb{R}^{s_*}$.
\end{lemma}

Recall
\begin{align}\label{JRn}
\bJ^{\calT}_{\calR_n}=([\mathbb{E}\{\nabla_{\bpsi_{\calS_*}}\bg^{\calT}_{i,\calR_n}(\bpsi_0)\}]^\T\{\bV^{\calT}_{\calR_n}(\bpsi_0)\}^{-1/2})^{\otimes2}\,.
\end{align}
For any $\balpha\in\mathbb{R}^{s_*}$ with unit $L_2$-norm, let $\bdelta=\{\bJ^{\calT}_{\calR_n}\}^{-1/2}\balpha$. Following the same arguments stated in the proof of Proposition \ref{pn:0}, we have $|\bdelta|_2=O(1)$. Lemma \ref{prop3} indicates that $\mathcal{R}_n\subset\mathcal{M}_{\hat{\bpsi}_{{\PEL}}}(\tilde{c})=\mathcal{I}\cup\mathcal{D}_{\hat{\bpsi}_{\PEL}}(\tilde{c})$ w.p.a.1 for some $\tilde{c}\in(C_*,1)$. As we have shown in the proof of Proposition \ref{thm1}, it holds that
\begin{align}
|\bar\bg^{\calT}_{\calR_n}(\hat\bpsi_{\PEL})-\hat\bfeta_{\calR_n}|_2
\le&~\Bigg|\begin{array}{c}\bar\bg^{\calI}(\hat\bpsi_{\PEL})\\\bar\bg^{\calT}_{\cald_{\hat\bpsi_{\PEL}}(\tilde{c})}(\hat\bpsi_{\PEL})-\nu\rho'_2(0^+)\sgn\{\bar\bg^{\calT}_{\cald_{\hat\bpsi_{\PEL}}(\tilde{c})}(\hat\bpsi_{\PEL})\}
\end{array}\Bigg|_2=O_\p(\ell_n^{1/2} \aleph_n)\,.\label{7.171}
\end{align}
Together with Lemmas \ref{lem1},  \ref{lem3}, and \ref{lem2}, (\ref{7.7}) implies that
\begin{align*}
&\bdelta^\T[\mathbb{E}\{\nabla_{\bpsi_{\calS_*}}\bg^{\calT}_{i,\calR_n}(\bpsi_0)\}]^\T\{\bV_{\calR_n}^{\calT}(\bpsi_{0})\}^{-1}\{\bar\bg^{\calT}_{\calR_n}(\hat\bpsi_{\PEL})-\hat\bfeta_{\calR_n}\}\\
&~~~~~~~~~~~=O_\p(\ell_n^{3/2}s^{1/2}\aleph_n^2)+O_\p(\ell_ns^{3/2}\phi_n\aleph_n)+O_\p(\ell_n^{3/2}n^{1/\gamma}\aleph_n^2)\,.
\end{align*}
Notice that $\mathbb{P}(\hat{\bpsi}_{{\PEL},\calS^{\c}}=\bzero)\rightarrow1$ and $\bpsi_{0,\mathcal{S}^{\c}}=\bzero$. By the Taylor expansion, we have
$
\bar\bg^{\calT}_{\calR_n}(\hat\bpsi_{\PEL})=\bar\bg^{\calT}_{\calR_n}(\bpsi_0)+\nabla_{\bpsi_{\calS}}\bar\bg^{\calT}_{\calR_n}(\tilde\bpsi)(\hat\bpsi_{{\PEL},{\calS}}-\bpsi_{0,{\calS}})$ w.p.a.1,
where $\tilde\bpsi$ is on the line joining $\bpsi_0$ and $\hat\bpsi_{\PEL}$. Recall that $\mathcal{R}_n=\mathcal{I}\cup\mathcal{A}_*\cup\mathcal{A}_{*,\c}$ with $\mathcal{A}_*=\{j\in\mathcal{A}:\hat{\lambda}_j\neq 0\}$ and $\mathcal{A}_{*,\c}=\{j\in\mathcal{A}^{\c}:\hat{\lambda}_j\neq 0\}$. Notice that $\mathcal{S}_*=\mathcal{P}\cup\mathcal{A}_{*,\c}\subset\mathcal{S}$, $\bpsi_{\calS}=(\btheta^\T,\bxi^\T_{\cala^{\c}})^\T$ and $\bpsi_{\calS_*}=(\btheta^\T,\bxi^\T_{\cala_{*,\c}})^\T$. For any $j\in\mathcal{R}_n$ and $k\in\mathcal{S}\backslash\mathcal{S}_*$, we know that $g^\calT_{i,j}(\bpsi)$ does not involve $\psi_k$, which implies that $\partial\bar{g}^{\calT}_j(\tilde{\bpsi})/\partial\psi_k=0$. Therefore, it holds that
$
\bar\bg^{\calT}_{\calR_n}(\hat\bpsi_{\PEL})=\bar\bg^{\calT}_{\calR_n}(\bpsi_0)+\nabla_{\bpsi_{\calS_*}}\bar\bg^{\calT}_{\calR_n}(\tilde\bpsi)(\hat\bpsi_{{\PEL},{\calS_*}}-\bpsi_{0,{\calS_*}})$ w.p.a.1,
which leads to
\begin{align}
&~\bdelta^\T[\mathbb{E}\{\nabla_{\bpsi_{\calS_*}}\bg^{\calT}_{i,\calR_n}(\bpsi_0)\}]^\T\{\bV_{\calR_n}^{\calT}(\bpsi_{0})\}^{-1}\{\nabla_{\bpsi_{\calS_*}}\bar\bg^{\calT}_{\calR_n}(\tilde\bpsi)(\hat\bpsi_{{\PEL},{\calS_*}}-\bpsi_{0,{\calS_*}})-\hat\bfeta_{\calR_n}\} \notag\\
=&-\bdelta^\T[\mathbb{E}\{\nabla_{\bpsi_{\calS_*}}\bg^{\calT}_{i,\calR_n}(\bpsi_0)\}]^\T\{\bV_{\calR_n}^{\calT}(\bpsi_{0})\}^{-1}\bar\bg^{\calT}_{\calR_n}(\bpsi_0)+O_\p(\ell_n^{3/2}s^{1/2}\aleph_n^2)\label{7.8}\\
&+O_\p(\ell_ns^{3/2}\phi_n\aleph_n)+O_\p(\ell_n^{3/2}n^{1/\gamma}\aleph_n^2)\,.\notag
\end{align}

Next, we will specify the convergence rate of $|[\mathbb{E}\{\nabla_{\bpsi_{\calS_*}}\bg^{\calT}_{i,\calR_n}(\bpsi_0)\}
-\nabla_{\bpsi_{\calS_*}}\bar\bg^{\calT}_{\calR_n}(\tilde\bpsi)](\hat\bpsi_{{\PEL},{\calS_*}}-\bpsi_{0,{\calS_*}})|_2$. Since $\ell_n \aleph_n=o(\nu)$, it follows from \eqref{7.171} that
$
|\bar\bg^{\calT}_{\calR_n}(\hat\bpsi_{\PEL})-\bar\bg^{\calT}_{\calR_n}(\bpsi_0)|_2\le|\bar\bg^{\calT}_{\calR_n}(\hat\bpsi_{\PEL})|_2+|\bar\bg^{\calT}_{\calR_n}(\bpsi_0)|_2=O_\p(\ell_n^{1/2}\nu)+O_\p(\ell_n^{1/2}\aleph_n)=O_\p(\ell_n^{1/2}\nu)$.
On the other hand, it holds that
$
|\bar\bg^{\calT}_{\calR_n}(\hat\bpsi_{\PEL})-\bar\bg^{\calT}_{\calR_n}(\bpsi_0)|_2\ge\lambda_{\rm min}^{1/2}([\{\nabla_{\bpsi_{\calS_*}}\bar\bg^{\calT}_{\calR_n}(\bar{\bpsi})\}^\T]^{\otimes2})|\hat\bpsi_{{\PEL},{\calS_*}}-\bpsi_{0,{\calS_*}}|_2$,
where $\bar{\bpsi}$ is on the line joining $\bpsi_0$ and $\hat\bpsi_{\PEL}$. Similar to Lemma \ref{lem3}, Condition \ref{A.7} implies that $|\hat\bpsi_{{\PEL},{\calS_*}}-\bpsi_{0,{\calS_*}}|_2=O_\p(\ell_n^{1/2}\nu)$. Hence, by similar arguments of Lemma \ref{lem3}, we have $|[\mathbb{E}\{\nabla_{\bpsi_{\calS_*}}\bg^{\calT}_{i,\calR_n}(\bpsi_0)\}
-\nabla_{\bpsi_{\calS_*}}\bar\bg^{\calT}_{\calR_n}(\tilde\bpsi)](\hat\bpsi_{{\PEL},{\calS_*}}-\bpsi_{0,{\calS_*}})|_2=O_\p(\ell_ns^{3/2}\phi_n\nu)+O_\p(\ell_ns^{1/2}\nu\aleph_n)$.
Recall $
\hat\bzeta_{\calR_n}=\{\bJ^{\calT}_{\calR_n}\}^{-1}[\mathbb{E}\{\nabla_{\bpsi_{\calS_*}}\bg^{\calT}_{i,\calR_n}(\bpsi_0)\}]^\T\{\bV^{\calT}_{\calR_n}(\bpsi_0)\}^{-1}\hat\bfeta_{\calR_n}$
with $\bJ^{\calT}_{\calR_n}$ defined in (\ref{JRn}). Recall $\bdelta=\{\bJ^{\calT}_{\calR_n}\}^{-1/2}\balpha$.
It follows from (\ref{7.8}) that
\begin{align*}
\bdelta^\T\bJ^{\calT}_{\calR_n}(\hat\bpsi_{{\PEL},{\calS_*}}-\bpsi_{0,{\calS_*}}-\hat\bzeta_{\calR_n}) =&-\balpha^\T\{\bJ^{\calT}_{\calR_n}\}^{-1/2}[\mathbb{E}\{\nabla_{\bpsi_{\calS_*}}\bg^{\calT}_{i,\calR_n}(\bpsi_0)\}]^\T\{\bV_{\calR_n}^{\calT}(\bpsi_{0})\}^{-1}\bar\bg^{\calT}_{\calR_n}(\bpsi_0)\notag\\
&+O_\p(\ell_n^{3/2}n^{1/\gamma}\aleph_n^2)+O_\p(\ell_ns^{3/2}\phi_n \nu)+O_\p(\ell_ns^{1/2}\nu\aleph_n)\,.
\end{align*}
Lemma 4 of \cite{ChangTangWu2018} yields
$
n^{1/2}\balpha^\T\{\bJ^{\calT}_{\calR_n}\}^{-1/2}[\mathbb{E}\{\nabla_{\bpsi_{\calS_*}}\bg^{\calT}_{i,\calR_n}(\bpsi_0)\}]^\T\{\bV_{\calR_n}^{\calT}(\bpsi_{0})\}^{-1}\bar\bg^{\calT}_{\calR_n}(\bpsi_0)
\stackrel{d}{\rightarrow}
\mathcal{N}(0,1)$. Then
$
n^{1/2}\balpha^\T\{\bJ^{\calT}_{\calR_n}\}^{1/2}(\hat\bpsi_{{\PEL},{\calS_*}}-\bpsi_{0,{\calS_*}}-\hat\bzeta_{\calR_n})
\stackrel{d}{\rightarrow}
\mathcal{N}(0,1)
$ as $n\to\infty$.

Notice that $\hat\bpsi_{\PEL,\calS_*}=(\hat{\btheta}_{\PEL}^\T,\hat{\bxi}_{\PEL,{\cala}_{*,\c}}^\T)^\T$. In the sequel, we will specify the limiting distribution of $\hat{\btheta}_{\PEL}$. Recall $\mathcal{R}_n=\mathcal{I}\cup\mathcal{A}_*\cup\mathcal{A}_{*,\c}$ and $\mathcal{I}^*=\mathcal{I}\cup\mathcal{A}_*$. We write $\{\bJ^{\calT}_{\calR_n}\}^{-1}$, $\mathbb{E}\{\nabla_{\bpsi_{\calS_*}}\bg^{\calT}_{i,\calR_n}(\bpsi_0)\}$ and $\{\bV_{\calR_n}^{\calT}(\bpsi_0)\}^{-1}$ with following blocks
\begin{align*}
&\{\bJ^{\calT}_{\calR_n}\}^{-1}=\left(\begin{array}{cc}
~[\{\bJ^{\calT}_{\calR_n}\}^{-1}]_{11}     &  [\{\bJ^{\calT}_{\calR_n}\}^{-1}]_{12}\\
~[\{\bJ^{\calT}_{\calR_n}\}^{-1}]_{21}   & [\{\bJ^{\calT}_{\calR_n}\}^{-1}]_{22}
\end{array}\right)\,,~~~\{\bV_{\calR_n}^{\calT}(\bpsi_0)\}^{-1}=\left(\begin{array}{cc} \bS_{11}&\bS_{12}\\ \bS_{21} & \bS_{22}\end{array} \right)\,,\\
&~~~~~~\mathbb{E}\{\nabla_{\bpsi_{\calS_*}}\bg^{\calT}_{i,\calR_n}(\bpsi_0)\}=\left(\begin{array}{cc}
\mathbb{E}\{\nabla_{\btheta}\bg^{\calT}_{i,\cali^*}(\btheta_0)\}     &  \bzero\\
\mathbb{E}\{\nabla_{\btheta}\bg^{\calT}_{i,\cala_{*,\c}}(\btheta_0)\}     & -{\bf I}
\end{array}\right)=:\left(\begin{array}{cc} \bG_{\cali^*}&\bzero\\ \bG_{\cala_{*,\c}}& -\bI\end{array} \right)\,,
\end{align*}
where $[\{\bJ^{\calT}_{\calR_n}\}^{-1}]_{11}$ is a $p\times p$ matrix, and $\bS_{11}$ is an $|\mathcal{I}^*|\times |\mathcal{I}^*|$ matrix. Recall $\bV^{\calT}_{\cali^*}(\btheta_0)=\mathbb{E}\{\bg_{i,\cali^*}^{\calT}(\btheta_0)^{\otimes2}\}$ and
$\bJ^{\calT}_{\cali^*}=([\mathbb{E}\{\nabla_{\btheta}\bg^{\calT}_{i,\cali^*}(\btheta_0)\}]^\T\{\bV^{\calT}_{\cali^*}(\btheta_0)\}^{-1/2})^{\otimes2}$.
Then $\bJ_{\cali^*}^{\calT}=\bG_{\cali^*}^\T(\bS_{11}-\bS_{12}\bS_{22}^{-1}\bS_{21})\bG_{\cali^*}$.
By \eqref{JRn}, we have
$
[\{\bJ^{\calT}_{\calR_n}\}^{-1}]_{11}=\{\bG_{\cali^*}^\T(\bS_{11}-\bS_{12}\bS_{22}^{-1}\bS_{21})\bG_{\cali^*}\}^{-1}=\{\bJ_{\cali^*}^{\calT}\}^{-1}$. For any $\tilde\balpha\in\mathbb{R}^p$ with unit $L_2$-norm, let
$
\balpha=\{\bJ_{\calR_n}^{\calT}\}^{-1/2}[
 \{\bJ_{\cali^*}^{\calT}\}^{1/2} , \bzero
]^\T\tilde\balpha$. Then $
|\balpha|_2^2=\tilde\balpha^\T\tilde\balpha
=1$.
Hence,
$
\tilde\balpha^\T \{\bJ_{\cali^*}^{\calT}\}^{1/2}\{\hat{\btheta}_{\PEL}-\btheta_0-\hat{\bzeta}_{\calR_n,(1)}\}=\balpha^\T\{\bJ_{\calR_n}^{\calT}\}^{1/2}(\hat\bpsi_{{\PEL},{\calS_*}}-\bpsi_{0,{\calS_*}}-\hat\bzeta_{\calR_n})
\stackrel{d}{\rightarrow}
\mathcal{N}(0,1)
$, where $
\hat{\bzeta}_{\calR_n,(1)}$ is the first $p$ components of $\hat{\bzeta}_{\calR_n}$. We complete the proof of Theorem \ref{thm2}. $\hfill\Box$

\subsection{Proof of Theorem \ref{thm3}}\label{se:pn4}
Recall $a_n=\sum_{k\in\mathcal{D}}P_{1,\pi}(|\xi_{0,k}|)+\sum_{l\in\mathcal{P}}P_{1,\pi}(|\theta_{0,l}|)$ and $\mathcal{S}=\mathcal{P}_{\sharp}\cup\mathcal{A}^{\c}$ with $s=|\mathcal{S}|$ in the current setting. Define $b_{1,n}=\max\{a_n,r_1\aleph_n^2\}$ and $b_{2,n}=\max\{b_{1,n},\nu^2 \}$. Then $
\phi_n=\max\{p_{\sharp} b_{1,n}^{1/2},b_{2,n}^{1/2}\}$. Notice that $\mathcal{M}^*_{\bpsi}=\mathcal{I}\cup\mathcal{D}^*_{\bpsi}$ with $\mathcal{D}^*_{\bpsi}=\{j\in\mathcal{D}:|\bar{g}^{\calT}_{j}(\bpsi)|\ge C_*\nu\rho'_2(0^+)\}$ for any $\bpsi\in\bPsi$, where $C_*\in(0,1)$ is a prescribed constant. Recall $\mathcal{D}_{\bpsi}({c})=\{j\in\mathcal{D}:|\bar g_j^{\calT}(\bpsi)|\ge {c}\nu\rho'_2(0^+)\}$ for any $c\in(C_*,1)$ and $\mathcal{M}_{\bpsi}({c})=\mathcal{I}\cup\mathcal{D}_{\bpsi}({c})$.  In this section, we redefine
\begin{align}\label{eq:newsn}
&S_n(\bpsi)=\max_{\blambda\in\hat\Lambda_{n}^{\calT}(\bpsi)}f(\blambda;\bpsi)+\sum_{k\in\mathcal{D}}P_{1,\pi}(|\xi_k|)+\sum_{l\in\mathcal{P}}P_{1,\pi}(|\theta_l|)
\end{align}
for any $\blambda=(\lambda_1,\ldots,\lambda_r)^\T$ and $\bpsi=(\theta_1,\ldots,\theta_p,\xi_1,\ldots,\xi_{r_2})^\T\in\bPsi$, where $f(\blambda;\bpsi)$ is defined as \eqref{eq:flampsi} in Section \ref{se:b1}.
In comparison to $S_n(\bpsi)$ defined in Section \ref{se:b1} for a low-dimensional $\btheta$, the newly defined $S_n(\bpsi)$ here for a high-dimensional $\btheta$ has an extra term $\sum_{l\in\mathcal{P}}P_{1,\pi}(|\theta_l|)$ which is caused by the penalty imposed on $\btheta$. Write $\bxi_0=(\xi_{0,1},\ldots,\xi_{0,r_2})^\T$ and $\btheta_0=(\theta_{0,1},\ldots,\theta_{0,p})^\T$. Recall $\mathcal{P}_{\sharp}=\{k\in\mathcal{P}:\theta_{0,k}\neq0\}$. Then $\bpsi_{\calS}=(\btheta_{\calp_{\sharp}}^\T,\bxi_{\cala^{\c}}^\T)^\T$ and $\bpsi_{0,\calS^{\c}}=\bzero$. Similar to that in Section \ref{se:b1}, we define
$\bPsi_{*}=\{\bpsi\in\bPsi:|\bpsi_{\calS}-\bpsi_{0,\calS}|_\infty\le\varepsilon,|\bpsi_{\calS^{\c}}|_1\le\aleph_n\}$
for some fixed $\varepsilon>0$. Consider
\begin{align}\label{eq:locest2}
\hat{\bpsi}=\arg\min_{\bpsi\in\bPsi_{*}}S_n(\bpsi)\,.
\end{align}
Analogously to Proposition \ref{thm1}, here Proposition \ref{sthm1} shows that such defined $\hat{\bpsi}$ is a sparse local minimizer for the nonconvex optimization \eqref{eq:est22}.

\begin{proposition} \label{sthm1}
Let $P_{1,\pi}(\cdot),P_{2,\nu}(\cdot)\in\mathscr{P}$ for $\mathscr{P}$ defined as {\rm(\ref{eq:classp})}, and $P_{2,\nu}(\cdot)$ be convex with bounded second derivative around $0$. Let $b_{1,n}=\max\{a_n,r_1\aleph_n^2\}$ with $a_n=\sum_{k\in\mathcal{D}}P_{1,\pi}(|\xi_{0,k}|)+\sum_{l\in\mathcal{P}}P_{1,\pi}(|\theta_{0,l}|)$, $b_{2,n}=\max\{b_{1,n},\nu^2\}$, and $\phi_n=\max\{p_{\sharp}b_{1,n}^{1/2},b_{2,n}^{1/2} \}$.                                                                                                                                                                                                 
For $\hat\bpsi$ defined as {\rm\eqref{eq:locest2}}, assume that there exists a constant $\tilde{c}\in(C_*,1)$ such that $\mathbb{P}[\cup_{j\in\mathcal{T}}\{|\bar{g}_j^\calT(\hat\bpsi)|\in[\tilde{c}\nu\rho'_2(0^+),\nu\rho'_2(0^+))\}]\rightarrow0$. Under Conditions {\rm \ref{A.1h}}, {\rm \ref{A.2}--\ref{A.3}} and \eqref{eq:signal:2}, if $\log r=o(n^{1/3})$, $s^2\ell_n\phi_n^2=o(1)$, $b_{2,n}=o(n^{-2/\gamma})$, and $\ell_n\aleph_n=o(\min\{\nu,\pi\})$, then w.p.a.1 such defined
$\hat{\bpsi}=(\hat{\btheta}^\T,\hat\bxi^\T)^\T$ provides a local minimizer for the nonconvex optimization {\rm\eqref{eq:est22}} such that {\rm(i)} $|\hat{\btheta}_{\calp_{\sharp}}-\btheta_{0,\calp_{\sharp}}|_\infty=O_\p(b_{1,n}^{1/2})$, {\rm(ii)} $\mathbb{P}(\hat{\btheta}_{\calp^{\c}_{\sharp}}=\bzero)\to 1$ as $n\to\infty$, {\rm(iii)} $|\hat{\bxi}_{\cala^{\c}}-\bxi_{0,\cala^{\c}}|_\infty=O_\p(\phi_n)$, and {\rm(iv)} $\mathbb{P}(\hat{\bxi}_{\cala}=\bzero)\to 1$ as $n\to\infty$.
\end{proposition}

Since $f(\blambda;\bpsi)$ involved here for high-dimensional $\btheta$ is identical to that used in Section \ref{se:b1} for low-dimensional $\btheta$, Lemmas \ref{prop3} and \ref{lem11} still hold in the the current setting. With the newly defined $\mathcal{S}$ for high-dimensional $\btheta$, Lemma \ref{lem1} also holds in the current setting. The proof of Proposition \ref{sthm1} is almost identical to that of Proposition \ref{thm1}. Using the same arguments as those in the proof of Proposition \ref{thm1}, we obtain $\max_{\blambda\in\hat\Lambda_{n}^{\calT}(\bpsi_0)}f(\blambda;\bpsi_0)=O_{\p}(r_1\aleph_n^2)$. Recall that $b_{1,n}=\max\{a_n,r_1\aleph_n^2\}$ and $a_n=\sum_{k\in\mathcal{D}}P_{1,\pi}(|\xi_{0,k}|)+\sum_{l\in\mathcal{P}}P_{1,\pi}(|\theta_{0,l}|)$. Then $S_n(\bpsi_0)=O_\p(r_1\aleph_n^2)+a_n=O_\p(b_{1,n})$. Notice that $\hat{\bpsi}=\arg\min_{\bpsi\in\bPsi_*}S_n(\bpsi)$ with $\bPsi_*=\{\bpsi=(\bpsi_{\calS}^\T,\bpsi_{\calS^{\c}}^\T)^\T:|\bpsi_{\calS}-\bpsi_{0,\calS}|_\infty\le\varepsilon,|\bpsi_{\calS^{\c}}|_1\le \aleph_n  \}$, and $\bpsi_{0,\calS^{\c}}=\bzero$. We then have $\bpsi_0\in\bPsi_*$ which implies $S_n(\hat\bpsi)\le S_n(\bpsi_0)=O_\p(b_{1,n})$. We need to show $\hat\bpsi\in{\rm int}(\bPsi_*)$ w.p.a.1, which indicates that $\hat{\bpsi}$ is a local minimizer of $S_n(\bpsi)$. 

We now follow a slightly different line of proof: (i) to show that for any $\epsilon_n\to\infty$ satisfying $b_{1,n}\epsilon_n^{2}n^{2/\gamma}=o(1)$ and any $\bpsi=(\btheta^\T,\bxi^\T)^\T\in\bPsi_*$ satisfying $| \btheta_{\calp_{\sharp}}-\btheta_{0,\calp_{\sharp}} |_\infty>\epsilon_nb_{1,n}^{1/2}$, there exists a universal constant $K>0$ independent of $\bpsi$ such that $\mathbb{P}\{S_n(\bpsi)>Kb_{1,n}\epsilon_n^{2}\}\to 1$ as $n\to\infty$. Due to $b_{1,n}=o(n^{-2/\gamma})$, we can select an arbitrary slowly diverging $\epsilon_n$ satisfying $b_{1,n}\epsilon_n^{2}n^{2/\gamma}=o(1)$. Thus, we have $|\hat\btheta_{\calp_{\sharp}}-\btheta_{0,\calp_{\sharp}}|_\infty=O_\p(b_{1,n}^{1/2})$; (ii) to show that for any $\varepsilon_n\to\infty$ satisfying $b_{2,n} \varepsilon_n^2 n^{2/\gamma}=o(1)$ and $\bpsi=(\btheta^\T,\bxi_\cala^\T,\bxi_{\cala^{\c}}^\T)^\T\in\bPsi_*$ satisfying $|\btheta_{\calp_{\sharp}}-\btheta_{0,\calp_{\sharp}}|_\infty \leq O(\varepsilon_n^{1/2}b_{1,n}^{1/2})$ and $|\bxi_{\cala^{\c}}-\bxi_{0,\cala^{\c}}|_\infty>\varepsilon_n\phi_n$, 
there exists a universal constant $M>0$ independent of $\bpsi$ such that $\mathbb{P}\{S_n(\bpsi)>M b_{2,n}\varepsilon_n^{2}\}\to 1$ as $n\to\infty$. Since $|\hat{\btheta}_{\calp_{\sharp}}-\btheta_{0,\calp_\sharp}|_\infty\leq O(\varepsilon_n^{1/2}b_{1,n}^{1/2})$ w.p.a.1 and we can select an arbitrary slowly diverging $\varepsilon_n$ satisfying $b_{2,n} \varepsilon_n^2 n^{2/\gamma}=o(1)$, it holds that $|\hat\bxi_{\cala^{\c}}-\bxi_{0,\cala^{\c}}|_\infty=O_\p(\phi_n)$; (iii) to show that $\hat\bpsi_{\calS^{\c}}=\bzero$ w.p.a.1.

The proof of Part (i) is similar to that of Proposition 2 in \cite{ChangTangWu2018}. 
For any $\bpsi=(\btheta^\T,\bxi^\T)^\T\in\bPsi_*$ with $\btheta=(\btheta^{\T}_{\calp_{\sharp}},\btheta^{\T}_{\calp_{\sharp}^{\c}})^{\T}$ satisfying $|\btheta_{\calp_{\sharp}}-\btheta_{0,\calp_{\sharp}}|_\infty>\epsilon_nb_{1,n}^{1/2}$, take $\btheta^*=(\btheta_{\calp_{\sharp}}^{\T},\bzero^{\T})^\T$ and $j_0=\arg\max_{j\in\mathcal{I}}|\mathbb{E}\{g_{i,j}^{\calI}(\btheta^*)\}|$. Let $\mu_{j_0}=\mathbb{E}\{g_{i,j_0}^{\calI}(\btheta)\}$ and $\mu^*_{j_0}=\mathbb{E}\{g_{i,j_0}^{\calI}(\btheta^*)\}$. Select $\tilde\blambda=\delta b_{1,n}^{1/2}\epsilon_n\bfe_{j_0}$, where $\delta>0$ is a constant to be determined later, and $\bfe_{j_0}$ is an $r$-dimensional vector with the $j_0$-th component being $1$ and other components being $0$. Then $\tilde\blambda\in\hat\Lambda_{n}^{\calT}(\bpsi)$ w.p.a.1. For the newly defined $S_n(\bpsi)$ in \eqref{eq:newsn}, applying the identical arguments for proof of Part (i) in Section \ref{se:b1}, we still have 
\begin{align*}
\mathbb{P}\{S_n(\bpsi)\le Kb_{1,n}\epsilon_n^{2}\}
\le&~\mathbb{P}\bigg[\bar{g}^{\calI}_{j_0}(\btheta)-\mu_{j_0}\le b_{1,n}^{1/2}\epsilon_n\bigg\{\frac{K}{\delta}+ \frac{\delta}{n}\sum_{i=1}^n |g_{i,j_0}^{\calI}(\btheta)|^2\bigg\}-\mu_{j_0}\bigg]+o(1)\,.
\end{align*}
Condition \ref{A.1h} implies that $\mu^*_{j_0}\ge K'_1\epsilon_n b_{1,n}^{1/2}$ with $K'_1$ specified in Condition \ref{A.1h}, and $|\mu_{j_0}-\mu^*_{j_0}|\le K_3'|\btheta_{\calp_{\sharp}^c}|_1\le K_3'\aleph_n\leq K'_1\epsilon_n b_{1,n}^{1/2}/2$ for sufficiently large $n$, which implies $\mu_{j_0}\ge K'_1\epsilon_n b_{1,n}^{1/2}/2$ for sufficiently large $n$.
Using the same arguments stated in the proof of Proposition \ref{pn:0}, we have $\mathbb{P}\{S_n(\bpsi)>Kb_{1,n}\epsilon_n^{2}\}\to1$ as $n\rightarrow\infty$. The proof of Part (ii) and Part (iii) are almost identical to that of Proposition \ref{thm1}, except some small adjustments.
The first difference is for deriving the lower bound of $\mu_{j_0}$ appeared in the proof of Part (ii). Notice that $|\mathbb{E}\{g_{i,j_0}^{\calD}(\btheta)\}-\mathbb{E}\{g_{i,j_0}^{\calD}(\btheta_0)\}|\leq |\mathbb{E}\{\nabla_{\btheta}g_{i,j_0}^{\calD}(\dot{\btheta})\}|_\infty|\btheta-\btheta_0|_1\leq O(\varepsilon_n^{1/2}p_{\sharp}b_{1,n}^{1/2})+O(\varepsilon_n^{1/2}\aleph_n)=o(\varepsilon_n\phi_n)$. Hence, identical to \eqref{eq:muj0}, we still have $\mu_{j_0}\ge \varepsilon_n b_{2,n}^{1/2}/2$ when $n$ is sufficiently large. The second difference is \eqref{eq:diff}. In the current setting, it should be
\begin{align*}
\max_{\blambda\in\hat\Lambda_{n}^{\calT}(\hat\bpsi)}f(\blambda;\hat\bpsi)
&\le \max_{\blambda\in\hat\Lambda_{n}^{\calT}(\bpsi_0)}f(\blambda;\bpsi_0)+\sum_{k\in\mathcal{S}}P_{1,\pi}(|\psi_{0,k}|)-\sum_{k\in\mathcal{S}}P_{1,\pi}(|\hat\psi_k|)= O_\p(r_1\aleph_n^2)\,,\notag
\end{align*}
where $\hat\bpsi=(\hat\psi_1,\ldots,\hat\psi_{p+r_2})^\T$ and $\bpsi_0=(\psi_{0,1},\ldots,\psi_{0,p+r_2})^\T$.
 The third difference is that the index set $\mathcal{A}$ in \eqref{eq:diff2} should be replaced by $\mathcal{S}^{\c}$ due to the newly defined $S_n(\bpsi)$ in the current setting. Then \eqref{eq:sn1} changes to
\begin{align*}
    S_n(\hat\bpsi^*)\le&~S_n(\hat\bpsi)-\underbrace{\frac{1}{n}\sum_{i=1}^n\frac{\hat\blambda^{*,\T}\nabla_{\bpsi_{\calS^{\c}}}\bg_{i}^{\calT}(\check{\bpsi})}{1+\hat\blambda^{*,\T}\bg_{i}^{\calT}(\check{\bpsi})}\hat\bpsi_{\calS^{\c}}}_{{\rm I}}
    -\underbrace{\sum_{k\in\mathcal{S}^{\c}}P_{1,\pi}(|\hat\psi_k|)}_{{\rm II}}\,.
\end{align*}
The last difference appears in the upper bound of $|{\rm I}|$. Since $\bg_i^{\calT}(\bpsi)=\{\bg_i^{\calI}(\btheta)^\T,\bg_i^{\calD}(\btheta)^\T-\bxi^\T\}^\T$ and $\bpsi_{\calS^{\c}}=(\btheta_{\calp^{\c}_{\sharp}}^\T,\bxi_\cala^\T)^\T$, we now have
\begin{align*}
    |{\rm I}|=\bigg|\frac{1}{n}\sum_{i=1}^n\frac{\hat\blambda^{*,\T}\nabla_{\bpsi_{\calS^{\c}}}\bg_{i}^{\calT}(\check{\bpsi})}{1+\hat\blambda^{*,\T}\bg_{i}^{\calT}(\check{\bpsi})}\hat\bpsi_{\calS^{\c}}\bigg| & \le \ell_n^{1/2} |\hat\blambda^*|_2 |\hat\bpsi_{\calS^{\c}}|_1\{1+o_\p(1)\}\,,
\end{align*}
where the upper bound has an extra factor $\ell_n^{1/2}$ in comparison to \eqref{I}.
Due to $|\hat{\blambda}^*|_2=O_\p(\ell_n^{1/2}\aleph_n)$, it then holds $|{\rm I}|\le|\hat\bpsi_{\calS^{\c}}|_1\cdot O_\p(\ell_n\aleph_n)$. Also notice that
$
{\rm II}=\sum_{k\in \calS^{\c}}\pi\rho'_1(c_k|\hat\psi_k|;\pi)|\hat\psi_k|\ge C\pi|\hat\bpsi_{\calS^{\c}}|_1
$
for some $c_k\in(0,1)$. Then $\ell_n\aleph_n=o(\pi)$ is required for Proposition \ref{sthm1} rather than $\ell_n^{1/2}\aleph_n=o(\pi)$ required in Proposition \ref{thm1}.


Select $\hat{\bpsi}_{\PEL}$ as the sparse local minimizer given in Proposition \ref{sthm1}. Recall $\mathcal{R}_n=\mathcal{I}\cup \supp(\hat\blambda_{\mathcal{D}})$, $\mathcal{A}_*=\{j\in\mathcal{A}:\hat{\lambda}_j\neq 0\}$ and $\mathcal{A}_{*,\c}=\{j\in\mathcal{A}^{\c}:\hat{\lambda}_j\neq 0\}$. Notice that $\mathcal{S}_*=\mathcal{P}_{\sharp}\cup\mathcal{A}_{*,\c}$ and $\mathcal{S}=\mathcal{P}_{\sharp}\cup\mathcal{A}^{\c}$. Then $\mathcal{S}_*\subset\mathcal{S}$ and $s_*:=|\mathcal{S}_*|\le |\mathcal{S}|=s$.  For any $\bpsi\in\bPsi$, we have $\bpsi_{\calS_*}=(\btheta_{\calp_{\sharp}}^\T,\bxi^\T_{{\cala}_{*,\c}})^\T$. Under the conditions of Proposition \ref{sthm1}, the results of Lemmas \ref{lem3} and \ref{lemenvlope} hold with the newly defined $\hat\bpsi_\PEL$, $\mathcal{S}_*$, $\mathcal{R}_n$, $\ell_n$, $s$, $s_*$ and $\phi_n$. The proof of Theorem \ref{thm3} is almost identical to that of Theorem \ref{thm2} stated in Section \ref{se:b2}. We only point out the difference here. The first difference is the definition of $H_n(\bpsi,\blambda)$. In comparison to $H_n(\bpsi,\blambda)$ given in \eqref{eq:Hpsilam} for the low-dimensional $\btheta$, we define
 \[
 H_n(\bpsi,\blambda)=\frac{1}{n}\sum_{i=1}^n\log\{1+\blambda^\T\bg^{\calT}_{i}(\bpsi)\}+\sum_{k\in\mathcal{D}}P_{1,\pi}(|\xi_k|)+ \sum_{l\in\mathcal{P}}P_{1,\pi}(|\theta_l|) -\sum_{j\in\mathcal{D}}P_{2,\nu}(|\lambda_j|)
 \]
 in current high-dimensional setting. Following the same arguments stated in Section \ref{se:b2}, \eqref{7.7} still holds with $\hat\bfvarsigma_{\calS_*}=\{\sum_{k=1}^{p+r_2}\nabla_{\bpsi_{\calS_*}}P_{1,\pi}(|\psi_k|)\}|_{\bpsi=\hat\bpsi_{{\PEL}}}$. It follows from Proposition \ref{sthm1} that $\hat\bfvarsigma_{\calS_*}=\bzero$ w.p.a.1. Notice that Lemma \ref{lem2} still holds in the current setting. Identical to the arguments below \eqref{7.7}, it holds that $
n^{1/2}\balpha^\T\{\bJ^{\calT}_{\calR_n}\}^{1/2}(\hat\bpsi_{{\PEL},{\calS_*}}-\bpsi_{0,{\calS_*}}-\hat\bzeta_{\calR_n})\stackrel{d}{\rightarrow} \mathcal{N}(0,1)
$ as $n\to\infty$. Notice that $\hat{\bpsi}_{\PEL,\calS_*}=(\hat{\btheta}_{\PEL,\calp_{\sharp}}^\T,\hat{\bxi}_{\cala_{*,\c}}^\T)^\T$. Recall $\mathcal{R}_n=\mathcal{I}\cup\mathcal{A}_*\cup\mathcal{A}_{*,\c}$ and $\mathcal{I}^*=\mathcal{I}\cup\mathcal{A}_*$. We write $\{\bJ^{\calT}_{\calR_n}\}^{-1}$, $\mathbb{E}\{\nabla_{\bpsi_{\calS_*}}\bg^{\calT}_{i,\calR_n}(\bpsi_0)\}$ and $\{\bV_{\calR_n}^{\calT}(\bpsi_0)\}^{-1}$ with following blocks
\begin{align*}
&\{\bJ^{\calT}_{\calR_n}\}^{-1}=\left(\begin{array}{cc}
 ~[\{\bJ^{\calT}_{\calR_n}\}^{-1}]_{11}    &  [\{\bJ^{\calT}_{\calR_n}\}^{-1}]_{12} \\
  ~[\{\bJ^{\calT}_{\calR_n}\}^{-1}]_{21}    & [\{\bJ^{\calT}_{\calR_n}\}^{-1}]_{22}
\end{array}\right)\,,~~
\{\bV^{\calT}_{\calR_n}(\bpsi_0)\}^{-1}=\left(\begin{array}{cc}
  \bS_{11}  &  \bS_{12}\\
  \bS_{21}   & \bS_{22}
\end{array}\right)\,,\\
&~~~~~\mathbb{E}\{\nabla_{\bpsi_{\calS_*}}\bg^{\calT}_{i,\calR_n}(\bpsi_0)\}=\left(\begin{array}{cc}
 \mathbb{E}\{\nabla_{\btheta_{\calp_{\sharp}}}\bg^{\calT}_{i,\cali^*}(\btheta_0)\}    &  \bzero\\
 \mathbb{E}\{\nabla_{\btheta_{\calp_{\sharp}}}\bg^{\calD}_{i,\cala_{*,\c}}(\btheta_0)\}    & -\bI
\end{array}\right)=\left(\begin{array}{cc}
\bG_{\cali^*}     &  \bzero\\
\bG_{\cala_{*,\c}}     & -\bI
\end{array}\right)\,,
\end{align*}
where $ [\{\bJ^{\calT}_{\calR_n}\}^{-1}]_{11} $ is a $p_{\sharp}\times p_{\sharp}$ matrix, and $\bS_{11}$ is an $|\mathcal{I}^*|\times|\mathcal{I}^*|$ matrix.
Recall $\bV^{\calT}_{\cali^*}(\btheta_0)=\mathbb{E}\{\bg^\calT_{i,\cali^*}(\btheta_0)^{\otimes2}\}$ and $\bW^{\calT}_{\cali^*}=([\mathbb{E}\{\nabla_{\btheta_{\calp_{\sharp}}}\bg^{\calT}_{i,\cali^*}(\btheta_0)\}]^\T\{\bV^{\calT}_{\cali^*}(\btheta_0)\}^{-1/2})^{\otimes2}$.
Then
$
[\{\bJ^{\calT}_{\calR_n}\}^{-1}]_{11}=\{\bG_{\cali^*}^\T(\bS_{11}-\bS_{12}\bS_{22}^{-1}\bS_{21})\bG_{\cali^*}\}^{-1}=\{\bW^{\calT}_{\cali^*}\}^{-1}$.
By the same arguments of Theorem \ref{thm2}, we complete the proof of Theorem \ref{thm3}. $\hfill\Box$

\subsection{Proof of Theorem \ref{thm:clt}}

To prove Theorem \ref{thm:clt}, we first present the following lemma whose proof is given in Section \ref{se:pflem7}.

\begin{lemma}\label{thm:consistency}
Let $|\bpsi_{\calM}^* - \bpsi_{0,\calM}|_1 = O_\p(\varpi_{1,n})$ and $|\bpsi_{\calM^{\c}}^* - \bpsi_{0,\calM^{\c}}|_1 = O_\p(\varpi_{2,n})$ for some $\varpi_{1,n}\to0$ and $\varpi_{2,n}\to0$. Under Conditions {\rm\ref{A.2}}--{\rm\ref{A.3}} and {\rm\ref{A.10}}, if $n\varpi_{2,n}^2(\varsigma^2+\varpi_{1,n}^2+\varpi_{2,n}^2)=O(1)$, $m (\omega_n+\varpi_{1,n}+\varpi_{2,n})=o(1)$, $m n^{-1/2+1/\gamma} = o(1)$ and $\omega_n^2\log r=O(1)$, then $| \tilde\bpsi_{\calM} - \bpsi_{0,\calM} |_2=O_\p(m^{1/2}n^{-1/2})$.
\end{lemma}

Now we begin to prove Theorem \ref{thm:clt}. Let $\hat\blambda^* = \arg\max_{\blambda\in\tilde\Lambda_n(\tilde\bpsi_{\calM})}n^{-1}\sum_{i=1}^n\log\{1+\blambda^\T \bff_i^{\bA_n}(\tilde\bpsi_{\calM},\bpsi_{\calM^{\c}}^*)\}$. By the same arguments in the proof of Lemma \ref{thm:consistency} for bounding $\tilde{\blambda}$ there, we have $|\hat\blambda^*|_2=O_\p(m^{1/2}n^{-1/2})$. Identical to \eqref{eq:exp1}, it holds that
$$
\bzero = \{ \bD^*(\tilde\bpsi_{\calM}) \}^\T \{\bC^*(\tilde\bpsi_{\calM})\}^{-1} \bar{\bff}^{\bA_n}(\tilde\bpsi_{\calM},\bpsi_{0,\calM^{\c}}^*)\,,
$$
where
\begin{align*}
\bD^*(\tilde\bpsi_{\calM}) =\frac{1}{n} \sum_{i=1}^n \frac{\nabla_{\bpsi_{\calM}}{\bff_i^{\bA_n}(\tilde\bpsi_{\calM},\bpsi_{\calM^{\c}}^*)}}{1+\hat\blambda^{*,\T} \bff_i^{\bA_n}(\tilde\bpsi_{\calM},\bpsi_{\calM^{\c}}^*)}~~\textrm{and}~~\bC^*(\tilde\bpsi_{\calM}) =\frac{1}{n} \sum_{i=1}^n \frac{\{\bff_i^{\bA_n}(\tilde\bpsi_{\calM},\bpsi_{\calM^{\c}}^*)\}^{\otimes2}}{ \{1+c \hat\blambda^{*,\T} \bff_i^{\bA_n}(\tilde\bpsi_{\calM},\bpsi_{\calM^{\c}}^*)\}^{2}}
\end{align*} for some $|c| < 1$. Write $\widehat{\bV}_{\bff^{\bA_n}}(\tilde\bpsi_{\calM},\bpsi^*_{\calM^{\c}}) = n^{-1} \sum_{i=1}^n \{\bff_i^{\bA_n}(\tilde\bpsi_{\calM},\bpsi_{\calM^{\c}}^*)\}^{\otimes2}$. Similar to Lemma \ref{lem2}, 
$
\| \bC^*(\tilde\bpsi_{\calM}) - \widehat{\bV}_{\bff^{\bA_n}}(\tilde\bpsi_{\calM},\bpsi_{\calM^{\c}}^*) \|_2 = O_\p(mn^{-1/2+1/\gamma})$, and
$
| \{\bD^*(\tilde\bpsi_{\calM}) -   \nabla_{\bpsi_{\calM}}{\bar\bff^{\bA_n}(\tilde\bpsi_{\calM},\bpsi_{\calM^{\c}}^*)} \} \bz|_2 = | \bz |_2 \cdot O_\p(m^{3/2}n^{-1/2})$
holds uniformly over $\bz \in \mathbb{R}^m$. Let $\hat{\bJ}^* = [ \{ \nabla_{\bpsi_{\calM}} \bar{\bff}^{\bA_n}(\tilde\bpsi_{\calM},\bpsi_{\calM^{\c}}^*) \}^\T \{ \widehat{\bV}_{\bff^{\bA_n}}(\tilde\bpsi_{\calM},\bpsi_{\calM^{\c}}^*) \}^{-1/2} ]^{\otimes2} $.
For any $\balpha \in \mathbb{R}^m$, let $\bdelta = (\hat{\bJ}^*)^{-1/2}\balpha$, and it holds that \[
\bdelta^\T \{ \nabla_{\bpsi_{\calM}}\bar{\bff}^{\bA_n}(\tilde\bpsi_{\calM},\bpsi_{\calM^{\c}}^*)\}^\T \{\widehat{\bV}_{\bff^{\bA_n}}(\tilde\bpsi_{\calM},\bpsi^*_{\calM^{\c}})\}^{-1} \bar{\bff}^{\bA_n}(\tilde\bpsi_{\calM},\bpsi_{\calM^{\c}}^*) = O_\p(m^{3/2}n^{-1+1/\gamma}) + O_\p(m^2n^{-1})\,.
\]
Expanding $\bar{\bff}^{\bA_n}(\tilde\bpsi_{\calM},\bpsi_{\calM^{\c}}^*)$ near $\bpsi_{\calM} = \bpsi_{0,\calM}$, we obtain
\begin{align}
&~\bdelta^\T \{ \nabla_{\bpsi_{\calM}}\bar{\bff}^{\bA_n}(\tilde\bpsi_{\calM},\bpsi_{\calM^{\c}}^*)\}^\T \{\widehat{\bV}_{\bff^{\bA_n}}(\tilde\bpsi_{\calM},\bpsi^*_{\calM^{\c}})\}^{-1} \{\nabla_{\bpsi_{\calM}}{\bar{\bff}^{\bA_n}(\check\bpsi_{\calM},\bpsi_{\calM^{\c}}^*)}\} (\tilde\bpsi_{\calM}-\bpsi_{0,\calM})\notag\\
= &- \bdelta^\T \{ \nabla_{\bpsi_{\calM}}\bar{\bff}^{\bA_n}(\tilde\bpsi_{\calM},\bpsi_{\calM^{\c}}^*)\}^\T \{\widehat{\bV}_{\bff^{\bA_n}}(\tilde\bpsi_{\calM},\bpsi^*_{\calM^{\c}})\}^{-1} \bar{\bff}^{\bA_n}(\bpsi_{0,\calM},\bpsi_{\calM^{\c}}^*)\label{eq:7.1}\\
 &+  O_\p(m^{3/2}n^{-1+1/\gamma}) + O_\p(m^2n^{-1})\,, \notag
\end{align}
where $\check\bpsi_{\calM}$ is on the line joining $\tilde\bpsi_{\calM}$ and $\bpsi_{0,\calM}$. By Condition \ref{A.4}, we have
$
 | \{\nabla_{\bpsi_{\calM}}{\bar{\bff}^{\bA_n}(\check\bpsi_{\calM},\bpsi_{\calM^{\c}}^*)} - \nabla_{\bpsi_{\calM}}{\bar{\bff}^{\bA_n}(\tilde\bpsi_{\calM},\bpsi_{\calM^{\c}}^*)}\} (\tilde\bpsi_{\calM}-\bpsi_{0,\calM})|_2 = O_\p(m^{5/2}n^{-1})$.
Moreover, by the proof of Theorem 1 in \cite{Chang2020}, if $m \omega_n^2 \log r = o(1)$ and $n m \varpi_{2,n}^2 (\varsigma^2 + \varpi_{1,n}^2 + \varpi_{2,n}^2) = o(1)$, we have
$
| \bar{\bff}^{\bA_n}(\bpsi_{0,\calM},\bpsi_{\calM^{\c}}^*) - \bar{\bff}^{\bA}(\bpsi_{0}) |_2 = o_\p(n^{-1/2})$.
Together with \eqref{eq:7.1}, it holds that
\begin{align*}
n^{1/2}\balpha^\T(\hat{\bJ}^*)^{1/2}(\tilde\bpsi_{\calM} - \bpsi_{0,\calM})=& - n^{1/2}\balpha^\T (\hat{\bJ}^*)^{-1/2} \{\nabla_{\bpsi_{\calM}}{\bar{\bff}^{\bA_n}(\tilde\bpsi_{\calM},\bpsi_{\calM^{\c}}^*)}\}^\T \{\widehat{\bV}_{\bff^{\bA_n}}(\tilde\bpsi_{\calM},\bpsi_{\calM^{\c}}^*)\}^{-1} \bar{\bff}^{\bA}(\bpsi_{0})\\
 & + O_\p(m^{3/2}n^{-1/2+1/\gamma}) + O_\p(m^{5/2}n^{-1/2})+ o_\p(1)\,.
\end{align*}
Let $\bJ = ([\mathbb{E}\{\nabla_{\bpsi_{\calM}} \bff^{\bA}_i(\bpsi_0) \}]^\T \{\bV_{\bff^{\bA}}(\bpsi_0)\}^{-1/2})^{\otimes2}$ with $\bV_{\bff^{\bA}}(\bpsi_0) = \mathbb{E}\{\bff^{\bA}_i(\bpsi_0)^{\otimes2} \}$. If $m^{5/2}n^{-1/2}=o(1)$ and $m^{3/2}(\omega_n+\varpi_{2,n})=o(1)$, then by
similar arguments in the proof of Lemma 4 of \cite{ChangTangWu2018}, we have \begin{align*}
n^{1/2}\balpha^\T(\hat{\bJ}^*)^{1/2}(\tilde\bpsi_{\calM} - \bpsi_{0,\calM})=&-n^{1/2}\balpha^\T \bJ^{-1/2} [\mathbb{E}\{\nabla_{\bpsi_{\calM}} \bff^{\bA}_i(\bpsi_0) \}]^\T \{\bV_{\bff^{\bA}}(\bpsi_0)\}^{-1} \bar{\bff}^{\bA}(\bpsi_{0})\\
&+ O_\p(m^{3/2}n^{-1/2+1/\gamma}) + O_\p(m^{5/2}n^{-1/2})\\
&+o_\p(1)+ O_\p\{m^{3/2}(\omega_n+\varpi_{2,n}) \}\\
\xrightarrow{d}&~\mathcal{N}(0,1)\,.
\end{align*} We complete the proof. $\hfill\Box$

\setcounter{lemma}{0}
\subsection{Proofs of auxiliary lemmas}

\subsubsection{Proof of Lemma \ref{prop3}}\label{sec:pflem1}
To simplify the notation, we write $\mathcal{M}_{\bpsi_n}({c})$ and $\mathcal{D}_{\bpsi_n}(c)$ as $\mathcal{M}_{\bpsi_n}$ and $\mathcal{D}_{\bpsi_n}$, respectively. Due to the convexity of $P_{2,\nu}(\cdot)$, we know that $f(\blambda;\bpsi_n)$ is a concave function w.r.t $\blambda$. We only need to show there exists a sparse local maximizer $\hat{\blambda}(\bpsi_n)$ satisfying the three results. By the definition of $\mathcal{M}_{\bpsi_n}$ and $\mathcal{M}_{\bpsi_n}^*$, we have $\mathcal{M}_{\bpsi_n}\subset\mathcal{M}_{\bpsi_n}^*$ which implies $|\mathcal{M}_{\bpsi_n}|\le m_n$ w.p.a.1. Notice that $m_n^{1/2}u_n n^{1/\gamma}=o(1)$. Given $\mathcal{M}_{\bpsi_n}$, we select $\delta_n$ satisfying $\delta_n=o(m_n^{-1/2}n^{-1/\gamma})$ and $u_n=o(\delta_n)$. Let $\bar\blambda_n=\arg\max_{\blambda\in\Lambda_n}f(\blambda;\bpsi_n)$ where $\Lambda_n=\{\blambda=(\blambda_{\calM_{\bpsi_n}}^\T,\blambda_{\calM^{\c}_{\bpsi_n}}^\T)^\T\in\mathbb{R}^r:|\blambda_{\calM_{\bpsi_n}}|_2\le\delta_n, \blambda_{\calM^{\c}_{\bpsi_n}}=\bzero\}$. It follows from $\max_{j\in\mathcal{T}}{n}^{-1}\sum_{i=1}^n|g^{\calT}_{i,j}(\bpsi_n)|^\gamma=O_\p(1)$ that $\max_{i\in[n]}|g^{\calT}_{i,j}(\bpsi_n)|=O_\p(n^{1/\gamma})$ holds uniformly over $j\in\mathcal{T}$, which implies that $\max_{i\in[n]}|
\bg_{i,\calM_{\bpsi_n}}^{\calT}(\bpsi_n)|_2=O_\p(m_n^{1/2}n^{1/\gamma})$. Thus, $\max_{i\in[n]}|\bar\blambda_n^\T\bg_{i}^{\calT}(\bpsi_n)|=o_\p(1)$. By the Taylor expansion, we have
\begin{equation}\label{eq:tyexp1}
\begin{split}
    0=f(\bzero;\bpsi_n)\le&~ f(\bar\blambda_n;\bpsi_n)\\
    =&~\frac{1}{n}\sum_{i=1}^n\bar\blambda_n^\T\bg_{i}^{\calT}(\bpsi_n)-\frac{1}{2n}\sum_{i=1}^{n}\frac{\bar\blambda_n^\T\bg_{i}^{\calT}(\bpsi_n)^{\otimes2}\bar\blambda_n}{\{1+\bar{c}\bar\blambda_n^\T\bg_{i}^{\calT}(\bpsi_n)\}^2}-\sum_{j\in\mathcal{D}}P_{2,\nu}(|\bar\lambda_{n,j}|)\,,
\end{split}
\end{equation}
where $\bar\blambda_n=(\bar\lambda_{n,1},\ldots,\bar\lambda_{n,r})^\T$ and $\bar{c}\in(0,1)$.  Recall $P_{2,\nu}(t)=\nu\rho_2(t;\nu)$. By the convexity of $P_{2,\nu}(\cdot)$, we have $\rho'_2(t;\nu)\ge\rho'_2(0^+)$ for any $t>0$. Notice that $\lambda_{\rm min}\{\widehat\bV^{\calT}_{\calM_{\bpsi_n}}(\bpsi_n)\}$ is uniformly bounded away from zero w.p.a.1, and $|\bar{\lambda}_{n,j}|\geq \bar{\lambda}_{n,j}\cdot\sgn\{\bar{g}_j^{\calT}(\bpsi_n)\}$. Thus, (\ref{eq:tyexp1}) leads to
\[
0\le\bar\blambda^\T_{n,\calM_{\bpsi_n}}\left( \begin{array}{c}\bar\bg^{\calI}(\bpsi_n)\\\bar\bg^{\calT}_{\cald_{\bpsi_n}}(\bpsi_n)-\nu\rho'_2(0^+)\sgn\{\bar\bg^{\calT}_{\cald_{\bpsi_n}}(\bpsi_n)\}
\end{array}\right)-C|\bar\blambda_{n,\calM_{\bpsi_n}}|_2^2\{1+o_\p(1)\}\,.
\]
Due to $|\bar\bg^{\calI}(\bpsi_n)|_2^2+|\bar{\bg}^{\calT}_{\cald_{\bpsi_n}}(\bpsi_n)-\nu\rho'_2(0^+)\sgn\{\bar\bg_{\cald_{\bpsi_n}}^{\calT}(\bpsi_n)\}|_2^2=O_\p(u_n^2)$, it holds that $|\bar\blambda_{n,\calM_{\bpsi_n}}|_2=O_\p(u_n)=o_\p(\delta_n)$. Recall $\mathcal{M}_{\bpsi_n}=\mathcal{I}\cup\mathcal{D}_{\bpsi_n}$. Then $|\bar\blambda_{n}|_2=|\bar\blambda_{n,\calM_{\bpsi_n}}|_2=O_\p(u_n)$ and $\{j\in\mathcal{D}:\bar{\lambda}_{n,j}\neq0\}\subset\mathcal{D}_{\bpsi_n}$. Write $\bar\blambda_{n,\cald_{\bpsi_n}}=(\bar\lambda_1,\ldots,\bar\lambda_{|\cald_{\bpsi_n}|})^\T$. We have w.p.a.1 that
\begin{equation}\label{7.4}
\bzero=\frac{1}{n}\sum_{i=1}^n\frac{\bg_{i,\cald_{\bpsi_n}}^{\calT}(\bpsi_n)}{1+\bar\blambda^\T_{n,\calM_{\bpsi_n}}\bg_{i,\calM_{\bpsi_n}}^{\calT}(\bpsi_n)}-\hat\bfeta\,,
\end{equation}
where $\hat\bfeta=(\hat\eta_1,\ldots,\hat\eta_{|\cald_{\bpsi_n}|})^\T$ with $\hat\eta_j=\nu\rho'_2(|\bar\lambda_j|;\nu)\sgn(\bar{\lambda}_j)$ for $\bar\lambda_j\neq0$ and $\hat\eta_j\in[-\nu\rho'_2(0^+),\nu\rho'_2(0^+)]$ for $\bar\lambda_j=0$. It follows from (\ref{7.4}) that $\hat\bfeta=\bar\bg_{\cald_{\bpsi_n}}^{\calT}(\bpsi_n)+\bR$ with
\[
\begin{split}
   |\bR|^2_\infty=&~\bigg|\frac{1}{n}\sum_{i=1}^n\frac{\bar\blambda^\T_{n,\calM_{\bpsi_n}}\bg_{i,\calM_{\bpsi_n}}^{\calT}(\bpsi_n)\bg_{i,\cald_{\bpsi_n}}^{\calT}(\bpsi_n)}{1+\bar\blambda^\T_{n,\calM_{\bpsi_n}}\bg_{i,\calM_{\bpsi_n}}^{\calT}(\bpsi_n)}\bigg|^2_\infty\\
   \leq&~\max_{j\in\cald_{\bpsi_n}}\bigg\{\frac{1}{n}\sum_{i=1}^n|\bar\blambda^\T_{n,\calM_{\bpsi_n}}\bg_{i,\calM_{\bpsi_n}}^{\calT}(\bpsi_n)||g_{i,j}^{\calT}(\bpsi_n)|\bigg\}^2\cdot\{1+o_\p(1)\}\\
   \le&~\max_{j\in\cald_{\bpsi_n}}\big\{\bar\blambda^\T_{n,\calM_{\bpsi_n}}\widehat{\bV}_{\calM_{\bpsi_n}}^{\calT}(\bpsi_n)\bar\blambda_{n,\calM_{\bpsi_n}}\big\}\bigg\{\frac{1}{n}\sum_{i=1}^n|g_{i,j}^{\calT}(\bpsi_n)|^2\bigg\}\cdot\{1+o_\p(1)\}\\
   =&~O_\p(|\bar\blambda_{n,\calM_{\bpsi_n}}|_2^2)\,,
\end{split}
\]
which indicates that $|\bR|_\infty=O_\p(u_n)=o_\p(\nu)$. Hence, w.p.a.1 we have $\sgn(\bar\lambda_{n,j})=\sgn\{\bar g^{\calT}_{j}(\bpsi_n)\}$ for any $j\in\mathcal{D}_{\bpsi_n}$ with $\bar\lambda_{n,j}\neq0$. To complete the proof, we need to show that $\bar\blambda_n$ is a local maximizer of $f(\blambda;\bpsi_n)$ w.p.a.1. Our proof includes two steps.

\underline{{\it Step 1.}} Define $\Lambda_n^*=\{\blambda=(\blambda^\T_{\calM^*_{\bpsi_n}},\blambda^\T_{\calM^{*,\c}_{\bpsi_n}})^\T\in\mathbb{R}^r:|\blambda_{\calM^*_{\bpsi_n}}|_2\le\varepsilon,\blambda_{\calM^{*,\c}_{\bpsi_n}}=\bzero\}$ for some sufficiently small $\varepsilon>0$, where $\mathcal{M}^*_{\bpsi_n}=\mathcal{I}\cup\mathcal{D}^*_{\bpsi_n}$ with $\mathcal{D}^*_{\bpsi_n}=\{j\in\mathcal{D}:|\bar{g}^{\calT}_{j}(\bpsi_n)|\ge C_*\nu\rho'_2(0^+)\}$ for some constant $C_*\in(0,1)$. For $\bar{\blambda}_n$ defined before, we will show in this step that $\bar\blambda_n=\arg\max_{\blambda\in\Lambda_n^*}f(\blambda;\bpsi_n)$ w.p.a.1. Due to $\bar{\blambda}_n\in\Lambda_n$ and $\mathcal{M}_{\bpsi_n}\subset\mathcal{M}_{\bpsi_n}^*$, we know $\bar{\blambda}_n\in\Lambda_n^*$ w.p.a.1. Restricted on $\blambda\in\Lambda_n^*$, by the concavity of $f(\blambda;\bpsi_n)$ w.r.t $\blambda_{\calM_{\bpsi_n}^*}$, it suffices to show that w.p.a.1 for any $j\in\mathcal{M}^*_{\bpsi_n}$ it holds that
\begin{equation}\label{7.41}
\frac{\partial f(\bar{\blambda}_n;\bpsi_n)}{\partial \lambda_j}
=0\,.
\end{equation}
Due to $\bar{\blambda}_n\in\Lambda_n$ and $|\bar{\blambda}_n|_2=o_\p(\delta_n)$, then $\bar{\blambda}_{n,\calM_{\bpsi_n}}$ is an interior point of the set $\{\blambda_{\calM_{\bpsi_n}}\in\mathbb{R}^{|\calM_{\bpsi_n}|}:|\blambda_{\calM_{\bpsi_n}}|_2\leq \delta_n\}$.  Restricted on $\blambda\in\Lambda_n$, we know $f(\blambda;\bpsi_n)$ is concave w.r.t $\blambda_{\calM_{\bpsi_n}}$. Notice that $\bar\blambda_n=\arg\max_{\blambda\in\Lambda_n}f(\blambda;\bpsi_n)$. Therefore, (\ref{7.41}) holds for any $j\in\mathcal{M}_{\bpsi_n}$. Recall $\bar{\blambda}_n=(\bar{\lambda}_{n,1},\ldots,\bar{\lambda}_{n,r})^\T$. For any $j\in\mathcal{M}_{\bpsi_n}^*\backslash\mathcal{M}_{\bpsi_n}$, we have $\bar{\lambda}_{n,j}=0$ and
\[
\frac{1}{n}\sum_{i=1}^n\frac{g_{i,j}^{\calT}(\bpsi_n)}{1+\bar\blambda^\T_{n,\calM^*_{\bpsi_n}}\bg_{i,\calM^*_{\bpsi_n}}^{\calT}(\bpsi_n)}=\bar g^{\calT}_{j}(\bpsi_n)+O_\p(u_n)\,,
\]
where the term $O_\p(u_n)=o_\p(\nu)$ is uniform over $j\in\mathcal{M}_{\bpsi_n}^*\backslash\mathcal{M}_{\bpsi_n}$. Such conclusion can be obtained by the same arguments for deriving the convergence rate of $|\bR|_\infty$ stated above. By the definition of $\mathcal{M}_{\bpsi_n}^*$ and $\mathcal{M}_{\bpsi_n}$, we know $\mathcal{M}_{\bpsi_n}^*\backslash\mathcal{M}_{\bpsi_n}=\mathcal{D}^*_{\bpsi_n}\backslash\mathcal{D}_{\bpsi_n}$. Then $C_*\nu\rho'_2(0^+)\le|\bar g^{\calT}_{j}(\bpsi_n)|<{c}\nu\rho'_2(0^+)$ for any $j\in\mathcal{M}_{\bpsi_n}^*\backslash\mathcal{M}_{\bpsi_n}$. Hence, we have w.p.a.1 that
\[
\bigg|\frac{1}{n}\sum_{i=1}^n\frac{g_{i,j}^{\calT}(\bpsi_n)}{1+\bar\blambda^\T_{n,\calM^*_{\bpsi_n}}\bg_{i,\calM^*_{\bpsi_n}}^{\calT}(\bpsi_n)}\bigg|\leq\nu\rho'_2(0^+)\,,
\]
which implies that there exists some $\hat\eta_j^*\in[-\nu\rho'_2(0^+),\nu\rho'_2(0^+)]$ such that
\[
0=\frac{1}{n}\sum_{i=1}^n\frac{g_{i,j}^{\calT}(\bpsi_n)}{1+\bar\blambda^\T_{n,\calM^*_{\bpsi_n}}\bg_{i,\calM^*_{\bpsi_n}}^{\calT}(\bpsi_n)}-\hat\eta_j^*\,.
\]
Hence, (\ref{7.41}) holds for any $j\in\mathcal{M}^*_{\bpsi_n}\backslash\mathcal{M}_{\bpsi_n}$.  Then we have $\bar\blambda_n=\arg\max_{\blambda\in\Lambda_n^*}f(\blambda;\bpsi_n)$ w.p.a.1.

\underline{{\it Step 2.}} Define $\tilde\Lambda_n=\{\blambda=(\blambda^\T_{\calM^*_{\bpsi_n}},\blambda^\T_{\calM^{*,\c}_{\bpsi_n}})^\T\in\mathbb{R}^r:|\blambda_{\calM^*_{\bpsi_n}}-\bar\blambda_{n,\calM^*_{\bpsi_n}}|_2\le o(u_n), |\blambda_{\calM^{*,\c}_{\bpsi_n}}|_1\leq \min\{O(m_n^{1/2}u_n), o(r_2^{-1/\gamma}n^{-1/\gamma})\}\}$. We will prove in this step that $\bar\blambda_n$ is the maximizer of $f(\blambda;\bpsi_n)$ over $\blambda\in\tilde\Lambda_n$.
Notice that $\max_{i\in[n],\blambda\in\tilde\Lambda_n}|\blambda^\T\bg_{i}^{\calT}(\bpsi_n)|=o_\p(1)$. For any $\blambda=(\lambda_1,\ldots,\lambda_r)^\T\in\tilde\Lambda_n$, denote by $\tilde\blambda=(\blambda^\T_{\calM^*_{\bpsi_n}},\bzero^\T)^\T$ the projection of $\blambda=(\blambda^\T_{\calM^*_{\bpsi_n}},\blambda^\T_{\calM^{*,\c}_{\bpsi_n}})^\T$ onto $\Lambda_n^*$ for $\Lambda_n^*$ defined in Step 1. Then it holds that
\[
\sup_{\blambda\in\tilde\Lambda_n}\{f(\blambda;\bpsi_n)-f(\tilde\blambda;\bpsi_n)\} =\sup_{\blambda\in\tilde\Lambda_n}\bigg\{\frac{1}{n}\sum_{i=1}^n\frac{\bg_{i}^{\calT}(\bpsi_n)^\T(\blambda-\tilde\blambda)}{1+\blambda_*^\T\bg_{i}^{\calT}(\bpsi_n)}-\sum_{j\in\calM^{*,\c}_{\bpsi_n}}P_{2,\nu}(|\lambda_j|)\bigg\}\,,
\]
where $\blambda_*$ is on the jointing line between $\blambda$ and $\tilde\blambda$. It follows from the Taylor expansion that
\begin{align*}
&~\frac{1}{n}\sum_{i=1}^n\frac{\bg_{i}^{\calT}(\bpsi_n)^\T(\blambda-\tilde\blambda)}{1+\blambda_*^\T\bg_{i}^{\calT}(\bpsi_n)}-\sum_{j\in\calM^{*,\c}_{\bpsi_n}}P_{2,\nu}(|\lambda_j|)\\
=&~\blambda_{\calM^{*,\c}_{\bpsi_n}}^\T\bar\bg^{\calT}_{\calM^{*,\c}_{\bpsi_n}}(\bpsi_n)-\bigg\{\frac{1}{n}\sum_{i=1}^n\blambda_*^\T\bg_{i}^{\calT}(\bpsi_n)\bg^{\calT}_{i,\calM^{*,\c}_{\bpsi_n}}(\bpsi_n)^\T\blambda_{\calM^{*,\c}_{\bpsi_n}}\bigg\}\{1+o_\p(1)\}
-\sum_{j\in\calM^{*,\c}_{\bpsi_n}}P_{2,\nu}(|\lambda_j|)\\
\le&~\big|\bar\bg^{\calT}_{\calM^{*,\c}_{\bpsi_n}}(\bpsi_n)\big|_\infty\big|\blambda_{\calM^{*,\c}_{\bpsi_n}}\big|_1+ \frac{1}{n}\sum_{i=1}^n\sum_{j\in\mathcal{T}}\sum_{k\in\calM^{*,\c}_{\bpsi_n}}|\lambda_{*,j}g^{\calT}_{i,j}(\bpsi_n)\lambda_kg^{\calT}_{i,k}(\bpsi_n)|\{1+o_\p(1)\}\\&~~~~~~~~~~~~~~~~~~~~~~~~~~~~~~~~
-\nu\rho'_2(0^+)\sum_{j\in\calM^{*,\c}_{\bpsi_n}}|\lambda_j|\\
\le&~C_*\nu\rho'_2(0^+)\sum_{j\in\calM^{*,\c}_{\bpsi_n}}|\lambda_j|+\max_{j\in\mathcal{T}}\bigg\{\frac{1}{n}\sum_{i=1}^n|g^{\calT}_{i,j}(\bpsi_n)|^2\bigg\}\bigg(\sum_{k\in\calM^{*,\c}_{\bpsi_n}}|\lambda_k|\bigg)|\blambda_*|_1 \{1+o_\p(1)\}\\
&~~~~~~~~~~~~~~~~~~~~~~~~~~~~~~~~-\nu\rho'_2(0^+)\sum_{j\in\calM^{*,\c}_{\bpsi_n}}|\lambda_j|\\
\le&~\big\{-(1-C_*)\nu\rho'_2(0^+)+O_\p(m_n^{1/2}u_n)\big\}\sum_{j\in\calM^{*,\c}_{\bpsi_n}}|\lambda_j|\,,
\end{align*}
where the term $O_\p(m_n^{1/2}u_n)$ holds uniformly over $\blambda\in\tilde{\Lambda}_n$. Since $m_n^{1/2}u_n=o(\nu)$, then $-(1-C_*)\nu\rho'_2(0^+)+O_\p(m_n^{1/2}u_n)<0$ w.p.a.1. Thus,
\[
\mathbb{P}\bigg[\sup_{\blambda\in\tilde\Lambda_n}\big\{f(\blambda;\bpsi_n)-f(\tilde\blambda;\bpsi_n)\big\}\leq0\bigg]\to1\,.
\]
Hence, $\bar\blambda_n$ is a local maximizer of $f(\blambda;\bpsi_n)$ w.p.a.1. We complete the proof of Lemma \ref{prop3}. $\hfill\Box$

\subsubsection{Proof of Lemma \ref{lem11}}\label{sec:pflem2}
To simplify the notation, we write $\mathcal{M}_{\bpsi_0}({c})$ and $\mathcal{D}_{\bpsi_0}(c)$ as $\mathcal{M}_{\bpsi_0}$ and $\mathcal{D}_{\bpsi_0}$, respectively. Recall  $\mathcal{M}_{\bpsi_0}^*=\mathcal{I}\cup\mathcal{D}_{\bpsi_0}^*$ with $|\mathcal{I}|=r_1$ and $\mathcal{D}_{\bpsi_0}^*=\{j\in\mathcal{D}:|\bar{g}_j^{\calT}(\bpsi_0)|\geq C_*\nu\rho_2'(0^+)\}$. Due to $\max_{j\in\mathcal{T}}n^{-1}\sum_{i=1}^n|g_{i,j}^{\calT}(\bpsi_0)|^2=O_\p(1)$, by the moderate deviation of self-normalized sums \citep{JingShaoWang2003}, it holds that
$|\bar{\bg}^{\calT}(\bpsi_0)|_{\infty}
=O_\p(\aleph_n)$ provided that $\log r=o(n^{1/3})$. Since $\nu\gg\aleph_n$, we know $\mathbb{P}(\mathcal{D}_{\bpsi_0}^*=\emptyset)\rightarrow1$ which implies $|\mathcal{M}_{\bpsi_0}^*|\leq 2r_1$ w.p.a.1. Pick $\delta_{n}$ satisfying  $\delta_{n}=o(r_1^{-1/2} n^{-1/\gamma})$ and $r_1^{1/2}\aleph_n=o(\delta_{n}) $ which can be guaranteed by $r_1\aleph_n=o(n^{-1/\gamma})$. Recall $\mathcal{M}_{\bpsi_0}=\mathcal{I}\cup\mathcal{D}_{\bpsi_0}$ with $\mathcal{D}_{\bpsi_0}=\{j\in\mathcal{D}:|\bar g_j^{\calT}(\bpsi_0)|\ge c\nu\rho'_2(0^+)\}$ for some $c\in(C_*,1)$. Then $|\mathcal{M}_{\bpsi_0}|\leq |\mathcal{M}_{\bpsi_0}^*|\leq 2r_1$ w.p.a.1. Let $\Lambda_{0}=\{\blambda=(\blambda^\T_{\calM_{\bpsi_0}},\blambda^\T_{\calM^{\c}_{\bpsi_0}})^\T \in \mathbb{R}^{r}:|\blambda_{\calM_{\bpsi_0}}|_{2}\leq \delta_{n}, \blambda_{\calM^{\c}_{\bpsi_0}}=\bzero \}$ and $\bar \blambda_{0}=\arg \max _{\blambda \in \Lambda_{0}} f(\blambda; \bpsi_0)$. Due to $\max_{i\in[n]}|g_{i,j}^{\calT}(\bpsi_0)|=O_\p(n^{1/\gamma})$ holds uniformly over $j\in\mathcal{T}$, we have $\max_{i\in[n]} |\bar \blambda^\T_{0} \bg^{\calT}_{i}(\bpsi_0)|=o_\p(1)$. Write $\bar \blambda_{0}=(\bar{\lambda}_{0,1},\ldots,\bar{\lambda}_{0,r})^\T$. Similar to (\ref{eq:tyexp1}), we have
\begin{equation*}
\begin{split}
0=f(\bzero;\bpsi_0) &\leq f(\bar \blambda_{0};\bpsi_0) \\
&=\frac{1}{n} \sum_{i=1}^{n} \bar \blambda^\T_{0} \bg^{\calT}_{i}(\bpsi_0)- \frac{1}{2n} \sum_{i=1}^{n} \frac{\bar \blambda^\T_{0} \bg^{\calT}_{i} (\bpsi_0)^{\otimes2} \bar \blambda_{0}}{\{1+\bar{c}\bar \blambda^\T_{0} \bg^{\calT}_{i} (\bpsi_0)\}^{2}} - \sum_{j\in\mathcal{D}} P_{2,\nu}(|\bar{\lambda}_{0,j}|)\\
&\leq  \bar \blambda_{0,\calM_{\bpsi_0}}^\T  \bar{\bg}^{\calT}_{\calM_{\bpsi_0}}(\bpsi_0) - \frac{1}{2} \lambda_{\min}\{\widehat{\bV}^{\calT}_{\calM_{\bpsi_0}} (\bpsi_0)\} |\bar \blambda_{0,\calM_{\bpsi_0}}|_{2}^{2}\{1+o_\p(1)\} \\
& \leq \bar \blambda_{0,\calM_{\bpsi_0}}^\T \bar{\bg}^{\calT}_{\calM_{\bpsi_0}}(\bpsi_0) - C |\bar \blambda_{0,\calM_{\bpsi_0}}|_{2}^{2}\{1+o_\p(1)\}
\end{split}
\end{equation*}
for some $\bar{c} \in (0,1)$.
Due to
$|\bar{\bg}^{\calT}(\bpsi_0)|_{\infty}
=O_\p(\aleph_n)$,
we have
$
|\bar{\bg}^{\calT}_{\calM_{\bpsi_0}}(\bpsi_0)|_{2}= O_\p(r_1^{1/2}\aleph_n)$.
Then $|\bar \blambda_{0,\calM_{\bpsi_0}}|_{2}=O_\p(r_1^{1/2}\aleph_n)=o_\p(\delta_{n})$. Using the same arguments to prove $\bar{\blambda}_n$ is a local maximizer of $f(\blambda;\bpsi_n)$ w.p.a.1 in the proof of Lemma \ref{prop3}, we can also show such defined $\bar{\blambda}_0$ is a local maximizer of $f(\blambda;\bpsi_0)$ w.p.a.1. Notice that $f(\blambda;\bpsi_0)$ is a concave function w.r.t $\blambda$. We complete the proof. $\hfill\Box$

\subsubsection{Proof of Lemma \ref{lemenvlope}}\label{sec:pflem5}


Recall
$
f(\blambda;\bpsi)=n^{-1}\sum_{i=1}^n\log\{1+\blambda^\T\bg^{\calT}_{i}(\bpsi)\}-\sum_{j\in\mathcal{D}}P_{2,\nu}(|\lambda_j|)
$
for any $\bpsi \in\bPsi$ and $\blambda = (\lambda_1, \ldots, \lambda_r)^\T$, and $\hat{\blambda}(\bpsi)=\arg\max_{\blambda\in\hat{\Lambda}_n^{\calT}(\bpsi)}f(\blambda;\bpsi)$. Then $\hat\bpsi_\PEL$ and its associated Lagrange multiplier $\hat\blambda(\hat\bpsi_\PEL)=(\hat{\lambda}_1,\ldots,\hat{\lambda}_r)^\T$ satisfy the score equation $\nabla_{\blambda} f\{\hat{\blambda}(\hat{\bpsi}_\PEL);\hat{\bpsi}_{\PEL}\}=\bzero$, that is,
\begin{align*}
\bzero=\frac{1}{n}\sum_{i=1}^n\frac{\bg_{i}^{\calT}(\hat\bpsi_{\PEL})}{1+\hat\blambda(\hat\bpsi_\PEL)^\T\bg_{i}^{\calT}(\hat\bpsi_{\PEL})}-\hat\bfeta\,,
\end{align*}
where $\hat\bfeta=(\hat\eta_1,\ldots,\hat\eta_r)^\T$ with $\hat\eta_j=0$ for $j\in\mathcal{I}$, $\hat\eta_j=\nu\rho'_2(|\hat\lambda_j|;\nu){\rm sgn}(\hat\lambda_j)$ for $j\in\mathcal{D}$ and $\hat\lambda_j\neq0$, and $\hat\eta_j\in[-\nu\rho'_2(0^+),\nu\rho'_2(0^+)]$ for $j\in\mathcal{D}$ and $\hat{\lambda}_j=0$. Recall $\mathcal{R}_n=\mathcal{I}\cup\supp\{\hat\blambda_{\mathcal{D}}(\hat\bpsi_\PEL)\}$. Restricted on $\mathcal{R}_n$, for any $\bpsi \in \mathbb{R}^{p+r_2}$ and $\bchi = (\chi_j)_{j\in\mathcal{R}_n}\in \mathbb{R}^{|\calR_n|}$ with $\chi_j\neq 0$ for any $j\notin\mathcal{I}$, define $\bm(\bchi,\bpsi) = n^{-1} \sum_{i=1}^n \bg^{\calT}_{i,\calR_n}(\bpsi) \{1+\bchi^\T\bg^{\calT}_{i,\calR_n}(\bpsi)\}^{-1} - \bw$, where $\bw = (w_j)_{j\in\mathcal{R}_n}$ with $w_j = 0$ for $j\in \mathcal{I}$ and $w_j = \nu\rho'_2(|\chi_j|;\nu){\sgn}(\chi_j)$ otherwise. Then $\hat{\blambda}_{\mathcal{R}_n}(\hat{\bpsi}_\PEL)$ and $\hat\bpsi_{\PEL}$ satisfy $\bm\{\hat{\blambda}_{\mathcal{R}_n}(\hat{\bpsi}_\PEL),\hat\bpsi_{\PEL}\} = \bzero$. By the implicit function theorem [Theorem 9.28 of \cite{Rudin1976}], for all $\bpsi$ in a $|\cdot|_2$-neighbourhood of $\hat\bpsi_{\PEL}$, there is a $\bchi(\bpsi)$ such that $\bm\{\bchi(\bpsi),\bpsi\}=\bzero$, $\bchi(\hat{\bpsi}_{\PEL})=\hat{\blambda}_{\mathcal{R}_n}(\hat{\bpsi}_{\PEL})$ and $\bchi(\bpsi)$ is continuously differentiable in $\bpsi$.

By Condition \ref{AA1}, we know the event $\mathcal{E}=\{\max_{j\in\mathcal{R}_n^{\c}}|\hat{\eta}_j|<\nu\rho_2'(0^+)\}$ holds w.p.a.1. Restricted on $\mathcal{E}$, let $\kappa_n=\nu\rho_2'(0^+)-\max_{j\in\mathcal{R}_n^{\c}}|\hat{\eta}_j|$. Define $\bPsi_{**}= \{\bpsi\in\mathbb{R}^{p+r_2}:|\bpsi-\hat\bpsi_\PEL|_1 \le o[\min\{\zeta_n,\kappa_n\}], |\bchi(\bpsi)-\bchi(\hat{\bpsi}_{\PEL})|_1\leq o(\kappa_n), |\bchi(\bpsi)-\bchi(\hat{\bpsi}_{\PEL})|_2\leq o(\ell_n^{-1/2}n^{-1/\gamma})\}$ for some $\zeta_n>0$.  Since $\chi_j(\hat{\bpsi}_{\PEL})\neq0$ for any $j\in\mathcal{R}_n\backslash\mathcal{I}$ and $\bchi(\bpsi)$ is continuously differentiable in $\hat{\bpsi}_{\PEL}$, we can select sufficiently small $\zeta_n$ such that $\chi_j(\bpsi)\neq0$ for any $\bpsi\in\bPsi_{**}$ and $j\in\mathcal{R}_n\backslash\mathcal{I}$.  For any $\bpsi\in\bPsi_{**}$, let $\tilde\blambda(\bpsi) \in \mathbb{R}^r$ satisfy $\tilde\blambda_{\mathcal{R}_n}(\bpsi) = \bchi(\bpsi)$ and $\tilde\blambda_{\mathcal{R}_n^{\c}}(\bpsi) = \bzero$. We will show that $\hat{\blambda}(\bpsi)=\tilde{\blambda}(\bpsi)$ for any $\bpsi \in \bPsi_{**}$ w.p.a.1. Restricted on $\mathcal{E}$, for any $j \in \mathcal{R}_n^{\c}$, we have
\begin{align*}
&\frac{1}{n} \sum _{i=1}^{n} \frac{g^{\calT}_{i,j}(\bpsi)}{1+\tilde{\blambda}(\bpsi)^{\T} \bg^{\calT}_{i}(\bpsi)} \\
&~~~~~~=\frac{1}{n} \sum _{i=1}^{n} \frac{g^{\calT}_{i,j}(\hat\bpsi_{\PEL})}{1+\tilde{\blambda}(\bpsi)^{\T} \bg^{\calT}_{i}(\hat\bpsi_{\PEL})}\\
&~~~~~~~~~~+
\bigg[\frac{1}{n} \sum_{i=1}^{n} \frac{\{\nabla_{\bpsi} g^{\calT}_{i,j}(\check{\bpsi})\}^{\T}}{1+\tilde{\blambda}(\bpsi)^{\T}\bg^{\calT}_{i}(\check{\bpsi})}-
\frac{1}{n} \sum_{i=1}^{n} \frac{g^{\calT}_{i,j}(\check{\bpsi})\tilde{\blambda}(\bpsi)^{\T}\nabla_{\bpsi}\bg^\calT_{i}(\check{\bpsi})}{\{1+\tilde{\blambda}(\bpsi)^{\T}\bg^{\calT}_{i}(\check{\bpsi})\}^2} \bigg] (\bpsi-\hat\bpsi_{\PEL} ) \\
&~~~~~~=\frac{1}{n} \sum _{i=1}^{n} \frac{g^{\calT}_{i,j}(\hat\bpsi_{\PEL})}{1+\hat{\blambda}(\hat\bpsi_\PEL)^{\T} \bg^{\calT}_{i}(\hat\bpsi_{\PEL})}+|\bpsi-\hat\bpsi_{\PEL}|_{1}\cdot O_\p(1) \\
&~~~~~~~~~~-\bigg[\frac{1}{n} \sum _{i=1}^{n} \frac{g^{\calT}_{i,j}(\hat\bpsi_{\PEL}) \bg^{\calT}_{i}(\hat\bpsi_{\PEL})^\T }{\{1+\check{\blambda}^{\T} \bg^{\calT}_{i}(\hat\bpsi_{\PEL})\}^2} \bigg] \{\tilde{\blambda}(\bpsi)-\hat{\blambda}(\hat\bpsi_\PEL)\} \\
&~~~~~~=\frac{1}{n} \sum _{i=1}^{n} \frac{g^{\calT}_{i,j}(\hat\bpsi_{\PEL})}{1+\hat{\blambda}(\hat\bpsi_\PEL)^{\T} \bg^{\calT}_{i}(\hat\bpsi_{\PEL})}+|\bchi(\bpsi)-\bchi(\hat\bpsi_\PEL)|_{1}\cdot O_\p(1) +|\bpsi-\hat\bpsi_{\PEL}|_{1}\cdot O_\p(1) \\
&~~~~~~=\hat{\eta}_j+\kappa_n\cdot o_\p(1) \,,
\end{align*}
where $\check{\bpsi}$ is on the line joining $\bpsi$ and $\hat\bpsi_{\PEL}$, $\check{\blambda}$ is on the line joining $\tilde\blambda(\bpsi)$ and $\hat{\blambda}(\hat\bpsi_\PEL)$, and the terms $O_\p(1)$ and $o_\p(1)$ hold uniformly over $j\in\mathcal{R}_n^c$. Notice that $\mathbb{P}(\mathcal{E})\rightarrow1$. Therefore, it holds w.p.a.1 that
\begin{align*}
\max_{j\in\mathcal{R}_n^{\c}}\bigg|\frac{1}{n} \sum _{i=1}^{n} \frac{g^{\calT}_{i,j}(\bpsi)}{1+\tilde{\blambda}(\bpsi)^{\T} \bg^{\calT}_{i}(\bpsi)}\bigg|\leq \nu\rho_2'(0^+)\,.
\end{align*}
On the other hand, since $\bm\{\bchi(\bpsi),\bpsi\}=\bzero$, $\tilde\blambda_{\mathcal{R}_n}(\bpsi) = \bchi(\bpsi)$ and $\tilde\blambda_{\mathcal{R}_n^{\c}}(\bpsi) = \bzero$ for any $\bpsi \in \bPsi_{**}$, then it holds that
\begin{align*}
0=\frac{1}{n}\sum_{i=1}^n\frac{g_{i,j}^{\calT}(\bpsi)}{1+\tilde{\blambda}(\bpsi)^\T\bg_{i}^{\calT}(\bpsi)}-\nu\rho_2'\{|\tilde{\lambda}_j(\bpsi)|;\nu\}{\sgn}\{\tilde{\lambda}_j(\bpsi)\}
\end{align*}
for any $j\in\mathcal{R}_n\setminus \mathcal{I}$, where $\tilde{\blambda}(\bpsi)=\{\tilde{\lambda}_1(\bpsi),\ldots,\tilde{\lambda}_r(\bpsi)\}^\T$, and $n^{-1}\sum_{i=1}^n g_{i,j}^{\calT}(\bpsi)\{1+\tilde{\blambda}(\bpsi)^\T\bg_{i}^{\calT}(\bpsi)\}^{-1}$ $ = 0$ for any $j\in\mathcal{I}$. By the concavity of $f(\blambda;\bpsi)$ with respect to $\blambda$, we know $\hat\blambda(\bpsi) = \tilde{\blambda}(\bpsi)$ for any $\bpsi\in\bPsi_{**}$ w.p.a.1. Therefore, $\hat\blambda(\bpsi)$ is continuously differentiable at $\hat{\bpsi}_{\PEL}$ and $\nabla_{\bpsi}\hat{\blambda}_{\mathcal{R}_n^{\c}}(\hat{\bpsi}_{\PEL})=\bzero$ w.p.a.1. We complete the proof. $\hfill\Box$

\subsubsection{Proof of Lemma \ref{thm:consistency}}\label{se:pflem7}

By the proof of Theorem 1 in \cite{Chang2020}, we have $|\bar{\bff}^{\bA_n}(\bpsi_{0,\calM},\bpsi_{\calM^{\c}}^*)|_2 = O_\p(m^{1/2}n^{-1/2})$ provided that $n\varpi_{2,n}^2(\varsigma^2+\varpi_{1,n}^2+\varpi_{2,n}^2)=O(1)$ and $\omega_n^2\log r=O(1)$. Next, we will specify the convergence rate of $|\bar{\bff}^{\bA_n}(\tilde\bpsi_{\calM},\bpsi_{\calM^{\c}}^*)|_2$. Define $B_n(\bpsi_{\calM},\blambda)=n^{-1}\sum_{i=1}^n\log\{1+\blambda^\T \bff^{\bA_n}_i(\bpsi_{\calM},\bpsi^*_{\calM^{\c}})\}$ for any $\bpsi_\calM\in\bPsi_\calM^*$ and $\blambda\in\tilde{\Lambda}_n(\bpsi_\calM)$. Let $\tilde{\blambda}=\arg\max_{\blambda\in\tilde{\Lambda}_n(\bpsi_{0,\calM})}B_n(\bpsi_{0,\calM},\blambda)$. Pick $\delta_n=o(m^{-1/2}n^{-1/\gamma})$ and $m^{1/2}n^{-1/2}=o(\delta_n)$. Let $\bar{\blambda}=\arg\min_{\blambda\in\Lambda_n}B_n(\bpsi_{0,\calM},\blambda)$ where $\Lambda_n=\{\blambda\in\mathbb{R}^m:|\blambda|_2\le\delta_n \}$. Conditions \ref{A.4} and \ref{A.10} imply that $\max_{i\in[n]} |\bff_i^{\bA_n}(\bpsi_{0,\calM}, \bpsi_{\calM^{\c}}^*)|_2= O_\p(m^{1/2} n^{1/\gamma})$, which implies $\max_{i\in[n],\blambda\in\Lambda_n}|\blambda^\T\bff^{\bA_n}_i(\bpsi_{0,\calM},\bpsi^*_{\calM^{\c}})|=o_\p(1)$. Under Conditions \ref{A.3} and \ref{A.10}, if $mn^{-1/2}=o(1)$ and $m(\omega_n^2+\varpi_{2,n}^2) = o(1)$,  Lemma 4 of \cite{Chang2020} implies that the eigenvalues of $n^{-1}\sum_{i=1}^n \bff^{\bA_n}_i(\bpsi_{0,\calM}, \bpsi^*_{\calM^{\c}})^{\otimes2}$ are uniformly bounded away from zero and infinity w.p.a.1. By the Taylor expansion, we have
\begin{align*}
0=B_n(\bpsi_{0,\calM},\bzero)\le&~ B_n(\bpsi_{0,\calM},\bar{\blambda})\notag\\
=&~ \bar{\blambda}^\T \bar{\bff}^{\bA_n}(\bpsi_{0,\calM}, \bpsi^*_{\calM^{\c}})-\frac{1}{2n}\sum_{i=1}^n\frac{\bar{\blambda}^\T \bff^{\bA_n}_i(\bpsi_{0,\calM}, \bpsi^*_{\calM^{\c}})^{\otimes2} \bar{\blambda} }{ \{ 1+c\bar{\blambda}^\T \bff^{\bA_n}_i(\bpsi_{0,\calM}, \bpsi^*_{\calM^{\c}}) \}^2 }\\
\le&~ |\bar{\blambda}|_2|\bar{\bff}^{\bA_n}(\bpsi_{0,\calM}, \bpsi^*_{\calM^{\c}})|_2 - C |\bar{\blambda}|_2^2 \{1+o_\p(1)\} \notag
\end{align*}
for some $c\in(0,1)$. Since $|\bar{\bff}^{\bA_n}(\bpsi_{0,\calM},\bpsi_{\calM^{\c}}^*)|_2 = O_\p(m^{1/2}n^{-1/2})$, then $|\bar{\blambda}|_2=O_\p(m^{1/2}n^{-1/2})=o_\p(\delta_n)$ which implies $\bar{\blambda}\in{\rm int}(\Lambda_n)$ w.p.a.1. Due to $\Lambda_n \subset \tilde{\Lambda}_n(\bpsi_{0,\calM})$ w.p.a.1 and the concavity of $B_n(\bpsi_{0,\calM},\blambda)$, $\tilde{\blambda} = \bar{\blambda}$ w.p.a.1 and $\max_{\blambda\in\tilde{\Lambda}_n(\bpsi_{0,\calM})}B_n(\bpsi_{0,\calM},\blambda)=O_\p(mn^{-1})$. For $\delta_n$ specified above, let $\blambda^*=\delta_n \bar{\bff}^{\bA_n}(\tilde{\bpsi}_\calM, {\bpsi}_{\calM^{\c}}^* ) /| \bar{\bff}^{\bA_n}( \tilde{\bpsi}_\calM, {\bpsi}_{\calM^{\c}}^* ) |_2 $ and then $  \blambda^*\in\Lambda_n $. Similar to Lemma 4 of \cite{Chang2020}, under Conditions \ref{A.3} and \ref{A.10}, if $mn^{-1/2}=o(1)$ and $m(\omega_n^2+\varpi_{1,n}^2 +\varpi_{2,n}^2) = o(1)$, we have the eigenvalues of $n^{-1}\sum_{i=1}^n \bff^{\bA_n}_i(\tilde\bpsi_{\calM}, \bpsi^*_{\calM^{\c}})^{\otimes2}$ are uniformly bounded away from zero and infinity w.p.a.1. Applying the Taylor expansion again, it holds that
\begin{align*}
O_\p(mn^{-1}) =&~\max_{\blambda\in\tilde{\Lambda}_n(\bpsi_{0,\calM})}B_n(\bpsi_{0,\calM},\blambda)\geq B_n(\tilde{\bpsi}_\calM, \blambda^*) \\
=&~ \blambda^{*,\T} \bar{\bff}^{\bA_n}( \tilde{\bpsi}_\calM, {\bpsi}_{\calM^{\c}}^* ) - \frac{1}{2n}\sum_{i=1}^n\frac{\blambda^{*,\T} \bff^{\bA_n}_i(\tilde{\bpsi}_{\calM}, \bpsi^*_{\calM^{\c}})^{\otimes2} \blambda^* }{ \{ 1+c \blambda^{*,\T} \bff^{\bA_n}_i(\tilde{\bpsi}_\calM, \bpsi^*_{\calM^{\c}}) \}^2 } \\
\ge&~ \delta_n | \bar{\bff}^{\bA_n}( \tilde{\bpsi}_\calM, {\bpsi}_{\calM^{\c}}^* ) |_2 - C \delta_n^2 \{ 1+o_\p(1) \}
\end{align*}
for some $c\in(0,1)$, which implies that $| \bar{\bff}^{\bA_n}( \tilde{\bpsi}_\calM, {\bpsi}_{\calM^{\c}}^* ) |_2 = O_\p(\delta_n)$. Given any $\epsilon_n\to0$, let $\blambda^{**} = \epsilon_n  \bar{\bff}^{\bA_n}( \tilde{\bpsi}_\calM, {\bpsi}_{\calM^{\c}}^* )$. Repeating above arguments again, we have $\epsilon_n |\bar{\bff}^{\bA_n}( \tilde{\bpsi}_\calM, {\bpsi}_{\calM^{\c}}^* ) |_2^2 = O_\p(mn^{-1}) $. Notice that we can select an arbitrary slow $\epsilon_n\to0$. Then $|\bar{\bff}^{\bA_n}( \tilde{\bpsi}_\calM, {\bpsi}_{\calM^{\c}}^* ) |_2 = O_\p(m^{1/2}n^{-1/2}) $.

Notice that $|\bar{\bff}^{\bA_n}(\bpsi_{0,\calM},\bpsi_{\calM^{\c}}^*) - \bar{\bff}^{\bA_n}(\tilde\bpsi_{\calM},\bpsi_{\calM^{\c}}^*)|_2 \ge \lambda^{1/2}_{\min}[\{ \nabla_{\bpsi_\calM} \bar\bff^{\bA_n}(\check\bpsi_{\calM},\bpsi^*_{\calM^{\c}}) \}^{\T,\otimes2}] |\tilde\bpsi_{\calM}-\bpsi_{0,\calM}|_2$, where $\check{\bpsi}_\calM$ is on the jointing line between $\bpsi_{0,\calM}$ and $\tilde{\bpsi}_\calM$. 
Recall $\bA = (\ba_j)_{j\in\calM}^\T$ and let $\bGamma = (\bgamma_j)_{j\in\calM}^\T $. Then $\bA \mathbb{E}\{\nabla_{\bpsi} \bg_i^\calT(\bpsi_0)\} = \bGamma$.  Write $\mathcal{M}=\{j_1,\ldots,j_m\}$ and denote by $\bGamma_{\cdot,\calM}$ the columns of $\bGamma$ that are indexed in $\mathcal{M}$. Recall $\bgamma_{j_k}$ is a $(p+r_2)$-dimensional vector with its $j_k$-th component being $1$ and all other components being $0$.  Then $\bGamma_{\cdot,\calM}=\bI_m$. Hence, the eigenvalues of $[\mathbb{E}\{\nabla_{\bpsi_\calM}\bff_i^{\bA}(\bpsi_{0})\}]^{\T,\otimes2} =\bGamma_{\cdot,\calM}^{\T,\otimes2}$ are uniformly bounded away from zero. Analogously to Lemma \ref{lem3}, we have $|[\nabla_{\bpsi_\calM} \bar\bff^{\bA_n}(\check\bpsi_{\calM},\bpsi^*_{\calM^{\c}}) - \mathbb{E}\{\nabla_{\bpsi_\calM}\bff_i^{\bA}(\bpsi_{0})\}]\bz |_2 = |\bz|_2 \cdot[ O_\p(mn^{-1/2}) + O_\p\{ m(\omega_n+\varpi_{1,n}+\varpi_{2,n}) \} ]$ holds uniformly over $\bz\in\mathbb{R}^m$. Then, if $mn^{-1/2}=o(1)$ and $m(\omega_n+\varpi_{1,n}+\varpi_{2,n})=o(1)$, it holds that $\lambda_{\min}[\{ \nabla_{\bpsi_\calM} \bar\bff^{\bA_n}(\check\bpsi_{\calM},\bpsi^*_{\calM^{\c}}) \}^{\T,\otimes2}] \ge C$ w.p.a.1, which implies $|\tilde\bpsi_{\calM}-\bpsi_{0,\calM}|_2 = O_\p(m^{1/2}n^{-1/2})$. We complete the proof.
$\hfill\Box$


%

%
%
%
%
%
%

\setcounter{equation}{0}
\section{Additional numerical results}

Due to limitations of space, in the main text we report the simulation results of the linear
IV model only. In this section, we display additional results for the linear IV model,
and also experiment with a nonlinear model.

\subsection{Linear IV model}
\label{sec2}

Although economists are primarily interested in the effect of the endogenous variable,
the exogenous variables in $\mathbf{z}_i$ control other sources of heterogeneity. 
To render a full picture of the performance of the estimation, 
we report the simulation results of $\beta_{\bz,1}$ and $\beta_{\bz,2}$---the coefficients for the exogenous variables $z_1$ and $z_2$---from the linear IV model in the main text.

\begin{table}[htbp]
\small
\centering \caption{RMSEs of the point estimations for $(\beta_{\bz,1},\beta_{\bz,2})$}
\label{tab1r}
\begin{spacing}{1.4}
\begin{tabular}{llccc}\hline
$(n, d_w, s)$  &  Correlation:     & weak   & moderate & strong \\\hline
\multicolumn{5}{c}{Panel A: low-dimensional setting}  \\
(100, 50, 6)     & PEL   & 0.172  & 0.176    & 0.167  \\
                 & 2SLS  & 0.219  & 0.219    & 0.219  \\
(200, 100, 6)    & PEL   & 0.114  & 0.115    & 0.115  \\
                 & 2SLS  & 0.152  & 0.152    & 0.152  \\\hline
\multicolumn{5}{c}{Panel B: high-dimensional setting} \\
(100, 120, 6)    & PEL   & 0.175  & 0.183    & 0.177  \\
                 & 2SLS  & 0.245  & 0.245    & 0.245  \\
(200, 240, 6)    & PEL   & 0.124  & 0.125    & 0.122  \\
                 & 2SLS  & 0.199  & 0.199    & 0.199  \\
(100, 120, 8)    & PEL   & 0.207  & 0.211    & 0.207  \\
                 & 2SLS  & 0.300  & 0.300    & 0.300  \\
(200, 240, 12)   & PEL   & 0.114  & 0.113    & 0.113  \\
                 & 2SLS  & 0.188  & 0.188    & 0.188  \\
(100, 120, 13)   & PEL   & 0.220  & 0.239    & 0.213  \\
                 & 2SLS  & 0.293  & 0.293    & 0.293  \\
(200, 240, 17)   & PEL   & 0.130  & 0.133    & 0.126  \\
                 & 2SLS  & 0.199  & 0.199    & 0.199  \\\hline
\end{tabular}
\end{spacing}
\end{table}

Table \ref{tab1r} displays the RMSE of PEL  and 2SLS for $(\beta_{\bz,1},\beta_{\bz,2})$.
Here,
$$
\text{RMSE} =
\sqrt{ \frac{1}{S}\sum_{s=1}^S 
\{(\hat{\beta}_{\bz,1}^{(s)} - \beta_{\bz,1}) ^2
+  (\hat{\beta}_{\bz,2}^{(s)} - \beta_{\bz,2}) ^2 \} }\,,
$$ where $S$ denotes the number of repetitions and $(\hat{\beta}_{\bz,1}^{(s)},\hat{\beta}_{\bz,2}^{(s)})$ denotes the estimate of $(\beta_{\bz,1},\beta_{\bz,2})$ in the $s$-th repetition.
The RMSE of 2SLS is significantly larger than that of PEL, showing that the advantages of PEL extend to the estimation for the coefficients of the exogenous variables.

\begin{table}[htbp]
\small
\setlength\tabcolsep{3pt}
\centering
\caption{Coverage probabilities for the CIs of $\beta_{\bz,1}$ and $\beta_{\bz,2}$ by PPEL}
\label{tab2r}
\begin{spacing}{1.4}
\begin{tabular}{lllllllllllll}\hline
               & Correlation: & \multicolumn{3}{c}{weak}                                                 &                      & \multicolumn{3}{c}{moderate}                                             &                      & \multicolumn{3}{c}{strong}                                               \\ 
$(n, d_w, s)$     & Method                   & \multicolumn{1}{c}{90} & \multicolumn{1}{c}{95} & \multicolumn{1}{c}{99} & \multicolumn{1}{c}{} & \multicolumn{1}{c}{90} & \multicolumn{1}{c}{95} & \multicolumn{1}{c}{99} & \multicolumn{1}{c}{} & \multicolumn{1}{c}{90} & \multicolumn{1}{c}{95} & \multicolumn{1}{c}{99} \\ \hline
\multicolumn{13}{c}{Panel A: low-dimensional setting}                                                                                                                                                                                                                                                                    \\ 
(100, 50, 6)   & $\beta_{\bz,1}$                    & 0.892                  & 0.948                  & 0.992                  &                      & 0.888                  & 0.948                  & 0.992                  &                      & 0.890                  & 0.950                  & 0.992                  \\
               & $\beta_{\bz,2}$                    & 0.874                  & 0.930                  & 0.990                  &                      & 0.868                  & 0.926                  & 0.990                  &                      & 0.870                  & 0.928                  & 0.990                  \\
(200, 100, 6)  & $\beta_{\bz,1}$                    & 0.898                  & 0.940                  & 0.988                  &                      & 0.896                  & 0.936                  & 0.988                  &                      & 0.894                  & 0.938                  & 0.988                  \\
               & $\beta_{\bz,2}$                    & 0.880                  & 0.938                  & 0.984                  &                      & 0.880                  & 0.936                  & 0.986                  &                      & 0.882                  & 0.936                  & 0.986                  \\ \hline
\multicolumn{13}{c}{Panel B: high-dimensional setting}                                                                                                                                                                                                                                                                   \\ 
(100, 120, 6)  & $\beta_{\bz,1}$                    & 0.878                  & 0.940                  & 0.984                  &                      & 0.870                  & 0.942                  & 0.980                  &                      & 0.874                  & 0.932                  & 0.980                  \\
               & $\beta_{\bz,2}$                    & 0.892                  & 0.930                  & 0.978                  &                      & 0.884                  & 0.926                  & 0.972                  &                      & 0.886                  & 0.926                  & 0.970                  \\
(200, 240, 6)  & $\beta_{\bz,1}$                    & 0.896                  & 0.936                  & 0.976                  &                      & 0.894                  & 0.936                  & 0.976                  &                      & 0.892                  & 0.938                  & 0.976                  \\
               & $\beta_{\bz,2}$                    & 0.878                  & 0.934                  & 0.986                  &                      & 0.872                  & 0.934                  & 0.986                  &                      & 0.868                  & 0.936                  & 0.986                  \\
(100, 120, 8)  & $\beta_{\bz,1}$                    & 0.870                  & 0.924                  & 0.976                  &                      & 0.866                  & 0.926                  & 0.974                  &                      & 0.864                  & 0.928                  & 0.974                  \\
               & $\beta_{\bz,2}$                    & 0.854                  & 0.928                  & 0.976                  &                      & 0.850                  & 0.924                  & 0.960                  &                      & 0.834                  & 0.916                  & 0.964                  \\
(200, 240, 12) & $\beta_{\bz,1}$                    & 0.886                  & 0.954                  & 0.994                  &                      & 0.884                  & 0.952                  & 0.994                  &                      & 0.874                  & 0.942                  & 0.994                  \\
               & $\beta_{\bz,2}$                    & 0.892                  & 0.940                  & 0.980                  &                      & 0.892                  & 0.938                  & 0.980                  &                      & 0.886                  & 0.938                  & 0.980                  \\
(100, 120, 13) & $\beta_{\bz,1}$                    & 0.872                  & 0.934                  & 0.990                  &                      & 0.862                  & 0.918                  & 0.988                  &                      & 0.860                  & 0.918                  & 0.984                  \\
               & $\beta_{\bz,2}$                    & 0.864                  & 0.930                  & 0.982                  &                      & 0.842                  & 0.916                  & 0.974                  &                      & 0.848                  & 0.916                  & 0.976                  \\
(200, 240, 17) & $\beta_{\bz,1}$                    & 0.898                  & 0.944                  & 0.988                  &                      & 0.896                  & 0.944                  & 0.990                  &                      & 0.892                  & 0.942                  & 0.988                  \\
               & $\beta_{\bz,2}$                    & 0.908                  & 0.948                  & 0.976                  &                      & 0.902                  & 0.948                  & 0.974                  &                      & 0.900                  & 0.940                  & 0.974                 \\ \hline
\end{tabular}
\end{spacing}
\end{table}

\begin{figure}[htbp]
  \centering
  \includegraphics[width=0.8\textwidth]{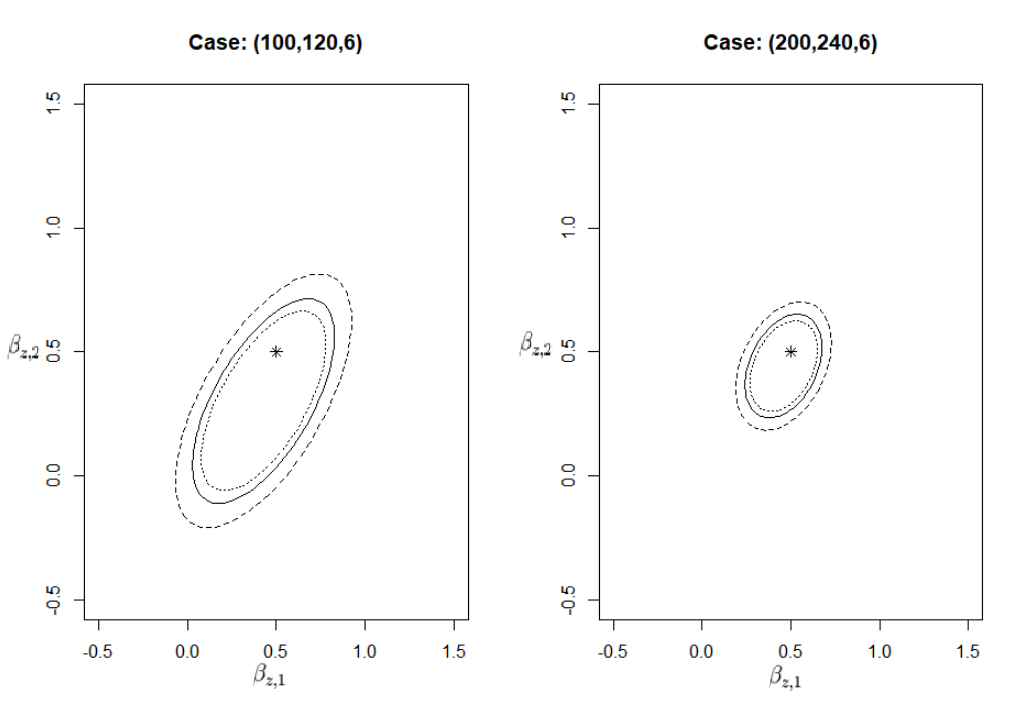}
  \caption{ 
  Confidence regions for $(\beta_{\bz,1}, \beta_{\bz,2})$ at levels 90\%, 95\% and 99\%}
  \label{fig1r}

\end{figure}

The coverage probabilities for the 
CIs of $\beta_{\bz,1}$ and $\beta_{\bz,2}$ are summarized in Table \ref{tab2r}.
It reveals the strength of PPEL for the inference of these coefficients.
Moreover, Figure \ref{fig1r} characterizes the shape of the confidence regions for the cases $(100, 120, 6)$ and $(200,240,6)$ with moderate correlation between $\epsilon_i$ and $\mathbf{w}_{3i}$.

\subsection{Dynamic panel data model}
\label{sec1}

Consider a simplified panel data model with time-varying individual
heterogeneity \citep{han2005estimation}:
\[
y_{i,j}=\lambda_{j}(\beta_{1})\alpha_{i}+\beta_{2}+e_{i,j}\,,
\]
where the zero mean error term $e_{i,j}$ may potentially correlate with the
individual-specific fixed effect $\alpha_{i}$,
and the nonlinear specification
$\lambda_{j}(\beta_{1})=2/\{1+\exp(\beta_{1}j)\}$, $j=0,1\ldots,h+s$,
is originated from \citet{kumbhakar1990production}.
Here we use $j$, instead of $t$, to  represent the panel's time dimension in order to
be consistent with the notations throughout the paper.

Notice that the fixed effect $\alpha_i$ can be canceled out by
\begin{align*}
e_{i,j} & =y_{i,j}-\lambda_{j}(\beta_{1})\alpha_{i}-\beta_{2} =y_{i,j}-\lambda_{j}(\beta_{1})(y_{i,0}-\beta_{2}-e_{i,0})-\beta_{2}  \\
& =y_{i,j}+\{\lambda_{j}(\beta_{1})-1\}\beta_{2}-\lambda_{j}(\beta_{1})y_{i,0}+\lambda_{j}(\beta_{1})e_{i,0}  = g_{i,j}(\btheta) +\lambda_{j}(\beta_{1})e_{i,0} \,,
\end{align*}
where $\btheta = (\beta_1, \beta_2)^\T$ and  $
g_{i,j}(\btheta)=y_{i,j}+\{\lambda_{j}(\beta_{1})-1\}\beta_{2}-\lambda_{j}(\beta_{1})y_{i,0}.
$
The above equation implies many moment conditions $\mathbb{E}\{g_{i,j}(\btheta)\}=\mathbb{E}\{e_{i,j}-\lambda_{j}(\beta_{1})e_{i,0}\}=0$ when $j$ varies.





The parameter of interest lies in $\beta_1 $, which determines the speed of decay of the individual-specific shock $\alpha_i$, whereas $\beta_2$ is an intercept.
In the simulation exercises,  the true parameter is set as
$\left(\beta_{1}^{0},\beta_{2}^{0}\right)=\left(0.5,-2\right)$.
Let $r=h+s$ and we specify $(n,r)=(100,120)$ and $(200,240)$,
$\alpha_i \sim \mathcal{N}(1,1) $,
and $e_{i,j} \sim 2^{-1/2} (\alpha_i - 1) + \mathcal{N}(0,1/2)$ i.i.d across time. All data across individuals are independent.
The number of known valid moments is fixed to be $r_1=5$, and then the number of validity-unknown moments is $r-r_1$.
After $h$ periods, there occurs a structural break in the mean where the intercept shifts from $\beta_2$ to $\beta_2 + \sigma$
and these corresponding moments become invalid at $j= h + 1, \ldots, h+ s$.
The number of invalid moments is $s=(6,8,13)$ and $(6,12,17)$ when $n=100$ and $200$, respectively, and
$\sigma = (0.2,0.3,0.4)$ for small, moderate, and large shifts, respectively.
The choice of the tuning parameters stays the same as that in the linear model.


Tables \ref{tab_nl_1}--\ref{tab_nl_3} are the counterparts of Tables \ref{tab1}--\ref{tab9} in the
high-dimensional setting.
In Table \ref{tab_nl_1}, correct moment selection improves with the sample size, and FN quickly goes to zero as $\sigma$ increases. As shown in Table \ref{tab_nl_2}, PEL offers reasonable estimation of the parameter while bias correction is difficult in this nonlinear model as well.
In Table \ref{tab_nl_3}, PPEL again exhibits more accurate coverage probability than DB-PEL.
Our proposed estimation and inference procedures are effective in terms of moment selection,
parameter estimation and coverage probability
in this nonlinear panel data model with many moments.

\begin{table}[htbp]
\small
\centering \caption{PEL's performance in moment selection}
\label{tab_nl_1}
\begin{spacing}{1.4}

\begin{tabular}{llllllllll} \hline
               & $\sigma$ & \multicolumn{2}{c}{0.2}                         &                      & \multicolumn{2}{c}{0.3}                         &                      & \multicolumn{2}{c}{0.4}                         \\
$(n, r, s) $    & Method                   & \multicolumn{1}{c}{FP} & \multicolumn{1}{c}{FN} & \multicolumn{1}{c}{} & \multicolumn{1}{c}{FP} & \multicolumn{1}{c}{FN} & \multicolumn{1}{c}{} & \multicolumn{1}{c}{FP} & \multicolumn{1}{c}{FN} \\ \hline
(100, 120, 6)  & PEL                      & 0.0874                 & 0.1600                 &                      & 0.0980                 & 0.0060                 &                      & 0.0984                 & 0.0000                 \\
               & DB-PEL                   & 0.0872                 & 0.1600                 &                      & 0.0980                 & 0.0060                 &                      & 0.0983                 & 0.0000                 \\
(200, 240, 6)  & PEL                      & 0.0168                 & 0.0607                 &                      & 0.0237                 & 0.0000                 &                      & 0.0166                 & 0.0000                 \\
               & DB-PEL                   & 0.0168                 & 0.0607                 &                      & 0.0237                 & 0.0000                 &                      & 0.0166                 & 0.0000                 \\ \hline
(100, 120, 8)  & PEL                      & 0.0986                 & 0.1218                 &                      & 0.0937                 & 0.0080                 &                      & 0.0914                 & 0.0000                 \\
               & DB-PEL                   & 0.0985                 & 0.1220                 &                      & 0.0936                 & 0.0080                 &                      & 0.0913                 & 0.0000                 \\
(200, 240, 12) & PEL                      & 0.0305                 & 0.0253                 &                      & 0.0303                 & 0.0000                 &                      & 0.0167                 & 0.0000                 \\
               & DB-PEL                   & 0.0305                 & 0.0253                 &                      & 0.0303                 & 0.0000                 &                      & 0.0167                 & 0.0000                 \\ \hline
(100, 120, 13) & PEL                      & 0.0988                 & 0.1292                 &                      & 0.1009                 & 0.0055                 &                      & 0.0926                 & 0.0000                 \\
               & DB-PEL                   & 0.0987                 & 0.1292                 &                      & 0.1007                 & 0.0055                 &                      & 0.0924                 & 0.0000                 \\
(200, 240, 17) & PEL                      & 0.0394                 & 0.0178                 &                      & 0.0305                 & 0.0000                 &                      & 0.0395                 & 0.0000                 \\
               & DB-PEL                   & 0.0394                 & 0.0178                 &                      & 0.0305                 & 0.0000                 &                      & 0.0395                 & 0.0000                \\ \hline
\end{tabular}

\end{spacing}
\end{table}

\begin{table}[htbp]
\setlength\tabcolsep{3pt}
\small
\centering
\caption{ Point estimations for $\beta_1$ }
\label{tab_nl_2}
\begin{spacing}{1.4}

\begin{tabular}{lllllllllllll} \hline
               & $\sigma$ & \multicolumn{3}{c}{0.2}                                                       &                      & \multicolumn{3}{c}{0.3}                                                       &                      & \multicolumn{3}{c}{0.4}                                                       \\
$(n, r, s)$     & Method                   & \multicolumn{1}{c}{RMSE} & \multicolumn{1}{c}{BIAS} & \multicolumn{1}{c}{STD} & \multicolumn{1}{c}{} & \multicolumn{1}{c}{RMSE} & \multicolumn{1}{c}{BIAS} & \multicolumn{1}{c}{STD} & \multicolumn{1}{c}{} & \multicolumn{1}{c}{RMSE} & \multicolumn{1}{c}{BIAS} & \multicolumn{1}{c}{STD} \\ \hline
(100, 120, 6)  & PEL                      & 0.053                    & 0.000                    & 0.053                   &                      & 0.053                    & 0.003                    & 0.053                   &                      & 0.053                    & 0.003                    & 0.053                   \\
               & DB-PEL                   & 0.062                    & 0.003                    & 0.062                   &                      & 0.058                    & 0.004                    & 0.058                   &                      & 0.058                    & 0.004                    & 0.058                   \\
(200, 240, 6)  & PEL                      & 0.038                    & 0.005                    & 0.038                   &                      & 0.039                    & 0.005                    & 0.038                   &                      & 0.039                    & 0.005                    & 0.038                   \\
               & DB-PEL                   & 0.042                    & 0.004                    & 0.041                   &                      & 0.043                    & 0.004                    & 0.042                   &                      & 0.042                    & 0.005                    & 0.041                   \\ \hline
(100, 120, 8)  & PEL                      & 0.053                    & 0.002                    & 0.053                   &                      & 0.053                    & 0.003                    & 0.053                   &                      & 0.053                    & 0.003                    & 0.053                   \\
               & DB-PEL                   & 0.058                    & 0.004                    & 0.058                   &                      & 0.058                    & 0.003                    & 0.058                   &                      & 0.058                    & 0.004                    & 0.058                   \\
(200, 240, 12) & PEL                      & 0.039                    & 0.005                    & 0.038                   &                      & 0.039                    & 0.005                    & 0.038                   &                      & 0.039                    & 0.005                    & 0.038                   \\
               & DB-PEL                   & 0.042                    & 0.004                    & 0.041                   &                      & 0.041                    & 0.005                    & 0.041                   &                      & 0.042                    & 0.004                    & 0.042                   \\ \hline
(100, 120, 13) & PEL                      & 0.053                    & 0.001                    & 0.053                   &                      & 0.053                    & 0.003                    & 0.053                   &                      & 0.053                    & 0.003                    & 0.053                   \\
               & DB-PEL                   & 0.056                    & 0.002                    & 0.056                   &                      & 0.056                    & 0.004                    & 0.056                   &                      & 0.056                    & 0.004                    & 0.056                   \\
(200, 240, 17) & PEL                      & 0.038                    & 0.005                    & 0.038                   &                      & 0.039                    & 0.005                    & 0.038                   &                      & 0.039                    & 0.005                    & 0.038                   \\
               & DB-PEL                   & 0.042                    & 0.004                    & 0.042                   &                      & 0.041                    & 0.004                    & 0.041                   &                      & 0.042                    & 0.004                    & 0.041                  \\ \hline
\end{tabular}

\end{spacing}
\bigskip

\setlength\tabcolsep{3pt}
\small
\centering
\caption{Coverage probabilities for the CIs of $\beta_1$}
\label{tab_nl_3}
\begin{spacing}{1.4}

\begin{tabular}{lllllllllllll} \hline
               & $\sigma$ & \multicolumn{3}{c}{0.2}                                                  &                      & \multicolumn{3}{c}{0.3}                                                  &                      & \multicolumn{3}{c}{0.4}                                                  \\
$(n, r, s)$     & Method                   & \multicolumn{1}{c}{90} & \multicolumn{1}{c}{95} & \multicolumn{1}{c}{99} & \multicolumn{1}{c}{} & \multicolumn{1}{c}{90} & \multicolumn{1}{c}{95} & \multicolumn{1}{c}{99} & \multicolumn{1}{c}{} & \multicolumn{1}{c}{90} & \multicolumn{1}{c}{95} & \multicolumn{1}{c}{99} \\ \hline
(100, 120, 6)  & PPEL                     & 0.894                  & 0.954                  & 0.994                  &                      & 0.882                  & 0.934                  & 0.986                  &                      & 0.914                  & 0.952                  & 0.990                  \\
               & DB-PEL                      & 0.836                  & 0.900                  & 0.974                  &                      & 0.862                  & 0.932                  & 0.988                  &                      & 0.844                  & 0.922                  & 0.986                  \\
(200, 240, 6)  & PPEL                     & 0.888                  & 0.930                  & 0.980                  &                      & 0.896                  & 0.952                  & 0.990                  &                      & 0.882                  & 0.926                  & 0.974                  \\
               & DB-PEL                      & 0.846                  & 0.904                  & 0.980                  &                      & 0.834                  & 0.906                  & 0.978                  &                      & 0.828                  & 0.904                  & 0.984                  \\ \hline
(100, 120, 8)  & PPEL                     & 0.886                  & 0.926                  & 0.986                  &                      & 0.902                  & 0.954                  & 0.980                  &                      & 0.908                  & 0.960                  & 0.984                  \\
               & DB-PEL                      & 0.868                  & 0.924                  & 0.978                  &                      & 0.848                  & 0.938                  & 0.974                  &                      & 0.834                  & 0.920                  & 0.982                  \\
(200, 240, 12) & PPEL                     & 0.894                  & 0.948                  & 0.986                  &                      & 0.896                  & 0.948                  & 0.990                  &                      & 0.898                  & 0.952                  & 0.978                  \\
               & DB-PEL                      & 0.846                  & 0.906                  & 0.980                  &                      & 0.848                  & 0.914                  & 0.978                  &                      & 0.834                  & 0.910                  & 0.974                  \\ \hline
(100, 120, 13) & PPEL                     & 0.906                  & 0.960                  & 0.984                  &                      & 0.896                  & 0.946                  & 0.978                  &                      & 0.876                  & 0.938                  & 0.984                  \\
               & DB-PEL                      & 0.878                  & 0.932                  & 0.982                  &                      & 0.870                  & 0.920                  & 0.984                  &                      & 0.864                  & 0.928                  & 0.980                  \\
(200, 240, 17) & PPEL                     & 0.926                  & 0.952                  & 0.990                  &                      & 0.894                  & 0.950                  & 0.992                  &                      & 0.912                  & 0.960                  & 0.990                  \\
               & DB-PEL                      & 0.844                  & 0.896                  & 0.982                  &                      & 0.844                  & 0.918                  & 0.980                  &                      & 0.838                  & 0.908                  & 0.978                 \\ \hline
\end{tabular}

\end{spacing}
\end{table}

\setcounter{equation}{0}
\section{Robustness of empirical application}

\begin{figure}[htbp]
  \centering
  \includegraphics[width=0.9\textwidth]{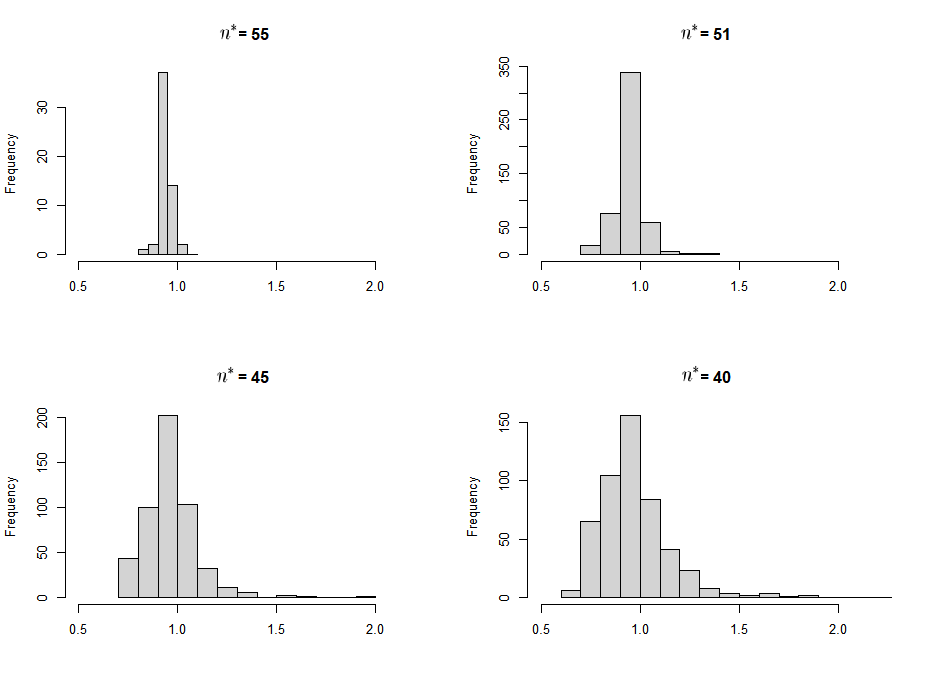}
  \includegraphics[width=0.7\textwidth]{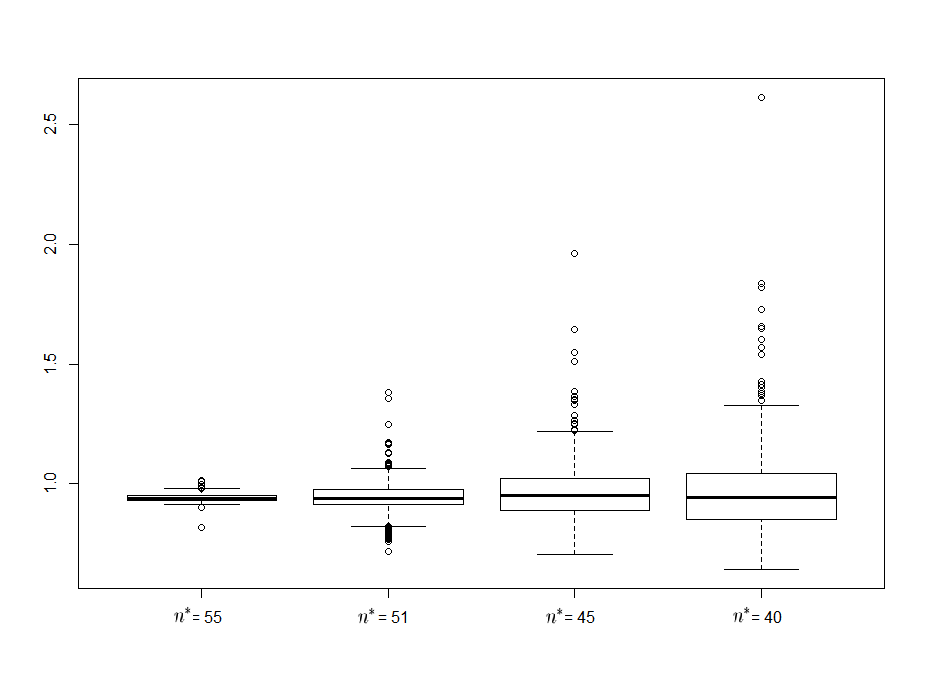}
  \caption{PEL estimation in bootstrap samples}
  \label{fig2r}
\end{figure}

To check the stability of PEL estimator in the real data analysis, we draw bootstrap samples from the original data and apply the proposed method to these sub-samples. 
Specifically, 
to witness the effect of sample sizes, we draw from the original sample with no replacement
bootstrap observations $n^* = 55, 51, 45$ and $40$,
which correspond to the sub-samples of size $(n-1)$, $\lceil 0.9n \rceil$, $\lceil 0.8n \rceil$ and $\lceil 0.7n \rceil$, respectively.
Given a sub-sample, the proposed PEL method is used to estimate the coefficient $\beta_x$.
We repeat the bootstrap exercise for 500 times. 
The results are summarized in Figure \ref{fig2r}. 
The point estimate from the true sample is 0.937, and the bootstrap point estimates are well centered around it. 
Regarding the histograms, despite a thin right tail when $n^*=40$, the dispersion quickly concentrates as $n^*$ increases.
The normal distribution appears to be an effective approximation when $n^*$ is above 50.

\end{spacing}

\end{document}